%% file: main.tex
\documentclass[preprint,12pt]{elsarticle}
\pdfoutput=1
\usepackage{graphicx}
\usepackage{caption}
\usepackage[labelformat=simple]{subcaption}

\usepackage{floatrow} % for caption beside the figure --- use the \floatbox and \capbeside commands provided by the floatrow package
\usepackage{amsmath}
\usepackage{amssymb}
\usepackage{amsthm}
\usepackage{setspace}
 \usepackage[normalem]{ulem} % strike through
\usepackage{cancel}
\usepackage{graphicx}
\usepackage{subcaption}
\usepackage{xspace}
\usepackage{tikz}
\usetikzlibrary{positioning}
\usetikzlibrary{arrows.meta, positioning, decorations.markings}
\usepackage{commath}
\usepackage{thm-restate}
\usepackage{pgfplots}
\usepackage{pgfplotstable}
\pgfplotsset{compat=1.7}

\usepackage[linesnumbered, ruled, vlined, noend]{algorithm2e}
\usepackage{algpseudocode}

\usepackage[]{color} 
\usepackage[]{url}
\usepackage[colorlinks]{hyperref}

\usepackage{tikz}
\usepackage{standalone}
\usetikzlibrary{arrows}
\usetikzlibrary{decorations} 
\usetikzlibrary{decorations.markings}
\usetikzlibrary{decorations,fit,shapes} 
\usetikzlibrary{arrows.meta,positioning}
\usetikzlibrary{shapes.misc}

\newtheorem{proposition}{Proposition}
\newtheorem{theorem}{Theorem}
\newtheorem{definition}[theorem]{Definition}
\newtheorem{property}{Property}
\newtheorem{lemma}{Lemma}
\newtheorem{corollary}{Corollary}
\newtheorem{remark}{Remark}
\newtheorem{obs}{Observation}
\newtheorem{claim}{Claim}

\usepackage{colortbl}
\newcommand{\boundellipse}[3]% center, xdim, ydim
{(#1) ellipse (#2 and #3)
}

\definecolor{cambridgeblue}{rgb}{0.64, 0.76, 0.68}
\definecolor{celadon}{rgb}{0.67, 0.88, 0.69}
\newcommand{\A}{\mathcal{A}\xspace}
\newcommand{\B}{\mathcal{P}\xspace}
\newcommand{\m}{\abs{\B}}%used in cost model
\newcommand{\correcting}{correcting\xspace}
\newcommand{\sfNP}{\textsf{NP}\xspace}
\newcommand{\RMMLQ}{{\sf RMM-LQ}\xspace}
\newcommand{\FAIRLQ}{{\sf FAIR-LQ}\xspace}

\newcommand{\CSMMD}{{\sf DOM-SIGN-MIN-MAX}\xspace}
\newcommand{\CSMTD}{{\sf DOM-SIGN-MIN-TOT}\xspace}

\newcommand{\OSMMD}{{\sf OPT-SIGN-MIN-MAX}\xspace}%{{\sf MIN-M-OS}\xspace}%
\newcommand{\OSMTD}{{\sf OPT-SIGN-MIN-TOT}\xspace}%{{\sf MIN-T-OS}\xspace}%
\newcommand{\RSMMD}{{\sf RMM-SIGN-MIN-MAX}\xspace} %{{\sf MIN-M-RS}\xspace}%
\newcommand{\RSMTD}{{\sf RMM-SIGN-MIN-TOT}\xspace} %{{\sf MIN-T-RS}\xspace}%
\newcommand{\FSMMD}{{\sf FAIR-SIGN-MIN-MAX}\xspace} %{{\sf MIN-M-FS}\xspace}%
\newcommand{\FSMTD}{{\sf FAIR-SIGN-MIN-TOT}\xspace} %{{\sf MIN-T-FS}\xspace}%
\newcommand{\OMMD}{{\sf OPT-MIN-MAX}\xspace}%{{\sf MIN-M-O}\xspace}%
\newcommand{\OMTD}{{\sf OPT-MIN-TOT}\xspace}%{{\sf MIN-T-O}\xspace}%
\newcommand{\RMMMD}{{\sf RMM-MIN-MAX}\xspace}%{{\sf MIN-M-R}\xspace}%
\newcommand{\RMMTD}{{\sf RMM-MIN-TOT}\xspace}%{{\sf MIN-T-R}\xspace}%
\newcommand{\FMMMD}{{\sf FAIR-MIN-MAX}\xspace}%{{\sf MIN-M-F}\xspace}%
\newcommand{\FMMTD}{{\sf FAIR-MIN-TOT}\xspace}%{{\sf MIN-T-F}\xspace}%
\newcommand{\RSMTM}{{\sf RMM-SIGN-MIN-TM}\xspace} 
\newcommand{\RSMTMD}{{\sf RMM-SIGN-MIN-$\textsf{T}_{\textsf{M}}$}\xspace}
\newcommand{\RSMMTD}{{\sf RMM-SIGN-MIN-$\textsf{M}_{\textsf{T}}$}\xspace}
\newcommand{\RMMTM}{{\sf RMM-MIN-TM}\xspace} 
\newcommand{\RMMTMD}{{\sf RMM-MIN-$\textsf{T}_{\textsf{M}}$}\xspace}
\newcommand{\RMMMTD}{{\sf RMM-MIN-$\textsf{M}_{\textsf{T}}$}\xspace}

\usepackage[capitalize,noabbrev]{cleveref}
\usepackage{nicematrix}
\usepackage{bm}
\usepackage{enumerate}
\usepackage{array}
\usepackage[framemethod=TikZ]{mdframed}
\usepackage[linguistics]{forest}
\usepackage{fancybox}
\usetikzlibrary{arrows, decorations.pathmorphing}
\newcommand \mydots{\hbox to 0.8em{.\hss .\hss .}}
\newcommand \myDots{\hbox to 0.8em{....}}

\tikzstyle{vertex}=[auto=left,circle,draw=black!80,fill=none,minimum size=15pt,inner sep=0pt]

\definecolor{lightgray}{rgb}{0.83, 0.83, 0.82}

\tikzset{cross/.style={cross out, draw=black, minimum size=2*(#1-\pgflinewidth), inner sep=0pt, outer sep=0pt},
	cross/.default={1pt}}
\usetikzlibrary{arrows, decorations.pathmorphing}

\tikzstyle{vertex}=[auto=left,circle,draw=black!80,fill=none,minimum size=15pt,inner sep=0pt]

\tikzset{
    photon/.style={decorate, decoration={snake}, draw=red}}

\newenvironment{appendix-lemma}[1]{\vspace{0.1in}\noindent{\bf Lemma~#1~} \em }{\vspace{0.1in}}
\newenvironment{appendix-theo}[1]{\vspace{0.1in}\noindent{\bf Theorem~#1~} \em }{\vspace{0.1in}}
\renewenvironment{proof}{{\textsc{Proof:}}}{\qed}
\newcommand{\proofofref}{}
\newproof{zproofof}{Proof of \proofofref}

\usepackage[margin = 0.8in]{geometry}

\title{Matchings under One-Sided Preferences with Soft Quotas\tnoteref{label1}}
\tnotetext[]{A preliminary version~\cite{santhinimatchings} of this work appeared in Thirty-Second International Joint Conference on Artificial Intelligence ({IJCAI} 2023).
Part of this work was done when the first author was a doctoral student and the second author was an undergraduate student at the Department of Computer Science and Engineering, IIT Madras}
\journal{arXiv}

\begin{document}

\begin{frontmatter}


\author[1]{Santhini K A}  
\ead{santhini@cse.iitm.ac.in}

% \ead[url]{http://www.cse.iitm.ac.in/~meghana}
\affiliation[1]{
% organization={Department of Computer Science},
            addressline={GPM Government College Manjeshwar},
            % city={Chennai},
            % postcode={600036}, 
            state={Kerala},
            country={India}}
    % \fntext[fn1]{Partially funded by SERB grant CRG/2019/004757.}
            
\author[2]{Raghu Raman Ravi}
\ead{raravi@student.ethz.ch}
% \ead[url]{http://www.cmi.ac.in/~prajakta/}

\affiliation[2]{
% organization={Department of CSE},
            addressline={ETH Zurich},
            % city={Chennai},
             % postcode={603103}, 
            % state={Tamilnadu},
            country={Switzerland}}
    % \fntext[fn2]{Partially funded by Infosys Grant and SERB grant CRG/2019/004757.}

\author[3]{Meghana Nasre}
\ead{meghana@cse.iitm.ac.in}
            
\affiliation[3]{
% organization={Department of CSE},
            addressline={Indian Institute of Technology Madras},
            % city={Chennai},
            % postcode={600036}, 
            state={Tamil Nadu},
            country={India}}

% \author[3,fn3]{Ankita Sarkar}
% \ead{ankita.sarkar.gr@dartmouth.edu}
% % \ead[url]{https://ankita.host.dartmouth.edu}
% \affiliation[3]{organization={Department of Computer Science},
%             addressline={Dartmouth College},
%             city={Hanover},
%             state={NH 03755},
%             country={USA}}
%     \fntext[fn3]{The work was partly done when the author was a Master's student at Chennai Mathematical Institute.}  
% \tnotetext[]{The first and fourth authors were partially supported by SERB grant CRG/2019/004757. The second author was partially supported by Infosys Grant and SERB grant CRG/2019/004757.}
% \author{}

% \affiliation{organization={},%Department and Organization

\begin{abstract}
Assigning applicants to posts in the presence of applicant preferences and quotas associated with posts is extensively investigated. For a post, a lower quota {\em guarantees}, and an upper quota {\em limits} the number of applicants assigned to it. Typically, quotas are assumed to be fixed, which need not be the case in practice. We address this by introducing a soft quota setting, in which every post is associated with two values – lower target and upper target - which together denote a range for the intended number of applicants in any assignment. Unlike the fixed quota setting, we allow the number of applicants assigned to a post to fall outside the range.  This leads to assignments with deviation. Here, we study the problem of computing an assignment that has two orthogonal optimization objectives – minimizing the deviation (maximum or total) with respect to soft quotas and ensuring optimality with respect to preferences of applicants (rank-maximality or fairness). The order in which these objectives are considered, the different possibilities to optimize deviation combined with the well-studied notions of optimality with respect to preferences open up a range of optimization problems of practical importance. We present efficient algorithms based on flow networks to solve these optimization problems. We also consider a generalization of the soft quota setting, where every deviation is associated with a cost. We show extensions of our algorithms for solving the problems in this setting.

\end{abstract}


\end{frontmatter}

\input{1-intro-R}

\input{3-Alg-target}

\input{3-deviation-first.tex}

\input{5-MIN-MAX}

\input{7-generalizedSQS}
\section{Discussion} \label{sec:dis}
In this work, we study the problem of computing optimal matchings with one-sided preferences in the soft quota setting. We consider two orthogonal optimization parameters, namely the deviation of the matching and preference optimality. For the preference optimality notions of rank-maximality and fairness, we present efficient  algorithms.  
Our algorithms rely on exponential weights on the edges of the flow network. It would be interesting to design algorithms that do not need large weights.

We also explore a natural generalization of the soft quota setting, where every deviation is associated with a cost. In this generalized soft quota setting, we present efficient algorithms for rank-maximality and fairness when the objective is to optimize the maximum deviation cost. For  minimizing total deviation cost, we provide an efficient algorithm for the case where no signature requirement is imposed. However, when  a signature requirement is imposed, we are not aware of any hardness or algorithmic results for  minimizing total deviation cost, leaving it as an open problem. 

%%%%%%%%%%%%%%%%%%%%%%%%%%%%%%%%%%%%%%%%%%%%%%%%%%%%%%%%%%%%%%%%%%%%%%%%
\section*{Acknowledgements} We thank Govind S Sankar for his valuable contributions and insightful discussions that greatly assisted  us in the initial stages of this work.
% We extend our gratitude to both Microsoft Research India and the IJCAI-AIJ 2023 travel grant committee for their generous contribution towards the partial funding of the travel grant.
\section*{Appendix}
\appendix
\costfm*
\begin{proof}
    Let $\sigma(M)=(x_1,x_2, \ldots,x_r,x_{r+1})$, and $\sigma(M')=(y_1,y_2, \ldots, y_r,y_{r+1})$. Let $i$ be the first rank at which $\sigma(M)$ and $\sigma(M')$ differ. Thus, $x_j=y_j$ for all $j>i$, and $x_i<y_i$.

   The cost difference between the matchings $M$ and $M'$ is 
   \[cost(M')-cost(M) = \sum\limits_{j=1}^{r+1} (y_j -x_j) (n+1)^{j} = (y_i-x_i)(n+1)^{i} + \sum\limits_{j=1}^{i-1} (y_j -x_j) (n+1)^{j}.\]
   Now, since $y_i - x_i \geq 1$, we have
   \begin{align*}
       cost(M')-cost(M) &\geq (n+1)^{i} + \sum\limits_{j=1}^{i-1} (y_j -x_j) (n+1)^{j}\\
       &\geq (n+1)^{i} - \sum\limits_{j=1}^{i-1} x_j (n+1)^{j}\\ 
       &\geq (n+1)^{i} - \left(\sum\limits_{j=1}^{i-1} x_j\right) (n+1)^{i-1}.
   \end{align*}
   Additionally, using the fact that $\sum\limits_{j=1}^{i-1} x_j \leq \sum\limits_{j=1}^{r+1} x_j = n$, we conclude that
   \[cost(M')-cost(M) \geq (n+1)^{i} - n (n+1)^{i-1} = (n+1)^{i-1} > 0.\]
\end{proof}

% \bibliography{refs}
 \bibliographystyle{elsarticle-num-names} 
 \bibliography{refs.bib}
% \appendix
% % \input{6-appendix}
% \input{supplementary}
\end{document}

%% file: 1-intro-R.tex
\section{Introduction}
\label{intro}
We study the many-to-one bipartite matching problem,
 where we have a bipartite graph $G=(\A \cup \B, E)$ of applicants $\A$ and posts $\B$, in the presence of preferences of applicants over the posts. 
 Typically, every post $p \in \B$ is associated with an input upper quota $q^+(p)$ that limits  the maximum number of applicants assigned to $p$ in any allocation.
Every applicant ranks all the posts in its neighbourhood, and this ranking can involve ties between the posts. We assume that there are no gaps in the preference list of an applicant. 
A matching $M$ is a collection of edges where an applicant has at most one edge incident to it in $M$, and a post $p$ has at most $q^+(p)$ many edges incident on it in $M$. In the literature, this setting is termed the  house allocation setting \cite{abraham2004pareto} or  the one-sided preference list setting, and can be used to model different problems such as assigning applicants to jobs,  students to courses~\cite{Bir2017ApplicationsOM}, 
papers to referees~\cite{DBLP:journals/algorithmica/GargKKMM10}, students to schools~\cite{abraham2009matching}, and  rental inventories to customers~\cite{abraham2006assignment}.

In certain applications, such as allocating students to courses or projects  \cite{biro2010college,DBLP:journals/algorithmica/ArulselvanCGMM18,10.5555/3535850.3535885}, a course may require a minimum number of students to be assigned in order to be operational. This can be readily captured by associating a lower quota with a course or post, which denotes the minimum number of applicants that need to be assigned to a post in any feasible allocation. 
That is, 
for a post $p$, in addition to the upper quota $q^+(p)$, a lower quota $q^-(p)$ is associated with it. 
For a matching $M$ and a post $p$, $M(p)$ denotes the set of applicants assigned to $p$ in $M$. In this setting, a matching $M$ is feasible
 if for every post $p$,  ${q^{-}(p) \le |M(p)| \le q^{+}(p)}$.
 We refer to this setting as the {\em fixed quota setting} throughout the paper. 
In the presence of lower quotas, there are simple instances  where no feasible matching exists. 
For example, consider the matching $M_1$ shown in Figure~\ref{fig:target}. In $M_1$,  the posts $p_3$ and $p_5$, marked with red circles, are not assigned any applicants and thus do not meet their lower quotas, and is  an infeasible matching in the fixed quota setting. 

\input{figures/fig_eg1}

In many practical applications, quotas need not be as rigid as assumed in the fixed quota setting, and relaxing lower quotas and augmenting upper quotas have been considered in the literature. Lower quotas have been  extensively investigated  in the two-sided preference list setting called the Hospital Residents problem \cite{DBLP:conf/stacs/GokoMMY22,goto2014strategy,hamada2016hospitals,fleiner2016matroid,monte2013matching,nasre2018popular,nasre2021popular,fragiadakis2016strategyproof}. Here,
 the lower quota of a hospital denotes the minimum number of residents (doctors)  needed for the smooth functioning of the hospital.
 As mentioned earlier, in the presence of lower quotas,  a feasible matching may not always exist. Therefore, relaxing these lower quotas has been considered as a solution.
For instance, instead of hospitals closing down due to the lack of a minimum number of doctors, they may simply reduce their service levels, as described in the work of \citet{DBLP:conf/stacs/GokoMMY22}. Similarly, in the context of school choice,
capacity planning to {\em augment} the number of available seats has recently been considered  in multiple works \cite{ijcai2022p1,DBLP:journals/corr/abs-2110-00734,https://doi.org/10.48550/arxiv.2205.01302}.
Motivated by such applications, we relax the fixed quota assumption by proposing and studying a {\em soft quota}
setting for the assignment problem in the one-sided preferences model.

\section{Soft Quota Setting}
An instance of the problem in the soft quota setting consists of a bipartite graph $G = (\A \cup \B, E)$ where applicants have preferences over posts and a post $p$ has two values -- a lower target $\ell(p)$ and an upper target $u(p)$. For a post $p$, the two values $\ell(p)$ and $u(p)$ define a range for the  intended number of  applicants for $p$ in any matching. 
There is no restriction on the number of applicants that can be assigned to a post in any matching. However, in order to control skew, we  define the  deviation of a post $p$ with respect to a matching $M$ as given below:
$$d_M(p) = \max \{0, \ell(p) - |M(p)|, |M(p)| - u(p)\}.$$ 
Our goal in the soft quota setting is to compute an assignment that has two orthogonal optimization objectives: 

\noindent{\bf (i) Minimizing deviation.}
For a matching $M$, we define the total deviation $d_{tot}(M)$ and maximum deviation  $ d_{max}(M)$ as below, and our goal is to minimize one of them.
$$d_{tot}(M)=\sum\limits_{p\in \B} d_M(p) \hspace{1.5 in} d_{max}(M)=\max_{p\in \B} \  d_M(p)$$ 

 \noindent{\bf (ii) Optimality with respect to preferences.} Two well-studied notions of optimality 
in the fixed quota setting  for the one-sided preference list model  are rank-maximality \cite{irving2006rank,paluch2013capacitated,DBLP:conf/aaai/Hosseini00S21,peters2022online}, and fairness \cite{huang2016fair}. A rank-maximal matching  matches the maximum number of applicants to rank $1$ posts, subject to that, the maximum number of applicants to rank $2$ posts, and so on. A fair matching minimizes the number of unmatched applicants, subject to that, minimizes the number of applicants matched to the last rank posts, and so on.  
The {\em signature} of a matching allows us to formalize these notions. Let $r$ be the maximum rank used by any applicant to order any post in its neighbourhood. The signature of a matching $M$, denoted as $\sigma(M)$, is an $r+1$ tuple $(x_1, x_2, \ldots,  x_{r+1})$ where $x_i$ denotes the number of applicants  matched to their rank $i$ post in $M$ for $i \in[r]$ and $x_{r+1}$ denotes the number of unmatched applicants in $M$.

\begin{definition}[Rank-maximality]
For two matchings $M$ and $M'$  with signatures 
\begin{align*}
  \sigma(M)=(x_1, x_2, \ldots, x_r, x_{r+1}) \text{ and }  \sigma(M')=(y_1, y_2, \ldots, y_r, y_{r+1}),
\end{align*}
 we say that $M \succ_{\mathcal{R}} M'$ ($M$ is more rank-maximal than $M'$)  if there exists an $i$ for $1 \leq i \leq r$ such that $x_{i}>y_{i}$ and, for $1 \leq j \leq i -1$, we have $x_j=y_j$.
A matching that has a maximal signature under the ordering $\succ_{\mathcal{R}}$ is a rank-maximal matching.
\end{definition}

\begin{definition}[Fairness]
For two matchings $M$ and $M'$ in $G$ with  signatures 
\begin{align*}
  \sigma(M)=(x_1, x_2, \ldots, x_r, x_{r+1}) \text{ and }  \sigma(M')=(y_1, y_2, \ldots, y_r, y_{r+1}),
\end{align*}
we say that $M\succ_{\mathcal{F}} M'$ ($M$ is fairer than $M'$) if there exists an $i$ for $1 <i \leq r+1 $ such that $x_{i}<y_{i}$ and, for $i+1 \leq j \leq r+1 $, we have $x_j=y_j$.
A matching that has a maximal signature under the ordering $\succ_{\mathcal{F}}$ is a fair matching. 
\end{definition}

  In the fixed quota setting, where the lower quota of every post is zero, optimal matchings under both these notions are guaranteed to exist and can be  efficiently computed using  the algorithms given in \citet{irving2006rank,paluch2013capacitated} and \citet{huang2016fair}. Even in the presence of lower quotas, an optimal matching among all  feasible matchings can be efficiently computed by  extending the algorithm  given in \citet{huang2016fair}.

By definition, rank-maximality and fairness  optimize the head and tail end of the signature, respectively, and one may perform poorly on the other criterion. Given an input signature $\rho$, a matching $M$ may have a signature that satisfies the optimality requirement with respect to rank-maximality, that is, $\sigma(M) \succeq_{\mathcal{R}} \rho$; however, this does not guarantee that $\sigma(M) \succeq_{\mathcal{F}} \rho$.   
A way of comparing two signatures that achieve a trade-off between these extreme notions of optimality is called {\em weak dominance}, and was introduced and studied in \citet{ka2022optimal,ASN25}. 

\begin{definition}[Weak dominance]
For two matchings $M$ and $M'$ in $G$ 
with  signatures 
\begin{align*}
  \sigma(M)=(x_1, x_2, \ldots, x_r, x_{r+1}) \text{ and }  \sigma(M')=(y_1, y_2, \ldots, y_r, y_{r+1}),
\end{align*}
we say that $M\succeq_{\mathcal{D}} M'$ ($M$ weakly dominates  $M'$)  if
 $$\sum\limits_{j=1}^{\ell} x_j \geq \sum\limits_{j=1}^{\ell}y_j \ \ \ \ \text{for all } \ell \in [r+1]. $$ 
\end{definition}

 In the fixed quota setting, 
 for  a given signature, deciding whether the instance admits a matching that weakly dominates the signature is {\sf NP-hard} \cite{ka2022optimal,ASN25}.
\subsection{Our Problems}
 In our soft quota setting, there is no restriction on the number of applicants that can be assigned to a post. Thus, we care about deviation as well as preference optimality and this defines our problems.

\noindent {\bf Deviation first, preference optimality next.}
There exist applications where having as low deviation as possible is required and, subject to that, optimality with respect to preferences needs to be respected. For example, in allocating projects to students \cite{cooper2020fair} or dividing political representatives to seats \cite{dhamala2009apportionment,pukelsheim2012network,serafini2012parametric} it is desirable to minimize the difference between input values and the assigned numbers. In a similar spirit, we consider minimizing the deviation from target values. Let $\mathcal{M}$ be the set of all matchings in the instance $G$ of the soft quota setting. 
Let $\mathcal{M}_t$ ($\mathcal{M}_m$) denote the set of matchings that have the minimum value of total deviation (respectively, the minimum value of maximum deviation). This is depicted in \cref{fig:omyd}. We define our first set of problems for the notions of rank-maximality and fairness:

\begin{itemize}
    \item \OMTD : The goal is to compute a matching that satisfies the notion OPT with respect to preferences in the set $\mathcal{M}_t$. For instance, if  the notion OPT is rank-maximality, then the problem \RMMTD outputs a rank-maximal matching in the set $\mathcal{M}_t$.
    \item \OMMD: The goal is to compute a matching that satisfies the notion OPT in the set $\mathcal{M}_m$.
\end{itemize}

\input{figures/fig3}

\noindent {\bf Preference optimality first, deviation next.} Analogously, there exist applications where preference optimality has to be respected first, and subject to that, we would like to minimize a deviation  parameter of interest. 
Since, in the soft quota setting, there is no bound on the upper quota of a post, it is not immediate how to determine the set of matchings in which we wish to compute a rank-maximal or fair matching. To address this, we first introduce
an additional input to our problem -- a signature requirement, $\rho$, for the output matching. We consider all matchings in  the soft quota setting that respect the signature requirement with respect to the preference optimality under consideration. Having a signature as a part of the input has been considered earlier in \citet{ka2022optimal,ASN25}, in the context of a different notion of preference optimality, namely \textit{weak dominance}.

For a fixed notion of preference optimality, a matching $M$ meets the signature requirement  if $\sigma(M) \succeq_{OPT} \rho$, and let $\mathcal{M}_{OPT}$ denote the set of all matchings that meet the signature requirement, as shown in \cref{fig:osmyd}.
Let $\mathcal{M}_t'$ ($\mathcal{M}_m'$) denote the set of matchings that have minimum value of total deviation (respectively, the minimum value of maximum  deviation)  in $\mathcal{M}_{OPT}$. 
We consider the following two problems for the notions of rank-maximality, fairness and weak dominance:
\begin{itemize}
    \item \OSMTD: The goal is to output any matching in $\mathcal{M}_t'$.
    \item \OSMMD:  The goal is to output any matching in $\mathcal{M}_m'$.
\end{itemize}

\begin{table*}[t]
    \centering
    
    \scalebox{0.75}{
    \begin{tabular}{|c|c|c|c|c|c|}
    % \toprule
    \hline
        Input & Optimality notion& &\multicolumn{3}{c|}{ Parameters of matching} \\
        \cline{4-6}
         & &   Matching&Signature &Max &Total\\
         & & & &dev&dev\\
         % \midrule
         \hline
         $G$&\RMMTD&$M_1=\{(a_1,p_1),(a_3,p_2),(a_4,p_7),(a_5,p_4),(a_6,p_6)\}$&$(2,2,1,2)$&$1$&$\textbf{2}$\\
         \hline
         $G$&\RMMMD&$M_2=\{(a_1,p_1),(a_2,p_1), (a_3,p_2),(a_4,p_7),(a_5,p_4),(a_6,p_6),(a_7,p_6)\}$&$(3,3,1,0)$&$\textbf{1}$&$4$\\
         \hline
         $G, \rho$&\RSMTD&$M_3=\{(a_1,p_1),(a_2,p_1),(a_3,p_2),(a_4,p_7),(a_5,p_4),(a_6,p_6),(a_7,p_1)\}$&$(4,2,1,0)$&$2$&$\textbf{4}$\\
         \hline
         $G, \rho$&\RSMMD&$M_4=\{(a_1,p_1),(a_2,p_1),(a_3,p_2),(a_4,p_1),(a_5,p_4),(a_6,p_6),(a_7,p_6)\}$&$(4,3,0,0)$&$\textbf{2}$&$5$\\
        %  $G,\rho$&{\sf EXACT-SIGN-MIN-Y-D}&$M_5=\{(a_1,p_1),(a_2,p_1), (a_3,p_3),(a_4,p_7),(a_5,p_5),(a_6,p_6),(a_7,p_1)\}$&$(4,0,3,0)$&$\textbf{2}$&$\textbf{4}$\\
        %  \hline
         \hline
         % \bottomrule
    \end{tabular}
    }
    \caption{Optimal matchings in the instance given in \cref{fig:target} corresponding to our problems. For the preference optimality first type problems, we consider the input signature as $(4, 0, 0, 3)$. We consider preference optimality as rank-maximality.}
    % \caption{Caption}
    \label{tab:my_label1}
\end{table*}

The order in which we consider the two orthogonal parameters for optimization provides a trade-off between deviation and preference optimality. As expected, when the deviation is optimized first, the output matching has a {\em weak} signature. In contrast, when preference optimality is given precedence, a higher deviation is incurred. 
Thus, our problems offer a  trade-off  based on the user requirement. To illustrate the optimal solutions for the different problems considered, we use our example instance $G$ given in \cref{fig:target}.

\subsection{Illustrative example}
We consider the example instance with seven applicants and seven posts, along with the preferences of applicants over posts, as shown in \cref{fig:eg1a}. For each post, two values -- a lower value and an upper value -- are given, as in \cref{fig:eg1b}. In the soft quota setting, these values represent the  lower and the upper  targets, respectively.  
First, we illustrate our problems for the optimality notion of rank-maximality.

\cref{tab:my_label1} shows the optimal solutions for each of the problems, along with parameters of the matching. 
We begin by considering the problems of the type deviation first, preference optimality next. See the first two rows of \cref{tab:my_label1}. Note that, in the example instance given in \cref{fig:target}, the posts $p_2$ and $p_3$ appear only in the preference list of applicant $a_3$. Therefore, any matching in the instance must leave  one of $p_2$ or $p_3$ unmatched. Similarly, any matching must leave  one of $p_4$ or $p_5$ unmatched. 
Thus, we make the following  useful observation.
\begin{obs}
    \label{obs:egfig1} Any matching in the example instance given in \cref{fig:target} must have a maximum deviation of at least one and a total deviation of at least two.
\end{obs}
 It can be verified that the matchings $M_1$ and $M_2$ shown in \cref{tab:my_label1} are optimal for the problems 
 \RMMTD and \RMMMD respectively. 
 
 Now we turn to problems where preference optimality is considered first and, subject to that, deviation is minimized. For concreteness, we fix the requirement for signature as $(4, 0, 0, 3)$. 
In order to meet the signature requirement, we must violate the upper target of the post $p_1$ by $2$, and hence the maximum deviation of any matching in $\mathcal{M}_{\mathcal{R}}$ is at least $2$. From Observation~\ref{obs:egfig1},  the sum of lower deviations over all posts (the total lower deviation) of any matching in the instance must be at least $2$, and therefore the total deviation of any matching in $\mathcal{M}_{\mathcal{R}}$ is at least $4$.
Furthermore, it can be verified that the matchings $M_3$ and $M_4$ shown in \cref{tab:my_label1} are optimal for the problems 
 \RSMTD and \RSMMD respectively.

 \begin{table*}[hbt!]
    \centering
    
    \scalebox{0.75}{
    \begin{tabular}{|c|c|c|c|c|c|}
    % \toprule
    \hline
        Input & Optimality notion& &\multicolumn{3}{c|}{ Parameters of matching} \\
        \cline{4-6}
         & &   Matching&Signature &Max &Total\\
         & & & &dev&dev\\
         % \midrule
         \hline
         $G$&\FMMTD&$M_5=\{(a_1,p_1),(a_3,p_2),(a_4,p_6),(a_5,p_4),(a_6,p_7)\}$&$(1,4,0,2)$&$1$&$\textbf{2}$\\
         \hline
         $G$&\FMMMD&$M_6=\{(a_1,p_1),(a_2,p_1),(a_3,p_2),(a_4,p_6),(a_5,p_4),(a_6,p_7),(a_7,p_6)\}$&$(2,5,0,0)$&$\textbf{1}$&$4$\\
         \hline
         $G, \rho$&\FSMTD&$ M_7=\{(a_1,p_1),(a_2,p_1),(a_3,p_2),(a_4,p_6),(a_5,p_4),(a_6,p_7),(a_7,p_1)\}$&$(3,4,0,0)$&$2$&$\textbf{4}$\\
         \hline
         $G, \rho$&\FSMMD&$M_6=\{(a_1,p_1),(a_2,p_1),(a_3,p_2),(a_4,p_6),(a_5,p_4),(a_6,p_7),(a_7,p_6)\}$&$(2,5,0,0)$&$\textbf{1}$&$4$\\ 
         \hline
        %  $G,\rho$&{\sf EXACT-SIGN-MIN-Y-D}&$M_5=\{(a_1,p_1),(a_2,p_1), (a_3,p_3),(a_4,p_7),(a_5,p_5),(a_6,p_6),(a_7,p_1)\}$&$(4,0,3,0)$&$\textbf{2}$&$\textbf{4}$\\
        %  \hline
        %  \hline
         % \bottomrule
    \end{tabular}
    }
    \caption{Optimal matchings in the instance given in \cref{fig:target} corresponding to our problems. For the preference optimality first type problems, we consider the input signature as $(4, 0, 3, 0)$. We consider preference optimality as fairness.}
    % \caption{Caption}
    \label{tab:my_label2}
\end{table*}

Note that the signatures of the matchings $M_3$ and $M_4$ weakly dominate the input signature, and hence both these matchings are candidate solutions for the problems {\sf DOM-SIGN-MIN-TOT} and {\sf DOM-SIGN-MIN-MAX} respectively, where DOM represents the optimality notion of weak dominance.

Next, we illustrate our problems and their solutions for the notion of fairness using the example instance $G$ given in \cref{fig:target}. 
 \cref{tab:my_label2} shows the optimal solutions for each of the problems, along with the parameters of the matching. Note that these matchings are different from the matchings given in \cref{tab:my_label1}. 
We first consider the problems of the type deviation first, preference optimality next. 
From Observation~\ref{obs:egfig1}, any matching in  $G$ must have a maximum deviation of at least one and a total deviation of at least two.
 It can be verified that the matchings $M_5$ and $M_6$ shown in \cref{tab:my_label2} are optimal for the problems 
 \FMMTD and \FMMMD respectively. 

 Now, we turn to problems where preference optimality is considered first and subject to that deviation is minimized. For concreteness, we fix the signature as $(4, 0, 3, 0)$. The input signature can be interpreted as follows:  the value of $0$ in the last component indicates that the required matching must assign all applicants, while  the value $3$ specifies that at most $3$ applicants can be matched to a rank $3$ post. Any matching whose signature is fairer than the input signature is a candidate solution.   
Observe that in order to  match all applicants, any matching must either violate the upper target of post $p_1$ by $2$ or violate the upper targets of the post $p_1$ and one of $p_6$ or $p_7$ by $1$ each. Thus, the maximum deviation of any matching in $\mathcal{M}_{\mathcal{F}}$ must be at least $1$. 
From Observation~\ref{obs:egfig1},
the total lower deviation of any matching in the instance must be at least $2$, making the total deviation of any matching in $\mathcal{M}_{\mathcal{F}}$  at least $4$.
Furthermore, it can be verified that matchings $M_7$ and $M_6$, shown in \cref{tab:my_label2}, are optimal for the \FSMTD and \FSMMD problems, 
 respectively.

\subsection{Our Results}
We show the following new results in our paper.
% state our results.
% \begin{restatable}{theorem}{stmm}
% \label{thm:stmm}
% %\begin{theorem}
% 	\ESMMD and \ESMTD are \sfNP-hard even when preferences of all applicants are strict with deviation $k=0$, $\ell(p)=u(p)=1$ for every post $p \in \B$.
% %\end{theorem}
% \end{restatable}

%\begin{theorem}
% \begin{restatable}{theorem}{cumd}
% \label{thm:cumd}
% 	\CSMMD and \CSMTD are \sfNP-hard even when preferences of all applicants are strict with deviation $k=0$, $\ell(p)=u(p)=1$ for every post $p \in \B$.
% % \end{theorem}
% \end{restatable}
\begin{restatable}{theorem}{algtsignopt} \OSMMD \label{algtsignopt}
and \OSMTD  admit polynomial time algorithms, where  {\sf OPT} can be one of rank-maximality ({\sf RMM}) or fairness ({\sf FAIR}). 
\end{restatable}
\begin{restatable}{theorem}{algtopt} \label{algopt}
\OMMD and \OMTD  admit polynomial time algorithms, where  {\sf OPT} can be one of rank-maximality or fairness. 
\end{restatable}

\begin{proposition}
   \label{thm:cumd}
	\CSMMD and \CSMTD are \sfNP-hard even when all applicants have strict preferences, the total deviation $k=0$, and for every post $p \in \B$, the lower target $\ell(p)=0$, and the upper target $u(p)=1$.
 
\end{proposition}
The above hardness is obtained from the hardness of {\sf DOM-Q} problem considered in \cite{ASN25} by viewing quotas as targets.

\paragraph{Our Techniques}
The algorithms for computing minimum total deviation matchings are based on a flow network
that we design. 
Flow networks are used  to compute Pareto-optimal matchings \cite{cechlarova2016pareto,cechlarova2017pareto,10.5555/3535850.3535885}, rank-maximal matchings \cite{nasre2019classified} and weakly dominant matchings \cite{ASN25,ka2022optimal} in the literature. 
For computing matchings with minimum value of maximum deviation we reduce the problem to an appropriate problem in the fixed quota setting, {which can be solved efficiently}. 
\paragraph{Generalized soft quota setting}%   
Given our positive results for the  problems in the soft quota setting for the optimality notions rank-maximality and fairness, it is natural to consider a generalization of our setting, where every deviation is associated with a cost. That is, every post $p$ in the {\em generalized soft quota setting} is associated with a lower deviation cost $c_{\ell}(p)$ and an upper deviation cost $c_{u}(p)$ that indicate the cost associated with deviating $1$ unit from the lower target and upper target, respectively. We show extensions of our  algorithms  in \cref{gsqs}.

\input{1-relatedworks}

%% file: figures/fig_eg1.tex
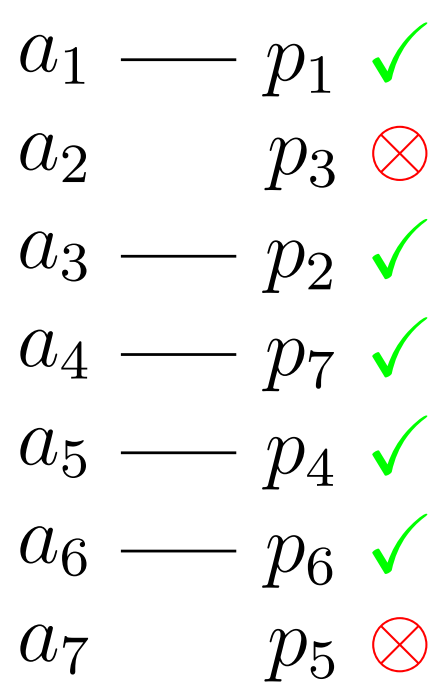
\begin{figure*}
     \centering
     \begin{subfigure}[b]{0.3\textwidth}
         \centering
         % \scalebox{0.6}{
         \begin{eqnarray*}
        a_1 & : & p_1 \\
        a_2 & : & p_1  \\
        a_3 & : & p_1, p_2, p_3 \\
        a_4 & : & p_1, p_6, p_7\\
        a_5 & : & p_1, p_4, p_5\\
        a_6 & : & p_6, p_7\\
        a_7 & : & p_1, p_6
    \end{eqnarray*}
         % }

         \caption{applicants}
         \label{fig:eg1a}
     \end{subfigure}      \hspace{0.05in}
     \begin{subfigure}[b]{0.3\textwidth}
         \centering
         % \scalebox{0.6}{
         \begin{eqnarray*}
p_1& [1, 1]\\
p_2& [1, 1]\\
p_3& [1, 1]\\
p_4& [1, 1]\\
p_5& [1, 1]\\
p_6& [0, 1]\\
p_7& [0, 1]
\end{eqnarray*}
         % }
         \caption{posts }
         \label{fig:eg1b}
     \end{subfigure}
      \hspace{0.05in}
     \begin{subfigure}[b]{0.3\textwidth}
         \centering
         \scalebox{1.3}{
\begin{tikzpicture}
            \begin{scope}[vertex style/.style={draw, circle, minimum size=0.0mm, inner sep=0pt, outer sep=0pt}]
                \path (2, -1) coordinate[vertex style] (ll);
            \end{scope}
            \node (a1) at (0, 2) {$a_1$};
            \node (a2) at (0, 1.5) {$a_2$};
            \node (a3) at (0, 1) {$a_3$};
            \node (a4) at (0, 0.5) {$a_4$};
            \node (a5) at (0, 0) {$a_5$};
            \node (a6) at (0, -0.5) {$a_6$};
            \node (a7) at (0, -1) {$a_7$};
            % \node (m) at (-0.25, -1.25) {$M_1$};
            \node (b1) at (1.5, 2) {$p_1$ \textcolor{green}{\checkmark}};
            \node (b2) at (1.5, 1.5) {$p_3$ {\textcolor{red}{$\otimes$}}};
            \node (b7) at (1.5, 0.5) {$p_7$ \textcolor{green}{\checkmark}};
            \node (b4) at (1.5, 0) {$p_4$ \textcolor{green}{\checkmark}};
            \node (b6) at (1.5, -0.5) {$p_6$ \textcolor{green}{\checkmark} };
            \node (b3) at (1.5, 1) {$p_2$ \textcolor{green}{\checkmark}};
            \node (b5) at (1.5, -1) {$p_5$ {\textcolor{red}{$\otimes$}}};
            \begin{scope} [edge style/.style={draw=black}]
                \draw[edge style] (a1)--(b1);
                \draw[edge style] (a3)--(b3);
                \draw[edge style] (a4)--(b7);
                \draw[edge style] (a5)--(b4);
                \draw[edge style] (a6)--(b6);
            \end{scope}
        \end{tikzpicture}
         }
         \caption{ $M_1$}
         \label{fig:eg1c}
     \end{subfigure}
         \caption{Example instance $G$ with $|\A| = |\B| = 7$. Applicants along with their preferences are given. Preference list of an applicant can be read as: $p_1$ is the rank 1 post of $a_3$, $p_2$ is the rank 2 post of $a_3$ and so on. 
    Each post has two values -- a {\em lower value } and an {\em upper value}. In the fixed quota setting these numbers represent rigid lower and upper quotas. In the soft quota setting these numbers are the lower and the upper targets, respectively. Matching $M_1$ shown in \cref{fig:eg1c} is an infeasible matching in the fixed quota setting since the posts $p_3$ and $p_5$ marked with red circles are not assigned their lower quota many applicants. The same matching $M_1$ is a valid assignment in the soft quota setting introduced in this paper.}   
    %In the fixed quota setting, for the post $p_7$, $[0,1]$  denotes that the lower quota $q^{-}({p_7})=0$ and the upper quota $q^{+}({p_7})=1$. In the soft quota setting,  for a post $p_7$, $[0,1]$  denotes that the lower target $\ell(p_7)$=0, and upper target $u(p_7)=1$. }
    \label{fig:target}
\end{figure*}

%% file: figures/fig3.tex
\begin{figure}
     \centering
          \subfloat[][{\sf OPT-MIN-Y}]{
     \scalebox{0.42}{
     \begin{tikzpicture}[
roundnode/.style={circle, draw=black!100,  inner sep=0pt, minimum size=3cm},
farrow/.style={-Latex}
]
\node[roundnode] (s) at (0,0) [label=above:{\huge $\mathcal{M}_t$}] {};
\node[roundnode] (t) at (2,0) [label=above:{\huge $\mathcal{M}_m$}] {};
% \draw \boundellipse{1,0}{3.5}{2.5};
\draw \boundellipse{1,0}{5}{4};
% \node (a) at (1,3.8) {$\mathcal{M}$};
% % \node (ap3)  at (-5,-1)  {$\vdots$};
%  \node (b) at (1,2.8) [label=above:{\huge $M \in \mathcal{M} : \sigma(M)\ge \rho$}] {};
 \node (a3) at (1,4) [label=above:{\huge $\mathcal{M}$}] {};
%  \node (a4) at (1,2.75) {\huge{$M \in \mathcal{M}$ \ : \ $\sigma(M) \ge \rho$}};
    \end{tikzpicture}}
     \label{fig:omyd}
     } \hspace{0.5in}
     \subfloat[][{\sf OPT-SIGN-MIN-Y}]{
     \scalebox{0.42}{
    \begin{tikzpicture}[
roundnode/.style={circle, draw=black!100,  inner sep=0pt, minimum size=3cm},
farrow/.style={-Latex}
]
\node[roundnode] (s) at (0,0) [label=above:{\huge $\mathcal{M}_t'$}] {};
\node[roundnode] (t) at (2,0) [label=above:{\huge $\mathcal{M}_m'$}] {};
\draw \boundellipse{1,0}{3.75}{2.75};
\draw \boundellipse{1,0}{5}{4};
% \node (a) at (1,3.8) {$\mathcal{M}$};
% % \node (ap3)  at (-5,-1)  {$\vdots$};
%  \node (b) at (1,2.8) [label=above:{\huge $M \in \mathcal{M} : \sigma(M)\ge \rho$}] {};
 \node (a3) at (1,4) [label=above:{\huge $\mathcal{M}$}] {};
 \node (a4) at (1,3.25) {\huge{$\mathcal{M}_{OPT}$}};
%  \node (a4) at (1,2.75) {\huge{$M \in \mathcal{M}$ \ : \ $\sigma(M) \ge \rho$}};
    \end{tikzpicture}
}
     \label{fig:osmyd}
     }
     \caption{{\sf OPT-MIN-Y} and {\sf OPT-SIGN-MIN-Y} illustration, where $Y \in \{tot,max\}$, and {\sf OPT} $\in \{${\sf RMM, FAIR}$\}$}
     \label{osmydomyd}
\end{figure}
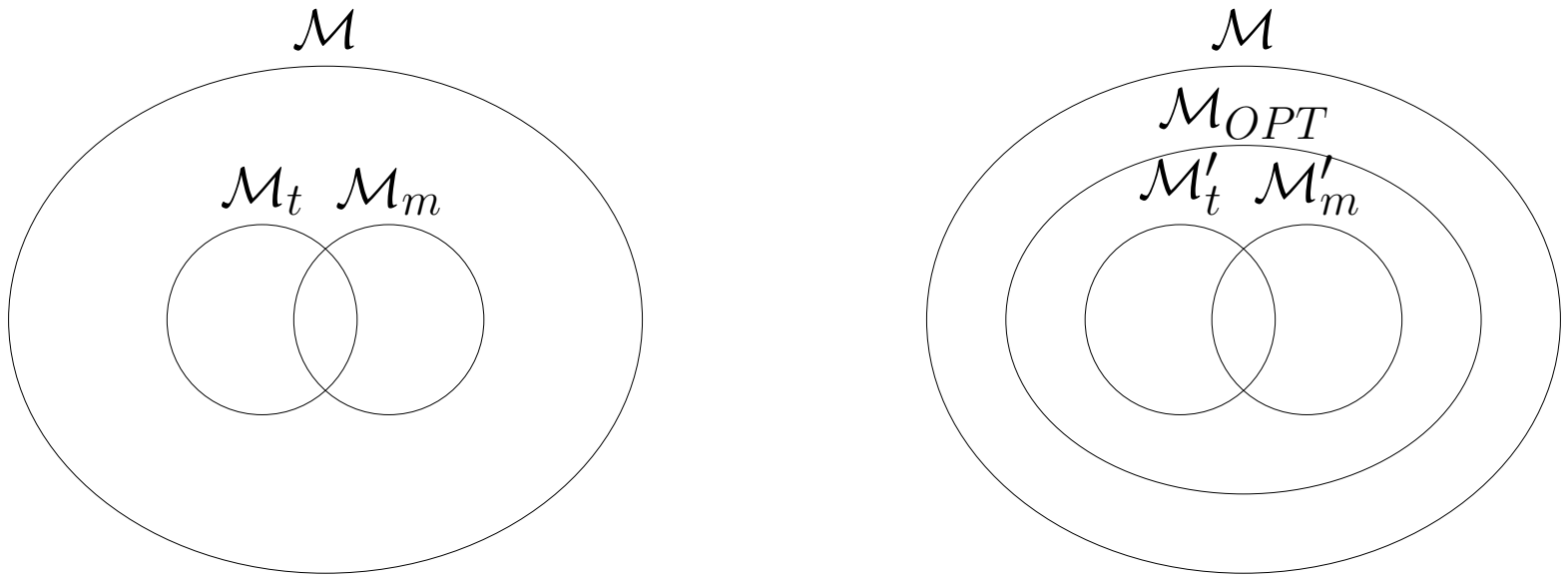

%% file: 1-relatedworks.tex
\subsection{Related Works}
\paragraph{Biproportional apportionment}
Our soft quota setting is similar to the biproportional apportionment problem \cite{dhamala2009apportionment,pukelsheim2012network,serafini2012parametric}, which deals with the fair division of political representatives among states. Political representatives correspond to applicants, and states  correspond to posts in our model. 
Many methods \cite{pukelsheim2012network,serafini2012parametric} for the biproportional apportionment problem reduce the problem to a flow network to minimize the absolute or relative min-max error. Note that the min-max error in the biproportional apportionment problem considered in the works of \citet{pukelsheim2012network} and  \citet{serafini2012parametric} corresponds to the min-max deviation in our model. However, the biproportional apportionment problem does not involve any preferences, whereas in our setting we have preferences associated with edges. 

\paragraph{Soft bounds}
The recent works \cite{ehlers2014school,echenique2015control,gonczarowski2019matching,aziz2021mechanism,kurata2017controlled} in the two-sided preferences model allow soft bounds on the capacity constraints. These works focus on the controlled school choice problem, in which students are associated with types and schools have soft bounds on the quota of each type. 
However, there are hard constraints on the number of students that can be assigned to each school or type. 
In the work by   \citet{aziz2021mechanism} and  \citet{echenique2015control}, lower-side deviation (or relaxation) from the target quotas or reserves is allowed. Our model has no hard capacity constraints, and target soft quotas are allowed to deviate on both sides. In the work by \citet{ehlers2014school}, students are associated with a single type, and for each type, there is a lower violation and upper violation w.r.t. an allocation, and these are analogous to the lower deviation and upper deviation in our model.  \citet{gonczarowski2019matching} considered a matching problem in the two-sided matching market with soft diversity constraints that allocate candidates to Pre-Military Academics programs (PMA). Here, each PMA defines a set of populations (or types) that they are interested, and specifies a soft minimum target and hard maximum quota for such populations.
 For more information, refer to the survey paper by \citet{Aziz_Biró_Yokoo_2022}.
% Recent work by \citet{Multi-rank2025} studies multi-ranked diversity goals with flexible reserves and develops polynomial-time algorithms based on constrained rank-maximal matchings.

\paragraph{Other models of flexibility}
In the context of school choice problem, capacity planning is studied in \cite{gajulapalli2020stability,rios2021improving,DBLP:journals/corr/abs-2110-00734,ijcai2022p1,chen2023optimal}. In the work by \citet{gajulapalli2020stability}, flexible quotas are used to increase the capacity of some schools. In the work by \citet{rios2021improving}, flexible quotas are used as a tie-breaker. In the work  by \citet{DBLP:journals/corr/abs-2110-00734}, the authors study the school choice problem in an expanded market, where some schools can increase their capacity to either allocate students who are previously unassigned or to improve the assignment of the previously assigned students. The budget used for capacity expansion in the work by \citet{DBLP:journals/corr/abs-2110-00734} is analogous to the total upper deviation of posts in our model. In \citet{ijcai2022p1}, Monte Carlo tree search based solutions for the capacity expansion problem are studied. 
In the work of \citet{limaye2021envy} and \citet{ASN25,ka2022optimal}, programs have costs instead of quotas, which allows flexibility. In all these works, the flexibility of upper quotas is considered, whereas in our work, we allow posts to deviate on both upper and lower targets.

To the best of our knowledge, this is the first work that investigates soft quotas in the one-sided preference list model.

Our problems consider optimizing two objective functions -- deviation and preference optimality -- and hence can be regarded as a special case of a multi-objective optimization problem \cite{esfandiari2016bi,korula2013bicriteria,aggarwal2014biobjective}. These works
differ from our work in terms of  the problem setup and the objectives  considered.

\paragraph{Organization of the paper}
Throughout the paper, we present the results with OPT set to rank-maximality (RMM). Analogous results for fairness can be obtained with minor modifications to our algorithms, as outlined in the relevant sections. In \cref{sec:algm-results-target} and \cref{sec:alg-min-y-d}, we focus on problems where the objective is to minimize the total deviation. The former section addresses preference-optimality-first type problems (the {\sf MIN-TOT} variant in \cref{algtsignopt}), while the latter discusses deviation-first type problems (the {\sf MIN-TOT} variant in \cref{algopt}). In \cref{minmax}, we present results for the {\sf MIN-MAX} problems by exploiting the connection between the fixed quota setting and the soft quota setting. \cref{sec:tm-both} explores the simultaneous optimization of total deviation and maximum deviation. Results for the generalized soft quota setting are described in \cref{gsqs}, and we conclude with a discussion in \cref{sec:dis}.

%% file: 3-Alg-target.tex
\section{Minimizing Total Deviation: Preference Optimality First}
%: Algorithm for \OSMTD
\label{sec:algm-results-target}
In this section, we present the results for minimizing total deviation for the preference optimality first, deviation next type of problems. As mentioned earlier, apart from the input instance these problems have a signature as part of the input.
Recall that OPT can be instantiated with either rank-maximality or fairness. 
We present the results for rank-maximality in detail and state the necessary modifications required for fairness.

\input{figures/fig_network2}

An instance of \RSMTD problem consists of a
graph $G$ with applicant preferences, soft quotas for posts, and an input signature $\rho$. Let $\mathcal{M}_{\mathcal{R}}$ denote the set of matchings such that, for every $M \in \mathcal{M}_{\mathcal{R}}$, we have 
$\sigma(M)\succeq_{\mathcal{R}} \rho$. Our goal is to output a matching $M$ that
has minimum total deviation in the set $\mathcal{M}_{\mathcal{R}}$.
We present a  flow-network based algorithm to solve the \RSMTD problem.

\paragraph{Overall idea} 
Consider an optimal matching $M^*$ for the \RSMTD problem. Let the total deviation of $M^*$ be $\Delta^*$. In $M^*$, a subset of posts incur deviation with respect to the {\em lower} target, whereas another subset of posts, disjoint from the previous set, incur deviation with respect to the upper target. Let $ld^*$ ($ud^*$) denote the sum of the deviation of posts that deviate with respect to the lower target (respectively, with respect to the upper target). Note that $\Delta^* = ld^*+ud^*$. Our algorithm considers all possible values of $ld$ and $ud$ such that $ld+ud=\Delta$. We show that there exists a flow network $H_{(ld^*, ud^*)}$ such that the matching $M^*$ has a feasible flow $f^*$ in $H_{(ld^*, ud^*)}$. For a fixed value of $\Delta$ and a  split of $\Delta$ as $ld+ud$, we construct a flow network with properties as specified below.

\subsection{Construction of the network}
For a given instance $G$, and a fixed value of  $\Delta$, which can be split as the sum of two non-negative integers $ld$ and $ud$, we construct the flow network $H_{(ld, ud)}=(V_H,E_H)$  with costs, demands and capacities on the edges.
The vertices of $H_{(ld, ud)}$ are: $$V_H=\big\{s, d',d'',t\big\}\cup \A \cup \big\{p',p'' \ : \ p \in \B\big\}.$$ 
For every applicant $a$ in $G$, we have a node $a$ in $V_H$. For every post $p$ in $G$, we have two nodes in $V_H$, namely, $p'$ (termed the original copy) and $p''$ (termed the extended copy). 
In addition, we have a source $s$, a sink $t$, and two special nodes $d'$ and $d''$.

\noindent{\bf Demand and capacity of the edges in $E_H$.} Every edge $e \in E_H$ has a demand and a capacity associated with it, and we use a  vector [$q^-(e),q^+(e)$] to denote these values.
The edge set $E_H$ is  the union of the following sets:   
\begin{align*}
\mathcal{E}_1=&\{(s,a) \ : \ a \in \A\} \text{ with vector [$0,1$]}\\
\mathcal{E}_2=&\{(a,p'), (a,p'') \ : \ (a,p) \in E\} \text{ with vector [$0,1$]}\\
\mathcal{E}_3=&\{(p',t) \ : \ p \in \B\} \text{ with vector [$\ell(p),u(p)$]} \\
\mathcal{E}_4=&\{(p'', d'') \ : \ p\in \B\} \text{ with vector [$0,ud$]}\\
\mathcal{E}_5=&\{(d',p') \ : \ p \in \B\}  \text{ with vector [$0,ld$]}\\
\mathcal{E}_6=&\{(s,d'), (d'',t)\} \text{ with vectors [$ld,ld$] and [$ud,ud$] respectively}  
\end{align*}

%%%%%%%%%%%%%%%%%
\noindent{\bf Cost of the edges in $E_H$.}
All the edges in $E_H$, except the edges in $\mathcal{E}_2$, have a cost of $0$. For an edge $(a,p)$ in $G$ of rank $i$, the corresponding two edges $(a,p')$ and $(a,p'')$ in $\mathcal{E}_2$ have a cost of $-(n+1)^{r-i}$, where $n=\abs{\A}$. 

As in standard reductions, there is a natural correspondence between edges in $\mathcal{E}_2$ carrying non-zero flow and a matching. With this, we complete the description of the flow network. See Figure~\ref{fig:network1} for an illustration.
A flow $f$ in $H_{(ld,ud)}$ is said to be a feasible flow if it meets the demands on every edge $e \in f \cap E_H$. The cost of a feasible flow $f$, denoted by $cost(f)$ is defined as $\sum\limits_{e\in E_H} f(e) \cdot cost(e)$, where $f(e)$ and $cost(e)$ denotes the flow across the edge $e$ and cost assigned on the edge $e$, respectively. For a matching $M$ with signature $\sigma(M)=(x_1, x_2, \ldots, x_r,x_{r+1})$, we define the cost of $M$, denoted by $cost(M)$ as $-\sum\limits_{j=1}^{r} x_j \cdot (n+1)^{r-j}$.
% The cost of a feasible flow $f$, denoted by $cost(f)$ is the sum of the costs of its edges. For every feasible flow $f$ and its corresponding matching $M_f$, we define  the cost of  $M_f$, denoted by $cost(M_f)$, to be equal to $cost(f)$.
% 
% \textcolor{red}{proof attempt below}
We have the following lemma regarding the cost assignment.
\begin{lemma} \label{lem:cost-rmm}
    Let $M, M'$ be matchings such that $\sigma(M) \succ_{\mathcal{R}} \sigma(M')$. Then, $cost(M) < cost(M')$, where each rank $i$ edge has a cost of $-(n+1)^{r-i}$.
\end{lemma}
\begin{proof}
    Let $\sigma(M)=(x_1,x_2, \ldots,x_r,x_{r+1})$, and $\sigma(M')=(y_1,y_2, \ldots, y_r,y_{r+1})$. Let $i$ be the first rank at which $\sigma(M)$ and $\sigma(M')$ differ. Thus, $x_j=y_j$ for all $j<i$, and $x_i>y_i$.

   The cost difference between the matchings $M$ and $M'$ is 
   \[cost(M')-cost(M) = \sum\limits_{j=1}^r (x_j -y_j) (n+1)^{r-j} = (x_i-y_i)(n+1)^{r-i} + \sum\limits_{j=i+1}^r (x_j -y_j) (n+1)^{r-j}.\]
   Now, since $x_i - y_i \geq 1$, we have
   \begin{align*}
       cost(M')-cost(M) &\geq (n+1)^{r-i} + \sum\limits_{j=i+1}^r (x_j -y_j) (n+1)^{r-j}\\
       &\geq (n+1)^{r-i} - \sum\limits_{j=i+1}^r y_j (n+1)^{r-j}\\ 
       &\geq (n+1)^{r-i} - \left(\sum\limits_{j=i+1}^r y_j\right) (n+1)^{r-i-1}.
   \end{align*}
   Additionally, using the fact that $\sum\limits_{j=i+1}^r y_j \leq \sum\limits_{j=1}^r y_j \leq n$, we conclude that
   \[cost(M')-cost(M) \geq (n+1)^{r-i} - n (n+1)^{r-i-1} = (n+1)^{r-i-1} > 0.\]
\end{proof}

We have the following remark regarding the cost assignment.
\begin{remark} \label{rem:1}
Let $\mathcal{M}_{(ld,ud)}$ be the set of matchings corresponding to feasible flows in the network. Then, a min-cost flow in $H_{(ld,ud)}$ corresponds to a rank-maximal matching in $\mathcal{M}_{(ld,ud)}$. 
\end{remark}
\begin{proof}
    Let $f$ be a min-cost feasible flow in $H_{(ld,ud)}$, and let $M_f$ be the corresponding matching in $\mathcal{M}_{(ld,ud)}$. Suppose, for contradiction, that $M_f$ is not a rank-maximal matching in $\mathcal{M}_{(ld,ud)}$. Then there exists another matching $M' \in \mathcal{M}_{(ld,ud)}$ such that $\sigma(M') \succ_{\mathcal{R}} \sigma(M_f)$. By \cref{{lem:cost-rmm}},  $cost(M')<cost(M_f)$.
    Since $M' \in \mathcal{M}_{(ld,ud)}$, there exists a feasible flow $f'$ corresponding to $M'$, with $cost(f')=cost(M')$, contradicting the minimality of $f$.
    % Let $f$ be a min-cost feasible flow in $H_{(ld,ud)}$, and let $M_f$ be the corresponding matching in $\mathcal{M}_{(ld,ud)}$. Suppose, for the sake of contradiction that $M_f$ is not a rank-maximal matching in $\mathcal{M}_{(ld,ud)}$. By \cref{{lem:cost-rmm}}, there exists another matching $M' \in \mathcal{M}_{(ld,ud)}$ such that $\sigma(M') \succ_{\mathcal{R}} \sigma(M_f)$, and $cost(M')<cost(M_f)$.
    % Since $M' \in \mathcal{M}_{(ld,ud)}$, there exists a feasible flow $f'$ in $H_{(ld,ud)}$ corresponding to $M'$. Furthermore, $cost(f')=cost(M')$. This contradicts the assumption that $f$ is a min-cost feasible flow in $H_{(ld,ud)}$. Thus, $M_f$ must be a rank-maximal matching in  $\mathcal{M}_{(ld,ud)}$.
\end{proof}

% The proof of Remark \ref{rem:1} follows from the cost assignment mentioned in \cite{irving2006rank}.
It is worth noting that our reductions apply to  any notion of optimality that can be encoded as a maximum weight matching problem.

\subsection{Total deviation of a flow  versus  total deviation of the corresponding matching }
The flow network $H_{(ld,ud)}$ has demands on its edges, and thus it may not admit a feasible flow. Assume that $H_{(ld,ud)}$ admits a feasible flow, say $f$. 
There is a natural correspondence between the flow $f$ in  $H_{(ld, ud)}$ and a matching $M$ in $G$ as follows:
\begin{align*}
    M=\{(a,p) \ | \ \text{$f\left(a,p'\right)=1$ or $f\left(a,p''\right)=1$,   $a\in \A$, $p \in \B$}\}.
\end{align*}
For any applicant $a$, the edge $(s,a)$ has unit capacity, and hence the incoming flow, if any, is routed either via the node $p'$ or via $p''$, but not both. 
For a feasible flow $f$, we define the deviation of a flow as follows.
% Recall that for a post $p \in \B$ in $G$, there are two nodes $p'$ and $p''$ in the network $H$. 
For the node $p'$, we define the lower deviation of $p'$ with respect to the flow $f$ as $f\left(d',p'\right)$, and the upper deviation is undefined for this node.
For the node $p''$, we define the upper deviation of $p''$ with respect to $f$ as $f\left(p'',d''\right)$, and the lower deviation is undefined for this node.
For a post $p \in \B$, we define the total deviation of $p$   with respect to $f$ as
$$d_f(p)=f\left(d',p'\right)+f\left(p'',d''\right).$$
Similar to the definition of total deviation of a matching, we define total deviation of the flow $f$ in $H_{(ld, ud)}$ as follows. 
\begin{definition}[{\bf Total deviation of a flow}] The  total deviation of a flow $f$ in $H_{(ld, ud)}$ is defined as
    $$d_{tot}(f)=\sum\limits_{p \in \B} f\left(d',p'\right)+f\left(p'',d''\right).$$
\end{definition}
 By the construction of the network $H_{(ld, ud)}$, and by the  definition of total deviation of a feasible flow $f$ in $H_{(ld, ud)}$, it is easy to verify that  $d_{tot}(f)=\Delta=ld+ud$.

\input{figures/apfig1.tex}
\paragraph{Some pathological cases} 
Having defined the correspondence between a flow in $H_{(ld,ud)}$ and a matching, it would be ideal to have a correspondence between the total deviation of a flow in $H_{(ld,ud)}$ and that of the matching. 
However, even for a single post, the total deviation of the post with respect to the flow need not be the same as the total deviation of the post with respect to the corresponding matching. This is  illustrated by three scenarios presented below.  
For all the three scenarios, we consider an input graph $G$  that has a single edge $(a,p)$,  and the post $p$ has $\ell(p)=0$ and $u(p)=1$.     

    \begin{itemize}
   \item Scenario 1:  
Consider the flow network $H_{(1,1)}$ for $\Delta=2$. The network $H_{(1,1)}$ admits a feasible flow $f$ with value $2$ that routes $1$ unit of flow through the path $\left<s, d', p', t\right>$ and $1$ unit of flow through the path $\left<s, a, p'', d'', t\right>$, and has a total deviation $d_{tot}(f)=2$ (see \cref{fig:sc1} for an illustration). Since $f\left(d',p'\right)>0$, the post $p$ has a lower deviation with respect to the flow $f$. Similarly, $p$ has an upper deviation due to the non-zero flow across the edge $\left(p'',d''\right)$. Thus, the flow $f$ has a total deviation $d_{tot}(f)=2$, as mentioned earlier. 
The matching corresponding to $f$ contains a single edge $(a,p)$ and has a total deviation of $0$. Thus, there is a mismatch between the total deviation of the flow and the total deviation of the corresponding matching. We remark that this issue arises because, for the same post $p$, both the edge corresponding to its original copy $\left(d',p'\right)$ and the edge corresponding to its extended copy $\left(p'',d''\right)$ carry non-zero flow. 
This issue can be avoided by ensuring that, for every post $p$, at most one of the edges $\left(d',p'\right)$ and $\left(p'',d''\right)$ carries non-zero flow in $f$.

    \item Scenario 2: Consider the flow network $H_{(0,1)}$ for $\Delta=1$. The network $H_{(0,1)}$ admits a feasible flow $f$ with value $1$ that routes $1$ unit of flow through the path $\left<s, a, p'', d'', t\right>$ and has a total deviation $d_{tot}(f)=1$. However, the matching corresponding to $f$ contains a single edge $(a,p)$ and has a total deviation of $0$. Note that here $f\left(p',t\right)=0 < u(p)$ and $f\left(p'',d''\right)>0$. For a feasible flow $f$, when $f\left(p'',d''\right)>0$, the post $p$ has an upper deviation, and we expect the flow across the edge $\left(p',t\right)$ to be saturated. However, this need not be true, as illustrated in the example. This situation can be avoided if the edge $(p',t)$ is saturated in $f$ whenever $f\left(p'',d''\right)>0$.

   \item Scenario 3: Consider the flow network $H_{(1,0)}$ for $\Delta=1$. The network $H_{(1,0)}$ admits a feasible flow $f$ with value $1$ that routes $1$ unit of flow through the path $\left<s, d', p', t\right>$ and has a total deviation $d_{tot}(f)=1$. Note that here $f\left(p',t\right)=1>\ell(p)$ and $f\left(d',p'\right)>0$. 
The matching corresponding to $f$ is an empty matching and has a total deviation of $0$. When we route flow along the edge $\left(d',p'\right)$, we expect the flow across the edge $\left(p',t\right)$ to be exactly equal to its demand $\ell(p)$. However, this need not be true, as illustrated in the example. This situation can be avoided if the flow across the edge $(p',t)$ in $f$ is exactly equal to its demand whenever $f\left(d', p'\right)>0$.
   
    \end{itemize}
\subsection{Good flow and correcting flow}    
    To address the discrepancies as illustrated in the pathological cases  and ensure that the total deviation of the flow corresponds to the total deviation of the matching, a feasible flow in our flow network should satisfy additional properties, which are listed below as Property~\ref{pro:1}. 

\begin{property}
\label{pro:1}
Let $f$ be a feasible flow in $H_{(ld, ud)}$. The flow $f$ is good 
if, for every post $p \in \B$, all of the following hold.
\begin{enumerate}[(i)]
    \item 
     \label{pro11} at most one of the edges $(d',p')$ and $(p'',d'')$ has a non-zero flow in $f$
    \item \label{pro12}if $f\left(p'',d''\right) > 0$ then $f\left(p',t\right) = u(p)$
    \item \label{pro13} if $f\left(d',p'\right) >0$ then $f\left(p',t\right)=\ell(p)$ 
\end{enumerate}
\end{property}
We note a useful observation below about good flows. 
\begin{obs} \label{obs:4}
{
Let $f$  be a good flow  in   $H_{(ld, ud)}$, and let $M$ be the corresponding matching in $G$. Then, for every post $p \in \B$, $d_f(p)=d_M(p)$. Therefore, $d_{tot}(f)=d_{tot}(M)$. }
\end{obs}
\begin{proof}
    Let $f$ be a good flow  in   $H_{(ld, ud)}$, and thus $f$ satisfies Property~\ref{pro:1}.
By  Property~\ref{pro:1}(\ref{pro11}), for a post $p$, at most one of the edges in $(d',p')$ and $(p'',d'')$ has a non-zero flow. 
Now consider the matching $M$ corresponding to $f$.
Observe that, for a post $p$, the number of applicants that are matched to $p$ in $M$ is $\abs{M(p)}=f\left(p'',d''\right)+f\left(p',t\right)-f\left(d',p'\right)$. Recall that the deviation of $p$ in $M$ is $d_M(p)=\max \{{0,\ell(p)-\abs{M(p)}, \abs{M(p)}- u(p)}\}$.
We consider the following cases.
\begin{itemize}
    \item \noindent{Case 1:} $p \in \B : f\left(d',p'\right)=f\left(p'',d''\right)=0$. Observe that, in this case, $d_f(p)=0$. Since $f\left(p'',d''\right)=0$, the node $p''$ does not have any non-zero flow into it in $f$, which implies that $\abs{M(p)}=f\left(p',t\right) \le u(p)$. Since $f\left(d',p'\right)=0$, all the flow along the edge $\left(p',t\right)$ comes from applicants in $\A$ (true applicants), and because $f$ is a feasible flow, $\abs{M(p)}\ge \ell(p)$. Therefore, $d_M(p)=0=d_f(p)$.

\item \noindent{Case 2:} $p \in \B : f\left(p'',d''\right)>0$, and $f\left(d',p'\right)=0$. Observe that $d_f(p)=f\left(p'',d''\right)$. In this case, by Property~\ref{pro:1}(\ref{pro12}),  $f\left(p',t\right)=u(p)$, and since $f\left(d',p'\right)=0$, all the flow along the edge $\left(p',t\right)$ is via the true applicants. Thus,   $\abs{M(p)}=u(p)+f\left(p'',d''\right)$, and this implies that $d_M(p)=f\left(p'',d''\right)=d_f(p)$.

\item \noindent{Case 3:} $p \in \B : f\left(d',p'\right)>0$, and  $f\left(p'',d''\right)=0$.  Observe that $d_f(p)=f\left(d',p'\right)$. In this case, by Property~\ref{pro:1}(\ref{pro13}),  $f\left(p',t\right)=\ell(p)$. Since, $f\left(p'',d''\right)=0$, $\abs{M(p)}=\ell(p)-f\left(d',p'\right)$. This implies that $d_M(p)=f\left(d',p'\right)=d_f(p)$.

\end{itemize}
Thus, for every post, the deviation with respect to a good flow and the deviation with respect to the corresponding matching are the same.
This immediately implies that the total deviation of a good flow  and that of the corresponding matching  are the same.
\end{proof}

It is not immediate how to compute a good flow in a flow network. However, given a feasible flow in $H_{(ld, ud)}$, we show that it is possible to construct a good flow in a {\em different} flow network. We need the following notation.
For an edge $(a,p)$ in $G$, we call the path $\left<s, a, p', t\right>$ in the network the {\em true path}, and the path $\left<s, a, p'', d'', t\right>$ the {\em extended path}. For a post $p$ in $G$, we call the path $\left<s, d', p', t\right>$ the {\em dummy path}$.$
Note that, corresponding to an applicant $a$ and a post $p$ in $G$, we have a true path and an extended path, and using either one of them, we can route a maximum of $1$ unit of flow in the network. Corresponding to a post $p$ in $G$, we have a unique dummy path, and the maximum flow that can be routed along this path is $\min{ld,u(p)}$.
Let $f$ be a feasible flow in $H_{(ld,ud)}$. We say that the status of a post $p\in \B$ is {\em good} with respect to $f$ if the corresponding pair of nodes $p', p''$ in the network satisfies the conditions given in Property~\ref{pro:1}; otherwise, the status of $p$ is {\em bad} with respect to the flow $f$.

Suppose we have a bad post $p$ in $G$ with respect to the flow $f$. We reroute the flow in a certain way to turn the post $p$ into a good post without changing the status of any other post. We now describe the different cases under which $p$ can be a bad post and the rerouting required to turn it into a good post.

\begin{definition}[\correcting flow]
Let $f$ be a feasible flow in the network $H_{(ld,ud)}$ that has a bad post $p$ with respect to $f$, and let $f'$ be another flow in $H_{(ld,ud)}$ that differs only in the flow across some of the $s$-$t$ paths corresponding to the post $p$. Then  $f'$ is called a {\em \correcting flow} for $f$ if it satisfies the following conditions.
\begin{itemize}
    \item[(i)] There exists a network $H_{(ld',ud')}$ such that $f'$ is feasible in the network $H_{(ld', ud')}$, where $ld'+ud'<ld+ud$.
    \item[(ii)] The matchings corresponding to $f$ and $f'$ are the same.
    \item[(iii)]  Status of any post $q \in \B \setminus \{p\}$ remains the same in $f$ and $f'$.
    % \item
\end{itemize}
\end{definition}
% Now we describe three types of \correcting flows. 
We describe the types of a \correcting flow $f'$.
\begin{itemize}
    \item {\em $(ld,ud)-$\correcting flow.} Assume that the post $p$  violates  Property~\ref{pro:1}(\ref{pro11}). Thus, $f\left(d',p'\right)>0$ and $f\left(p'',d''\right)>0$. We  reduce the flow along the dummy path corresponding to $p$ by $1$. Following this,  we  select an arbitrary extended path carrying unit flow corresponding to the post $p$, say $\left<s, a, p'',d'',t\right>$ and reroute the flow along the path $\left<s, a, p',t\right>$. Note that the resulting flow $f'$ is feasible in the network $H_{(ld', ud')}$, where $ld'=ld-1$ and $ud'=ud-1$.

\item {\em $ud-$\correcting flow.} Assume that the post $p$ violates Property~\ref{pro:1}(\ref{pro12}). Thus, $f\left(p'',d''\right)>0$ and $f\left(p',t\right)<u(p)$. 
We select an arbitrary extended path carrying unit flow corresponding to the post $p$, say $\left<s, a, p'',d'',t\right>$ and reroute the flow along the path $\left<s, a, p',t\right>$. Note that the resulting flow  $f'$ is feasible in a network $H_{(ld', ud')}$ where, $ld'=ld$ and $ud'=ud-1$.

\item {\em $ld-$\correcting flow.} Assume that  $p$ violates Property~\ref{pro:1}(\ref{pro13}). Thus, $f\left(d',p'\right)>0$ and $f\left(p',t\right)>\ell(p)$. 
We reduce the flow along the dummy path corresponding to $p$ by $1$. 
Note that the resulting flow $f'$  is feasible in a network $H_{(ld', ud')}$, where $ld'=ld-1$ and $ud'=ud$.
\end{itemize}
Note that in each of the above cases, if applicant $a$ is matched to the post $p$ prior to the rerouting, then it continues to remain matched to $p$ after the rerouting.  Since we have not changed the flow with respect to any post $q \in \B \setminus \{p\}$, the status of the post $q$ remains unchanged. Thus $f'$ is a \correcting flow.  
For illustration, the correcting flow for the scenario 1 is given in  \cref{fig:cf1}.

\begin{lemma}
If $f$ is a feasible flow in $H_{(ld,ud)}$ that does not satisfy Property~\ref{pro:1}, then there exists a \correcting flow $f^*$ for $f$ in $H_{(ld,ud)}$ 
that satisfies Property~\ref{pro:1}.
\end{lemma}
\begin{proof}
Let $p$ be a bad post with respect to $f$. We apply a sequence of \correcting flows to change the status of the post $p$ from bad to good. Assume that the pair of nodes $p', p''$ corresponding to $p$ violates  Property~\ref{pro:1}(\ref{pro11}), and thus  $f\left(d',p'\right)>0$ and $f\left(p'',d''\right)>0$. Let $$x_p=\min\{f\left(d',p'\right), f\left(p'',d''\right)\}.$$ We apply $x_p$ many $(ld,ud)-$\correcting flows.
        Let the resulting flow be $f'$. After the above changes, at least one of the edges $(d',p')$ or $(p'',d'')$  has a zero flow in $f'$. That is, the nodes $p'$ and $p''$ satisfy Property~\ref{pro:1}(\ref{pro11}) with respect to $f'$. However, the nodes  $p'$ and $p''$ may still violate either Property~\ref{pro:1}(\ref{pro12}) or Property~\ref{pro:1}(\ref{pro13}), but not both. Depending on which property is violated, we consider the following two cases.

       \noindent{\bf Case 1:} $f'\left(d',p'\right)>0$. In this case, because $f'\left(p'',d''\right)=0$, the pair of nodes $p',p''$ trivially satisfies Property~\ref{pro:1}(\ref{pro12}). Now suppose $f'\left(d',p'\right)>0$ and $f'\left(p',t\right)=u(p)>\ell(p)$, which implies that the pair of nodes $p',p''$ violates Property~\ref{pro:1}(\ref{pro13}). Let
$$y_p=\min\{f'\left(d',p'\right), f'\left(p',t\right)-\ell(p)\}.$$
In this case, we apply $y_p$ many $ld$-\correcting flows.
     
     \noindent{\bf Case 2:} $f'\left(p'',d''\right)>0$. In this case, because $f'\left(d',p'\right)=0$, the pair of nodes $p',p''$ trivially satisfies Property~\ref{pro:1}(\ref{pro13}). Now suppose $f'\left(p'',d''\right)>0$ and $f'\left(p',t\right)=\ell(p)<u(p)$, which implies that the pair of nodes $p',p''$ violates Property~\ref{pro:1}(\ref{pro12}). Let
$$z_p=\min\{f'\left(p'', d''\right), u(p)-f'\left(p', t\right)\}.$$
In this case, we apply $z_p$ many $ud$-\correcting flows.
    
    Let $\Tilde{f}$ be the resulting flow. Observe that the pair of nodes $p',p''$ satisfies all the conditions in Property~\ref{pro:1}, and thus the post $p$ is a good post with respect to the flow $\Tilde{f}$.

Recall that, by applying correcting flows, the status of the other posts remains unchanged. Therefore, we can apply these corrections independently. For all the bad posts with respect to $f$, we apply a sequence of \correcting flows as described above.
Observe that the resulting flow $f^*$ is a feasible flow that satisfies Property~\ref{pro:1} in another network $H_{(ld',ud')}$, where $ld'+ud'<ld+ud$.
\end{proof}

We are ready to design our flow network based algorithm, which employs linear search over  a specific range  to compute an optimal matching. For an optimal matching $M^*$ in $G$, we have the following claims.

\begin{claim}
    \label{cl:sec3cl1}
   The total deviation of $M^*$ is  at least $ \sum\limits_{p \in \B} \ell(p) - \abs{\A}$. 
\end{claim}
\begin{proof} Let $ld^*$ be the total lower deviation of the matching $M^*$. 
For every matched applicant, it may contribute to satisfying the lower target of one post. Thus, $ld^* \ge \sum\limits_{p \in \B} \ell(p) - \abs{M^*}$. 
Since the total deviation of $M^*$, $d_{tot}(M^*)$ is at least $ld^*$, and  $\abs{M^*}\leq \abs{\A}$, it is clear that $d_{tot}(M^*)$ is at least  $ \sum\limits_{p \in \B} \ell(p) - \abs{\A}$.
\end{proof}
\begin{claim}
    \label{cl:sec3cl2}
    The total deviation of $M^*$ is at most $ \sum\limits_{p \in \B} \ell(p) + \abs{\A}$.
\end{claim}
\begin{proof}
Observe that for any matching,  the maximum value possible for upper deviation is $\abs{\A}$, and the maximum possible value for lower deviation is $\sum\limits_{p \in \B} \ell(p)$. Thus, the  total deviation of $M^*$ is at most $ \sum\limits_{p \in \B} \ell(p) + \abs{\A}$.
\end{proof}

Thus, it is enough to search for the value of total deviation $\Delta$ in the range $[\max \{ 0, \sum\limits_{p \in \B} \ell(p)-\abs{A}\}, \sum\limits_{p \in \B} \ell(p)+\abs{A}]$. 
Note that, even if for a particular value of $\Delta$, and a split of $\Delta$ as $ld$ and $ud$, we find a feasible flow in $H_{(ld,ud)}$, it is required to ensure that the signature of the corresponding matching satisfies the signature requirement. We now present the algorithm. 

\begin{algorithm}
\SetAlCapNameFnt{}
    \caption{Algorithm for \RSMTD }\label{algo:1}
    \DontPrintSemicolon
    \SetAlgoLined
    \begin{spacing}{1.2}
    set low= $\max \{ 0, \sum\limits_{p \in \B} \ell(p)-\abs{A}\}$, high=$\sum\limits_{p \in \B} \ell(p)+\abs{A}$, $\Delta=low$ \;
    \end{spacing}
    \While{ $\Delta \le high $}{
        set found=0\;
        \For{each of the $\Delta+1$ ways to split $\Delta$ as $ld$ and $ud$ where $\Delta=ld+ud$}{
        construct $H_{(ld,ud)}$ \label{stepb}\;
        \If {$H_{(ld,ud)}$ admits a feasible flow } { 
        compute a min-cost flow $f$ and corresponding matching $M$\;
            \If{ \label{stepa} $\sigma(M)\succeq_{\mathcal{R}} \rho$}{
            set  found=1 \;
            }
        }
        }
        \If {found==0} {set $\Delta=\Delta+1$\;}
        \Else{return $f^*=f$ and $M^*=M$\;}
    }
\end{algorithm}

 To compute \RSMTD, we use linear search to find the optimum value of $\Delta$. Note that,  for a fixed value of $\Delta$, there are $\Delta+1$ ways to split the total deviation as sum of lower deviation $ld$ and upper deviation $ud$. For each of these values, we construct the network $H_{(ld,ud)}$ and compute a min-cost flow  $f$ in the network. Let $M$ be the matching corresponding to the flow $f$. If $M$ has a signature $\sigma(M)\succeq_{\mathcal{R}} \rho$, then the optimal value of total deviation is at most $\Delta$.
 The pseudocode is presented  as \cref{algo:1}.

\noindent\textbf{Correctness of the Algorithm for \RSMTD.}
Given the lower and upper bound for the deviation of the optimal matching $M^*$, it is clear that a linear search over the range works. Now, we will show that, at \cref{stepa} of our algorithm, if $\sigma(M)\prec_{\mathcal{R}} \rho$, then we can discard the split $(ld,ud)$, and this is described in Claim \ref{cl:sec3cl4}.

\begin{claim}
    \label{cl:sec3cl4}
    If at \cref{stepa}, $\sigma(M)\prec_{\mathcal{R}} \rho$,  then there does not exist any matching $M'$ with total lower deviation $ld$ and total upper deviation $ud$ such that $\sigma(M') \succeq_{\mathcal{R}} \rho$.
\end{claim}
\begin{proof}
    Note that any matching in $G$ with a total lower deviation $ld$ and total upper deviation $ud$ has a feasible flow in the network $H_{(ld,ud)}$. In \cref{stepa} of our algorithm, we compute a matching $M$ corresponding to a min-cost flow $f$ in the network. By the cost assignment of $\mathcal{E}_2$ edges in $H_{(ld,ud)}$, the matching $M$ is a rank-maximal matching among all matchings with total lower deviation $ld$ and total upper deviation $ud$. Thus if at \cref{stepa}, $\sigma(M)\prec_{\mathcal{R}} \rho$, then the signature of any matching   with  total lower deviation $ld$ and total upper deviation $ud$ is strictly less rank-maximal than the signature requirement $\rho$.
\end{proof}
% \textcolor{blue}{Suppose that our network $H_{(ld,ud)}$ admits a feasible flow $f$ such that the corresponding matching $M$ satisfies the signature requirement. We show that corresponding to any $\Delta'=ld'+ud'>ld+ud$, there exist at least one feasible flow $f'$ in $H_{(ld',ud')}$ such that the matching $M'$ corresponding to $f'$ meets the signature requirement.Note that for a particular value of $\Delta$, and for all the $(ld,ud)$ split such that $ld+ud=\Delta$, if the network $H_{(ld,ud)}$ does not admit any feasible flow, then we will move to a value $\Delta'>\Delta$.
% }

% Note that any matching in $G$ with a total lower deviation $ld^*$ and total upper deviation $ud^*$ has a feasible flow in the network $H_{(ld^*,ud^*)}$. The feasible flows corresponding to matchings that have a total lower deviation $ld^*$ and total upper deviation $ud^*$ are good flows in  $H_{(ld^*,ud^*)}$. By the cost assignment of $\mathcal{E}_2$ edges in $H_{(ld^*,ud^*)}$, the matching $M^*$ corresponding to the flow $f^*$ is a rank-maximal matching among all matchings with total lower deviation $ld^*$ and total upper deviation $ud^*$. 

% Since our algorithm computes rank-maximal matching that has a total lower deviation  $ld^*$, and a total upper deviation $ud^*$, if $G$ admits a  matching that has a signature at least as good as $\rho$, then it will appear in our computation.
Now assume that our algorithm outputs a solution. 
Let $\Delta^*$ be the value of $\Delta$ for which our algorithm stops, and let $ld^*$ and $ud^*$ be the corresponding split for lower deviation and upper deviation, respectively. Our flow network $H_{(ld^*,ud^*)}$ computes a min-cost flow, say $f^*$, which has a total deviation of $\Delta^*$. We have the following claim.

\begin{claim}
    \label{cl:sec3cl5}
    The flow $f^*$ is a good flow in $H_{(ld^*,ud^*)}$.
\end{claim}
\begin{proof}
If this is not the case, then $f^*$ violates at least one of the conditions of Property~\ref{pro:1}. Corresponding to the violated condition, we have a correcting flow in $H_{(ld^*,ud^*)}$, and after applying it, we get a good flow $f'$ in another network $H_{(ld',ud')}$, where $ld'+ud'<ld^*+ud^*$. Further, the matchings corresponding to $f^*$ and $f'$ are the same. This contradicts the fact that $\Delta^*$ is the smallest value such that we obtain a matching satisfying the signature requirement.    
\end{proof}
Since $f^*$ is a good flow in $H_{(ld^*,ud^*)}$, the corresponding matching $M^*$ is an optimal solution to the \RSMTD problem.
 \noindent\textbf{Time complexity of \cref{algo:1}.}
For a flow network $H=(V_H,E_H)$, the running time required to compute a min-cost flow is  $T_H=O((\ \abs{E_H} \log \ \abs{V_H} \ )( \ \abs{E_H} + \abs{V_H} \log \ \abs{V_H}\ ))$ using the enhanced capacity scaling algorithm \cite{ahuja1988network}. Given an optimal total deviation value $\Delta^*$, the number of calls to the min-cost flow network $H$ is $\Delta^*+1$. To compute $\Delta^*$, we apply linear search within the range [low, high], where low= $\max \{ 0, \sum\limits_{p \in \B} \ell(p)-\abs{\A}\}$, and high=$\sum\limits_{p \in \B} \ell(p)+\abs{\A}$. Therefore, the time complexity of \cref{algo:1}
 is $O(( (2\abs{\A}) (\sum\limits_{p \in \B} \ell(p)+\abs{\A} + 1) \cdot T_H)$. Note that the running time depends on the value of $\ell(p)$. 
 For a post $p$ with $\ell(p)>\abs{\A}$, we can  preprocess and adjust its $\ell(p)$ value before applying \cref{algo:1} to the modified network. The minimum total deviation value  in the modified network can be translated back to the original network, thereby eliminating dependency on the term $\sum\limits_{p \in \B} \ell(p)$. This ensures  that the time complexity of our \cref{algo:1} is polynomial in the size of the network. In the remaining sections of the paper, we assume that  for any post $p \in \B$, $\ell(p) \le \abs{\A}$.

\paragraph{Illustration of algorithm for \RSMTD} Given the instance $G$ in \cref{fig:target} and an input signature $(4,0,0,3)$, we perform linear search over the range $[low,high]$, where $low=0$, and $high=12$.
For $\Delta=0$ and $\Delta=1$, the network constructed in \cref{stepb} of \cref{algo:1}  does not admit any feasible flow. For $\Delta=3$, none of the splits gives a feasible solution. Although, corresponding to the splits $(2,1)$ and $(3,0)$, the network admits feasible flows, their corresponding matchings fail to meet the signature requirement.

For $\Delta=4$, we construct the network $H_{(ld,ud)}$ for each $(ld,ud)$ pair such that $ld+ud=4$. For the network $H_{(2,2)}$, consider the flow $f$ defined below. For the pairs $(a_1,p_1)$, $(a_3,p_2)$, $(a_4, p_7)$, $(a_5,p_4)$, and $(a_6,p_6)$, the flow $f$ routes $1$ unit of flow through the corresponding true path ($\left< s,a,p',t\right>$ path). It also routes $1$ unit of flow along extended paths for pairs $(a_2,p_1)$ and $(a_7,p_1)$, and along dummy paths for posts $p_3$ and $p_5$.

Observe that $f$ is a min-cost feasible flow  in $H_{(2,2)}$ and satisfies Property~\ref{pro:1}.  Corresponding to the flow $f$, we have a matching $M_3$ listed in \cref{tab:my_label1}. Since $\sigma(M_3)\succeq_{\mathcal{R}} (4,0,0,3)$, the matching is a feasible solution for the problem \RSMTD. 
It is easy to verify that  any matching that meets the signature requirement has a total deviation of at least $4$, and thus the matching $M_3$ is an optimal solution for the \RSMTD problem.

\subsection{Modification for Fairness}
For solving the problem \FSMTD, we construct the same network $H_{(ld, ud)}$ as in \cref{fig:network1}, where $ld+ud=\Delta$. The network $H_{(ld, ud)}$ for  \FSMTD differs from that of \RSMTD in the assignment of cost of the edges in $\mathcal{E}_2$. 
\begin{itemize}
    \item For an edge $(a,p')$ or $(a, p'')$ in $\mathcal{E}_2$, if the corresponding edge $(a,p)$ in $G$ is of rank $i$, then we set cost of that edge as $(n+1)^i$.
\end{itemize}
% \textcolor{red}{newly added}
We have the following lemma regarding the above cost assignment.
\begin{restatable}{lemma}{costfm} \label{lem:cost-fm}
    Let $M, M'$ be matchings such that $\sigma(M) \succ_{\mathcal{F}} \sigma(M')$. Then, $cost(M) < cost(M')$, where each rank $i$ edge has a cost of $(n+1)^{i}$, and every unmatched applicant is treated as being matched to a dummy post of rank $r+1$ with cost $(n+1)^{r+1}$.
\end{restatable}

% \begin{proof}
%     Let $\sigma(M)=(x_1,x_2, \ldots,x_r,x_{r+1})$, and $\sigma(M')=(y_1,y_2, \ldots, y_r,y_{r+1})$. Let $i$ be the first rank at which $\sigma(M)$ and $\sigma(M')$ differ. Thus, $x_j=y_j$ for all $j>i$, and $x_i<y_i$.

%    The cost difference between the matchings $M$ and $M'$ is 
%    \[cost(M')-cost(M) = \sum\limits_{j=1}^{r+1} (y_j -x_j) (n+1)^{j} = (y_i-x_i)(n+1)^{i} + \sum\limits_{j=1}^{i-1} (y_j -x_j) (n+1)^{j}.\]
%    Now, since $y_i - x_i \geq 1$, we have
%    \begin{align*}
%        cost(M')-cost(M) &\geq (n+1)^{i} + \sum\limits_{j=1}^{i-1} (y_j -x_j) (n+1)^{j}\\
%        &\geq (n+1)^{i} - \sum\limits_{j=1}^{i-1} x_j (n+1)^{j}\\ 
%        &\geq (n+1)^{i} - \left(\sum\limits_{j=1}^{i-1} x_j\right) (n+1)^{i-1}.
%    \end{align*}
%    Additionally, using the fact that $\sum\limits_{j=1}^{i-1} x_j \leq \sum\limits_{j=1}^{r+1} x_j = n$, we conclude that
%    \[cost(M')-cost(M) \geq (n+1)^{i} - n (n+1)^{i-1} = (n+1)^{i-1} > 0.\]
% \end{proof}

Note that our algorithm and proofs can be easily modified with respect to the notion of fairness.

%% file: figures/fig_network2.tex
\tikzset{middlearrow/.style={
        decoration={markings,
            mark= at position 0.5 with {\arrow{#1};} ,
        },
        postaction={decorate}
    }
}
\begin{figure}[t] 
\begin{center}
%\begin{subfigure}.55 %initially size was 0.6
\scalebox{0.7}{\begin{tikzpicture} [
roundnode/.style={circle, draw=black!100,  fill=black, inner sep=0pt, minimum size=4pt},
farrow/.style={-Latex}
]
\node[roundnode] (s) at (-8,0) [label=above:{\Large $s$}] {};

\node[roundnode] (t) at (6,4) [label=above:{\Large $t$}] {};
\node[roundnode] (ud) at (6,-2) [label=below:{\Large $d''$}] {};

 \node[roundnode] (ld) at (-5,4) [label=above:{\Large $d'$}] {};
 \node[roundnode] (ap1) at (-5,2) [label=above:{\Large $a_1$}] {};
\node[roundnode] (ap2) at (-5,0) [label=above:{\Large $a_2$}] {};
%\node[roundnode] (r3) at (1,-2.5) [label=below:$r$] {};
\node[roundnode] (ap4) at (-5,-3.5) [label=below:{\Large $a_n$}] {};%(-5,-4)

  \node (ap3)  at (-5,-1)  {$\vdots$};

\draw \boundellipse{-1,2.75}{.65}{2.2};
\node[roundnode] (a1) at (-1,4) [label=above:{\Large $p_1'$}] {};
\node[roundnode] (a2) at (-1,3) [label=above:{\Large $p_2'$}] {};
%\node[roundnode] (r3) at (1,-2.5) [label=below:$r$] {};
\node[roundnode] (a4) at (-1,1) [label=above:{\Large $p_{\m}'$}] {};

  \node (a3)  at (-1,3)  {$\vdots$};

\draw \boundellipse{-1,-2.25}{.65}{2.2};
\node[roundnode] (b1) at (-1,-1) [label=above:{\Large $p_1''$}] {};
\node[roundnode] (b2) at (-1,-2) [label=above:{\Large $p_2''$}] {};
%\node[roundnode] (r3) at (1,-2.5) [label=below:$r$] {};
\node[roundnode] (b4) at (-1,-4) [label=above:{\Large $p_{\m}''$}] {};%(-2,-1)

  \node (b3)  at (-1,-4)  {$\vdots$};

%   \node (d3)  at (-1,-2.2)  {$\vdots$};

 \draw[thick,line width=0.5mm, middlearrow={>}] (s.east) -- (ap1.west) node[midway, above left] {};
 % \draw[->] (s) -- node [above left] {[0,0,1]} (ap1);
  \draw[thick,line width=0.5mm, middlearrow={>}] (s) -- node [above ] {{\Large $[0,1]$}} (ap2);
  \draw[thick,line width=0.5mm, middlearrow={>}] (s) -- node [left] {} (ap4);
  \draw[thick, loosely dashed,line width=0.5mm, middlearrow={>}] (s) -- node [left] {{\Large $[ld,ld]$}} (ld);
  
   \draw[thick,line width=0.5mm, middlearrow={>}] (ap1) -- node [above left] {} (a1);
%   \draw[thick,line width=0.5mm, middlearrow={>}] (ap1) -- node [above left] {} (a4);
  \draw[ thick,line width=0.5mm, middlearrow={>}] (ap2) -- node [above left] {{\Large $[0,1]$}} (a2);
  \draw[thick, line width=0.5mm, middlearrow={>}] (ap4) -- node [left] {} (a4);
  \draw[thick, line width=0.5mm, middlearrow={>}] (ap2) -- node [above] {} (b2);
 \draw[thick, line width=0.5mm, middlearrow={>}] (ap1) -- node [above] {} (b1);
 \draw[thick, line width=0.5mm, middlearrow={>}] (ap2) -- node [above] {} (b2);
 \draw[thick, line width=0.5mm, middlearrow={>}] (ap4) -- node [above] {} (b4);
\draw[thick, line width=0.5mm, middlearrow={>}] (a4) -- node [below right ] {{}} (t);%\Large $[\ell(p_{\m}),u(p_{\m})]$
\draw[thick, line width=0.5mm, middlearrow={>}] (a1) -- node [above ] {{\Large $[\ell(p_{1}),u(p_{1})]$}} (t);
\draw[thick, line width=0.5mm, middlearrow={>}] (a2) -- node [below] {{}} (t);%\Large $[\ell(p_{2}),u(p_{2})]$

 \draw[thick, loosely dashed, line width=0.5mm, middlearrow={>}] (ld) -- node [above] {\Large $[0,ld]$} (a1);
 \draw[thick,  loosely dashed,line width=0.5mm, middlearrow={>}] (ld) -- node [above] {} (a2);
 \draw[thick,  loosely dashed,line width=0.5mm, middlearrow={>}] (ld) -- node [above] {} (a4);

\draw[thick, loosely dashed, line width=0.5mm, middlearrow={>}] (b1) -- node [above] {\Large $[0,ud]$} (ud);
 \draw[thick,  loosely dashed,line width=0.5mm, middlearrow={>}] (b2) -- node [above] {} (ud);
 \draw[thick,  loosely dashed,line width=0.5mm, middlearrow={>}] (b4) -- node [above] {} (ud);
\draw[thick, loosely dashed, line width=0.5mm, middlearrow={>}] (ud) -- node [right] {\Large $[ud,ud]$} (t);
\end{tikzpicture}}
\caption{ Flow network $H_{(ld, ud)}$ for solving \RSMTD, where $\Delta=ld+ud$} 
\label{fig:network1}
   \end{center}
\end{figure}
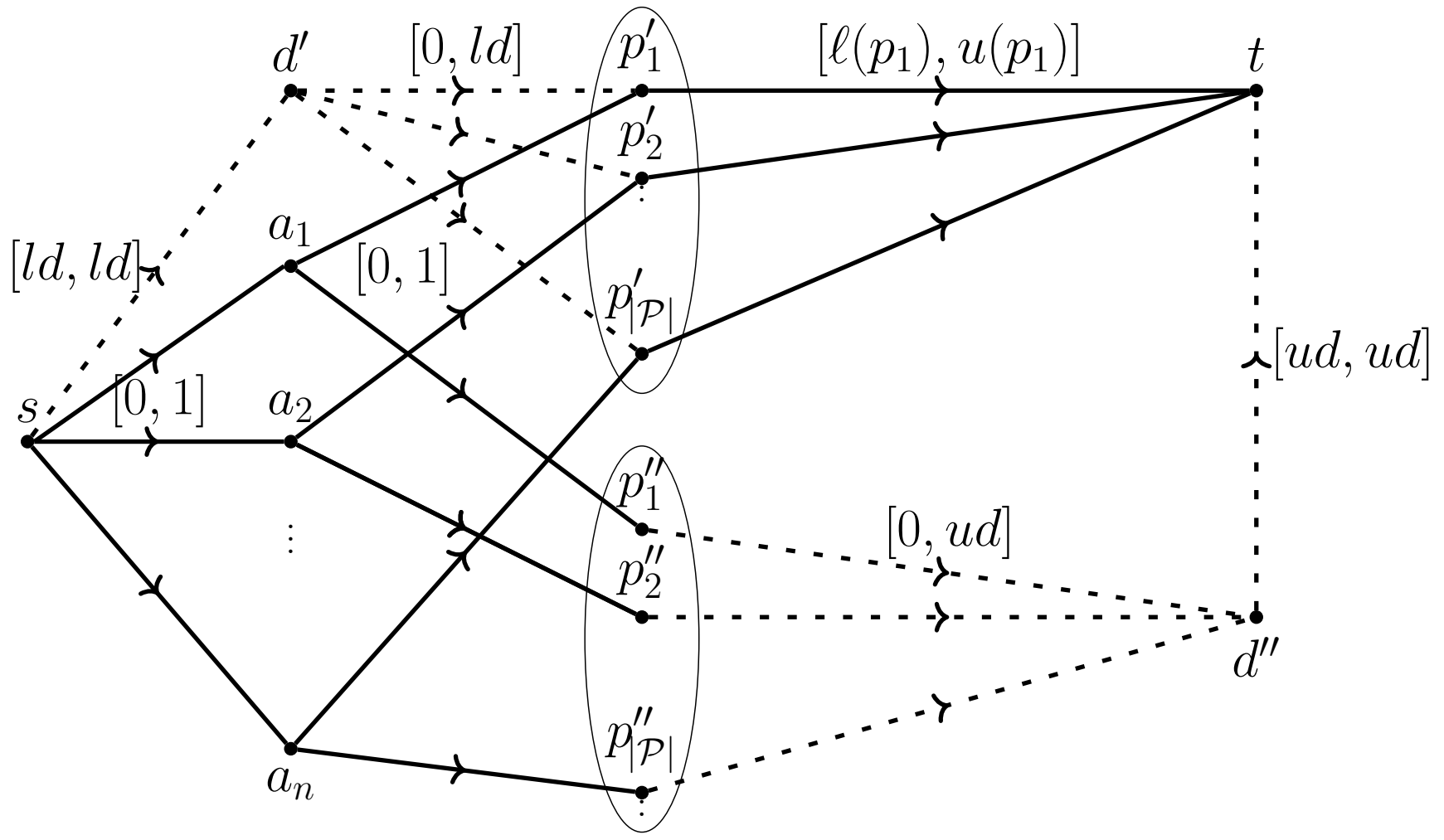

%% file: figures/apfig1.tex
\tikzset{middlearrow/.style={
        decoration={markings,
            mark= at position 0.5 with {\arrow{#1};} ,
        },
        postaction={decorate}
    }
}
\tikzstyle{selected edge} = [draw,line width=7pt,-,blue!20]
\tikzstyle{selected edge1} = [draw,line width=7pt,-,red!20]
\tikzstyle{selected edge2} = [draw,line width=7pt,-,green!20]
\tikzstyle{selected edge3} = [draw,line width=7pt,-,gray!20]

\begin{figure*}
     \centering
     \begin{subfigure}[b]{0.45\textwidth}
         \centering
         \scalebox{0.6}{\begin{tikzpicture} [roundnode/.style={circle, draw=black!100,  fill=black, inner sep=0pt, minimum size=4pt},farrow/.style={-Latex}]
\node[roundnode] (s) at (-8,2) [label=left:{\large $s$}] {};
\node[roundnode] (t) at (1,4) [label=above:{\large $t$}] {};
\node[roundnode] (ud) at (1,0) [label=below:{\large $d''$}] {};
 \node[roundnode] (ld) at (-6,4) [label=above:{\large $d'$}] {};
 \node[roundnode] (ap1) at (-6,0) [label=below:{\large $a$}] {};
\node[roundnode] (a1) at (-3,4) [label=above:{\large $p'$}] {};
\node[roundnode] (b1) at (-3,0) [label=below:{\large $p''$}] {};
\draw[selected edge1] (s.center) -- (ap1.center);
\draw[selected edge1] (ap1.center) -- (b1.center);
\draw[selected edge1] (b1.center) -- (ud.center);
\draw[selected edge1] (ud.center) -- (t.center);
\draw[selected edge1] (s.center) -- (ld.center);
\draw[selected edge1] (ld.center) -- (a1.center);
\draw[selected edge1] (a1.center) -- (t.center);
% \draw[selected edge] (ud.center) -- (t.center);
 \draw[thick,line width=0.5mm, middlearrow={>}] (s.center) -- (ap1.center) node[midway, below left] {{\Large $[0,1]$}};
  \draw[thick, loosely dashed,line width=0.5mm, middlearrow={>}] (s) -- node [left] {{\Large $[1,1]$}} (ld);
  
   \draw[thick,line width=0.5mm, middlearrow={>}] (ap1) -- node [above left] {{\Large $[0,1]$}} (a1);
 \draw[thick, line width=0.5mm, middlearrow={>}] (ap1) -- node [above] {{\Large $[0,1]$}} (b1);
\draw[thick, line width=0.5mm, middlearrow={>}] (a1) -- node [above ] {{\Large $[0,1]$}} (t);

 \draw[thick, loosely dashed, line width=0.5mm, middlearrow={>}] (ld) -- node [above] {\Large $[0,1]$} (a1);

\draw[thick, loosely dashed, line width=0.5mm, middlearrow={>}] (b1) -- node [above] {\Large $[0,1]$} (ud);
\draw[thick, loosely dashed, line width=0.5mm, middlearrow={>}] (ud) -- node [right] {\Large $[1,1]$} (t);

\node[roundnode] (s) at (-8,2) [label=left:{\large $s$}] {};
\node[roundnode] (t) at (1,4) [label=above:{\large $t$}] {};
\node[roundnode] (ud) at (1,0) [label=below:{\large $d''$}] {};
 \node[roundnode] (ld) at (-6,4) [label=above:{\large $d'$}] {};
 \node[roundnode] (ap1) at (-6,0) [label=below:{\large $a$}] {};
\node[roundnode] (a1) at (-3,4) [label=above:{\large $p'$}] {};
\node[roundnode] (b1) at (-3,0) [label=below:{\large $p''$}] {};

\end{tikzpicture}}

         \caption{Scenario 1: Total deviation of the flow $f$ (marked in red) is $2$ and that of the corresponding matching $M=\{(a,p)\}$ is $0$. The example illustrates the violation of Property \ref{pro:1}.(i) in the network $H_{(1,1)}$ and thus $f$ is not a good flow.}
         \label{fig:sc1}
     \end{subfigure}      \hspace{0.25in}
     \begin{subfigure}[b]{0.45\textwidth}
         \centering
         \scalebox{0.6}{\begin{tikzpicture} [
roundnode/.style={circle, draw=black!100,  fill=black, inner sep=0pt, minimum size=4pt},
farrow/.style={-Latex}
]
\node[roundnode] (s) at (-8,2) [label=left:{\large $s$}] {};
\node[roundnode] (t) at (1,4) [label=above:{\large $t$}] {};
\node[roundnode] (ud) at (1,0) [label=below:{\large $d''$}] {};
 \node[roundnode] (ld) at (-6,4) [label=above:{\large $d'$}] {};
 \node[roundnode] (ap1) at (-6,0) [label=below:{\large $a$}] {};
\node[roundnode] (a1) at (-3,4) [label=above:{\large $p'$}] {};
\node[roundnode] (b1) at (-3,0) [label=below:{\large $p''$}] {};
\draw[selected edge2] (s.center) -- (ap1.center);
\draw[selected edge2] (ap1.center) -- (a1.center);
% \draw[selected edge] (b1.center) -- (ud.center);
% \draw[selected edge] (ud.center) -- (t.center);
% \draw[selected edge] (s.center) -- (ld.center);
% \draw[selected edge] (ld.center) -- (a1.center);
\draw[selected edge2] (a1.center) -- (t.center);
% \draw[selected edge] (ud.center) -- (t.center);
 \draw[thick,line width=0.5mm, middlearrow={>}] (s.center) -- (ap1.center) node[midway, below left] {{\Large $[0,1]$}};
  \draw[thick, loosely dashed,line width=0.5mm, middlearrow={>}] (s) -- node [left] {{\Large $[0,0]$}} (ld);
  
   \draw[thick,line width=0.5mm, middlearrow={>}] (ap1) -- node [above left] {{\Large $[0,1]$}} (a1);
 \draw[thick, line width=0.5mm, middlearrow={>}] (ap1) -- node [above] {{\Large $[0,1]$}} (b1);
\draw[thick, line width=0.5mm, middlearrow={>}] (a1) -- node [above ] {{\Large $[0,1]$}} (t);

 \draw[thick, loosely dashed, line width=0.5mm, middlearrow={>}] (ld) -- node [above] {\Large $[0,0]$} (a1);

\draw[thick, loosely dashed, line width=0.5mm, middlearrow={>}] (b1) -- node [above] {\Large $[0,0]$} (ud);
\draw[thick, loosely dashed, line width=0.5mm, middlearrow={>}] (ud) -- node [right] {\Large $[0,0]$} (t);

\node[roundnode] (s) at (-8,2) [label=left:{\large $s$}] {};
\node[roundnode] (t) at (1,4) [label=above:{\large $t$}] {};
\node[roundnode] (ud) at (1,0) [label=below:{\large $d''$}] {};
 \node[roundnode] (ld) at (-6,4) [label=above:{\large $d'$}] {};
 \node[roundnode] (ap1) at (-6,0) [label=below:{\large $a$}] {};
\node[roundnode] (a1) at (-3,4) [label=above:{\large $p'$}] {};
\node[roundnode] (b1) at (-3,0) [label=below:{\large $p''$}] {};

\end{tikzpicture}}
         \caption{An $(ld,ud)$-correcting flow $f'$ (marked in green) in $H_{(1,1)}$ which is feasible in the network $H_{(0,0)}$ and fixes the violation of Property \ref{pro:1}.(i). Note that the total deviation of $f'$ is $0$ and that of the corresponding matching $M=\{(a,p)\}$ is $0$. }
         \label{fig:cf1}
     \end{subfigure}
        \caption{Scenario 1 where $d_{tot}(f) \neq d_{tot}(M)$ and the $(ld, ud)$-correcting flow $f'$.}
        \label{fig:sc1cf1}
\end{figure*}
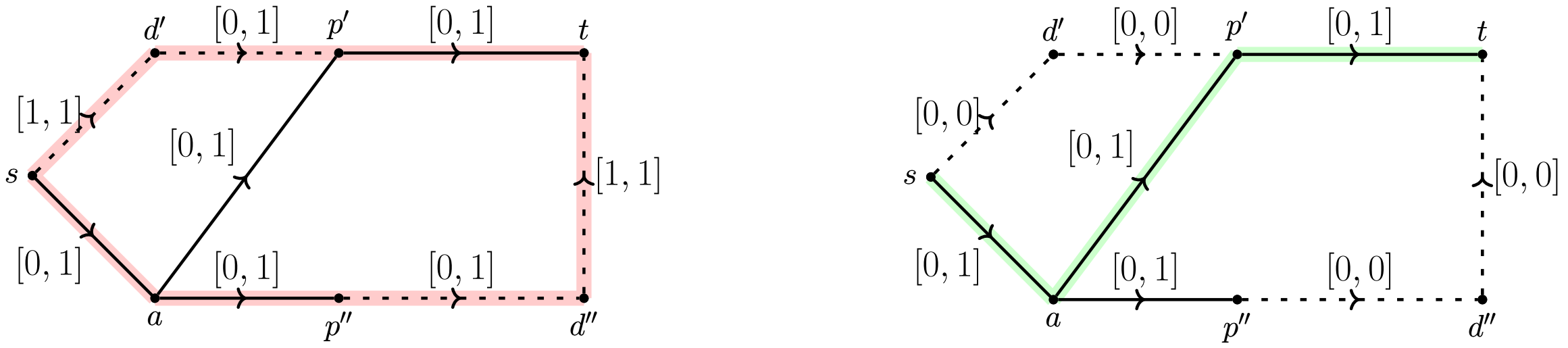

%% file: 3-deviation-first.tex
\section{Minimizing Total Deviation: Deviation First }
In this section, we present the results for minimizing total deviation for the deviation first, preference optimality next type of problems. 
\label{sec:alg-min-y-d}
 \subsection{Algorithm for \RMMTD} \label{sec:alg-min-tot-d}
 Given an instance of the soft quota setting, our goal is to compute a matching that is rank-maximal among all matchings that have a minimum value of total deviation. First, we consider the simpler problem of computing a minimum total deviation matching in $G$ {\em ignoring} applicant preferences.

 Given an instance $G$ of the soft quota setting, we denote the same instance viewed as an instance of the fixed quota setting  by $G_{x}$. Recall that in the fixed quota setting, for a post $p$, we have a fixed lower quota $q^-(p)$ and a fixed upper quota $q^+(p)$ associated with it. A matching $M$ is feasible in the fixed quota setting if, for every post $p \in \B$, $q^-(p) \leq \abs{M(p)} \leq q^+(p)$. We note  that with respect to the problem of minimizing total deviation, there is a useful connection between matchings in  $G$ and  $G_{x}$.
 If the fixed quota setting instance $G_{x}$ admits a feasible matching $M$, then  $M$ has a total deviation of $0$ in the soft quota setting instance $G$. Thus, $M$ is an optimal matching in $G$ as far as deviation is concerned. 
 However, if $G_{x}$ does not admit a feasible matching, any minimum total deviation matching in $G$ has a positive value for the total deviation.

 We observe that in an optimal matching $M^*$ in $G$, no post incurs a violation on the upper target.
 To see this, suppose that for a post $p$, we have more than $u(p)$ applicants matched to it in $M^*$. Then, we can remove the excess number of applicants, namely $\abs{M^*(p)}-u(p)$, matched to $p$ from $M^*$ and obtain a matching with a smaller value of total deviation. We use the above observation and the connection between $G$ and $G_x$ to compute a matching with minimum total deviation without applicant preferences.

\paragraph{Minimizing total deviation}  
 Given the instance $G$, we construct an instance $G'$ of the soft quota setting,  where for every post $p$, $\ell'(p)=u'(p)=\ell(p)$. Consider the fixed quota setting instance $G'_{x}$, where we set $q^-(p)=0$ and $q^+(p)=\ell(p)$ for every post $p$. We show a connection between a maximum matching in $G'_x$ and a minimum total deviation matching in $G'$, as stated in \cref{lem:mmsq}. 
 % We show that any maximum matching in $G'_x$ is a minimum total deviation matching in $G'$, as stated in \cref{lem:mmsq}.

 \begin{lemma} \label{lem:mmsq}
     Any maximum matching in $G'_x$ is a minimum total deviation matching in $G'$.
 \end{lemma}
 \begin{proof}
Let $M'$ be an arbitrary maximum  matching in $G'_{x}$. Using the maximality of $M'$ in the instance $G'_{x}$,
 we show that $M'$ is a minimum total deviation matching in $G'$.
 Suppose not. Let $M$ be a minimum total deviation matching in $G'$. That is, $d_{tot}(M) \text{ in $G'$}$ is less than  $d_{tot}(M')  \text{ in $G'$}$. Since $M$ is an optimal matching in $G'$, for every post $p$, $\abs{M(p)} \leq \ell(p)$, and thus it is a valid matching in $G'_x$. Now we compare the cardinalities of $M$ and $M'$.
 Since $u'(p)=\ell(p)$ for every post $p$, $\abs{M}=\sum\limits_{p \in \B} \ell(p)-d_{tot}(M)$, and $\abs{M'}=\sum\limits_{p \in \B} \ell(p)-d_{tot}(M')$.
 This  implies that $\abs{M} > \abs{M'}$, which is a contradiction to the fact that $M'$ is a maximum matching in $G'_x$. 
 \end{proof}

 Now we establish an interesting connection between the  minimum value of total deviation in the soft quota instance $G'$ and that of the original soft quota setting instance $G$, as stated in Lemma~\ref{cl:clgg'}.
 
 \begin{restatable}{lemma}{cl:clgg'}
\label{cl:clgg'}
The minimum value of total deviation of $G'$ and $G$ are the same.
\end{restatable}
\begin{proof}
Let $k'$ be the total deviation of a minimum total deviation matching in $G'$. Note that $k'$ is the sum of the lower deviations of all posts in $G'$. Since $G' \subseteq G$, the minimum value of the total deviation in $G$, denoted by $k$, satisfies $k \leq k'$.
Suppose that  $k < k'$. Let  $M$ be a matching that achieves the minimum value of total deviation in $G$. Note that $k$ is the sum of lower deviations of all posts in $G$.

In the soft quota setting instance $G'$,  with respect to the matching $M$, some posts may incur violation on the upper target. This is because, in $G'$, the upper target of a post $p$ is $u'(p)=\ell(p)$ which is at most the corresponding upper target $u(p)$ in $G$. Now we construct the matching $M'$ in $G'$, where for every post $p$, we remove $\max\{0, \abs{M(p)}-\ell(p) \}$ number of applicants which are matched to $p$. Note that in $G'$, with respect to the matching $M'$, no post incurs  a violation on the upper target. Thus, the total deviation of the matching $M'$ in $G'$ is $k$, which is the total deviation of $M$ in $G$.  
This contradicts the fact that $k'$ is the minimum value of total deviation in $G'$.
\end{proof}

 By \cref{lem:mmsq} and \cref{cl:clgg'}, any maximum matching $M'$ in $G'_x$ is also a minimum total deviation matching in the original soft quota setting instance $G$. 
 However, $M'$ need not be a maximum matching among all matchings with minimum value of total deviation in $G$.
 To obtain such a matching, 
 we use the  instance $G_{x}$ (recall that every post $p$ in $G_x$ has $q^+(p)=u(p)$)  and augment the matching $M'$ in $G_{x}$. We remark that augmenting a matching  never decreases the number of applicants matched to any post. Thus, augmenting $M'$ in $G_x$ increases the size of the matching while preserving its total deviation.

For an illustration, consider the soft quota instance $G$ given in Figure~\ref{fig:target}. We construct the soft quota instance $G'$, and let $G'_{x}$ be the corresponding fixed quota instance. Consider the
 maximum  matching $M'=\{(a_1,p_1),(a_3,p_3),(a_5,p_5)\}$ in $G'_x$.
 We augment the matching $M'$ in $G_{x}$ to obtain a matching $M=\{(a_1,p_1),(a_3,p_3),(a_5,p_5),(a_6,p_7), (a_7,p_6)\}$. Note that $M$ is a  maximum matching in $G_{x}$, and it has a total deviation of $2$ in $G$, which is the optimum value of total deviation.

 Since we disregarded preferences, the above procedure does not guarantee a matching that is optimal with respect to the preferences. For example, the matching $M$ computed above is not a solution of \RMMTD problem in $G$, since the matching $M_1$ (illustrated in \cref{tab:my_label1}) is more rank-maximal than $M$. 
In the rest of the section, we refine the above ideas to compute an \RMMTD matching in $G$.

\begin{itemize}
     \item We construct the instance $G'$  from $G$ as discussed above. We compute an \RMMTD matching $M$ in $G'$ using  the procedure given in Step~(A) below. 
     \item We augment the matching $M$ in $G_x$ to obtain  an \RMMTD matching $M^*$ in $G$. The procedure for computing $M^*$ is given in Step~(B).
    \end{itemize}
   \paragraph{Step~(A): Computing \RMMTD in $G'$}
Given the instance $G'$, we reduce the problem to computing a min-cost flow in a suitably designed flow network $H$. We show that a matching $M$ corresponding to a  min-cost flow in $H$ has the following properties:
\begin{itemize}
    \item $M$ is a minimum total deviation matching in $G'$ 
    and also in $G$.
    \item $M$ is a rank-maximal matching among all minimum total deviation matchings in $G'$. Note that $M$ need not be a solution to \RMMTD in $G$.
\end{itemize}

We construct the flow network $H$ as follows. For every post $p$, we introduce $\ell(p)$ many dummy applicants and denote this set by $\mathcal{D}_p$.
Let $\mathcal{D}=\bigcup_{p \in \B} \mathcal{D}_p $.
 The vertex set of $H$ is defined as below:
$$V(H)=\{s,t\} \cup (\A \cup \mathcal{D}) \cup \B.$$

The edge set  $E(H)$ is the union of the following sets:
\begin{align*}
    \mathcal{E}_{sA}=&\{(s,a) \ : \ a \in \A\} \\
    \mathcal{E}_{sD}=&\{(s,d) \ : \ d  \in \mathcal{D}_p, p \in \B \}\\ 
    \mathcal{E}_{AP}=&\{(a,p) \ : \ a \in \A, p \in \B\}\\
      \mathcal{E}_{DP}=&\{(d,p) \ : \ d \in \mathcal{D}_p, p \in \B\} \\
    \mathcal{E}_{Pt}=&\{(p,t) \ : \ p \in \B\} 
\end{align*}

\noindent{\bf Cost of the edges in $H$.}
For an edge $(a,p)$  in $G'$ having rank $i$, we set  cost of the corresponding edge in $\mathcal{E}_{AP}$ as $-((n+1)^r+(n+1)^{r-i})$. For all the other edges in $H$, we set the cost as $0$.

\noindent{\bf Capacity of the edges in $H$.}
For an edge $(p,t)$ in $\mathcal{E}_{Pt}$, we set its capacity as $\ell(p)$. All other edges in $H$ have a unit capacity.

Consider the cut $\{t, V(H)\setminus \{t\}\}$, and observe that the capacity of this cut is $\sum\limits_{p \in \B} \ell(p)$. Thus, by max-flow min-cut theorem, the value of any flow is at most $\sum\limits_{p \in \B} \ell(p)$. Note that for every post $p$, we can send $1$ unit of flow via the path $\langle s, d, p,t\rangle$, where $d\in \mathcal{D}_p$, which implies that the value of a max-flow is at least $\sum\limits_{p \in \B} \ell(p)$.
Thus, any max-flow in $H$ has a value of $\sum\limits_{p \in \B} \ell(p)$. We  define the deviation of a max-flow $f$ in $H$ as follows.

\begin{definition}[{\bf Total deviation of a max-flow}]
 The total deviation of a max-flow $f$ in $H$ is defined as 
   $$d_{tot}(f)=\sum\limits_{p\in \B} d_f(p) $$  where 
   $d_f(p)=\sum\limits_{\ d \in D_p} f\left(d,p\right)$
 is the  incoming flow to the vertex $p$ through the dummy applicants.
\end{definition}

There is a natural correspondence between a flow $f$ in $H$ and a matching $M$ in $G'$ as follows:
\begin{align*}
    M_f=\{(a,p) \ | \ f(a,p)=1 , \ (a,p)\in \mathcal{E}_{AP} \}.
\end{align*}

\noindent{\bf Cost of a max-flow in $H$ and total deviation of a matching in $G'$. }
For a max-flow $f$ in $H$, we define the total cost of the flow, denoted by $cost(f)$, as follows:
\begin{align*}
    cost(f)=\sum\limits_{e \in E(H)} f(e) \cdot c(e)
\end{align*} where $f(e)$ and $c(e)$ represent the flow and the cost on  the edge $e$, respectively. By the design of our network $H$, we ensure that the cost of a max-flow in $H$ corresponds to the total deviation of the corresponding matching in $G'$. 
This is proved using  \cref{cl:td-rmmlq4-1}.

\begin{lemma}
\label{cl:td-rmmlq4-1}
Let $f, f'$ be two flows in $H$, and let $M_f, M_{f'}$ be the corresponding matchings in $G'$. If \ $\abs{M_f}>\abs{M_{f'}}$, then $cost(f)<cost(f')$
\end{lemma}
\begin{proof}
For a post $p$, the incoming flow is via edges in $\mathcal{E}_{DP}$, and $\mathcal{E}_{AP}$. In $H$, only the edges in  $\mathcal{E}_{AP}$ have a non-zero cost, and recall the cost of an edge in $\mathcal{E}_{AP}$. Observe that cost of the flow $f$ can be upper bounded by
$$cost(f) \leq \abs{M_f}(-(n+1)^r) - \abs{M_f}$$ where the equality is achieved  when all edges are of rank $r$. Similarly, the  cost of the flow $f'$ can be lower bounded by  $$cost(f') \geq \abs{M_{f'}}(-(n+1)^r) - \abs{M_{f'}}((n+1)^{r-1})$$ and the equality is achieved when all edges are of rank $1$.
To show that $cost(f)<cost(f')$, we show the equivalent statement
$$\abs{M_f}((n+1)^r) + \abs{M_f} > \abs{M_{f'}}((n+1)^r) + \abs{M_{f'}}((n+1)^{r-1}).$$
Since $\abs{M_f} - \abs{M_{f'}} \ge 1$ and $\abs{M_{f'}} \le n$,

\[\begin{aligned}\left(\abs{M_f} - \abs{M_{f'}}\right) (n+1)^r + \abs{M_f} - \abs{M_{f'}}(n+1)^{r-1} &\ge (n+1)^r + \abs{M_f} - n (n+1)^{r-1} \\&= (n+1)^{r-1}+\abs{M_f} > 0.\end{aligned}\]
 Thus, $cost(f)<cost(f')$.
 \end{proof}

\begin{lemma}
\label{cl:td-new2}
    Let $f$ be a min-cost flow in $H$ and let $M_f$ be the corresponding matching in $G'$. Then $M_f$ is a minimum total deviation matching in $G'$.
\end{lemma}
\begin{proof}
Suppose not, let $M^*$ be a minimum total deviation matching in $G'$, and let $f^*$ be the corresponding flow. Recall that in $G'$, every post $p$ has $\ell'(p)=u'(p)=\ell(p)$, and thus $M^*$ is a maximum cardinality matching. Since $\abs{M^*}>\abs{M_f}$, by \cref{cl:td-rmmlq4-1}, we have $cost(f^*)<cost(f)$, which is a contradiction. 
\end{proof}

Given the connection between the total deviations of $G$ and $G'$ established in \cref{cl:clgg'}, which is independent of the preferences, we know that the minimum total deviation in $G$ is equal to that in $G'$. This leads to the following corollary.

\begin{corollary}\label{cor:cor1}
Any min-cost flow in $H$ corresponds to a minimum total deviation matching in $G$.
\end{corollary}

We now prove the preference optimality of the matching obtained in Step~(A).
\begin{lemma}
\label{cl:cl22}
Any min-cost flow in $H$ corresponds to a solution to \RMMTD in $G'$.
\end{lemma}
\begin{proof}
Let $f$ be a min-cost flow in $H$, and let $M_f$ be the corresponding matching in $G'$. By \cref{cor:cor1},  $M_f$ is a minimum total deviation matching in $G'$.
Suppose $M_f$ is not an \RMMTD matching in $G'$. Then there exists another matching $M_{f'}$ that is an \RMMTD matching in $G'$, and let $f'$ be the corresponding flow in $H$. Since $M_{f'}$ is a minimum total deviation matching, $\abs{M_f}=\abs{M_{f'}}$. 

Since $\sigma(M_{f'})\succ_{\mathcal{R}} \sigma(M_f)$,  by the definition of rank-maximality, there exists an index $i$ such that $\sigma(M_{f'})_i \succ \sigma(M_f)_i$ and $\sigma(M_{f'})_j = \sigma(M_f)_j$ for $j< i$.  Recall that, corresponding to a rank $i$ edge in $G'$, we have an edge in $\mathcal{E}_{AP}$ with a cost of $-((n+1)^r+(n+1)^{r-i})$ in $H$. 

Since $\abs{M_f}=\abs{M_{f'}}$, both flows have the same contribution from the term $-(n+1)^r$. 
By \cref{lem:cost-rmm}, the rank-dependent contribution to $cost(f')$ is smaller than that of $cost(f)$. Thus, $cost(f')<cost(f)$, which is a contradiction.
\end{proof}

 We observe that every minimum total deviation matching in $G$ has cardinality of at least $\abs{M}$.
\begin{lemma}\label{lem:cardinality}
Let $N$ be a matching in $G$ such that $d_{tot}(N)=d_{tot}(M)$. Then $\abs{N} \ge \abs{M}$.
\end{lemma}
\begin{proof}
    Since $M$ is a maximum matching in the restricted instance $G'$, we have $d_{tot}(M)=\sum\limits_{p \in \mathcal{P}} \ell(p) - \abs{M}$. Now let $N$ be a matching in $G$. Observe that $d_{tot}(N)\ge \sum\limits_{p \in \mathcal{P}} \ell(p) - \abs{N}.$ Since $d_{tot}(N)=d_{tot}(M)$, we obtain $\sum\limits_{p \in \mathcal{P}} \ell(p) - \abs{M} \ge \sum\limits_{p \in \mathcal{P}} \ell(p) - \abs{N}$, which implies $\abs{N} \ge \abs{M}$.
\end{proof}

In Step~(A), we compute a matching $M$ that is an \RMMTD matching in $G'$. In Step~(B), we show how to compute a matching $M^*$ that is an \RMMTD matching in $G$ by augmenting the matching $M$ in the fixed quota setting instance $G_{x}$. We augment the matching $M$ such that, in the $j$-th augmentation, we obtain a rank-maximal matching $M_j$ among all matchings of cardinality $\abs{M}+j$ in $G_x$ such that $d_{tot}(M)=d_{tot}(M_j)$.

\paragraph{Step~(B): Computing a solution for \RMMTD in $G$ by augmenting $M$ in $G_x$}
Our algorithm is iterative, and we begin with the matching $M_0=M$. Suppose that after the iteration $j-1$, we have computed a matching $M_{j-1}$ having cardinality $\abs{M}+j-1$, and total deviation $d_{tot}(M)$. In iteration $j$, we augment $M_{j-1}$ in $G_x$ by computing an $s$-$t$ path in a directed graph $H_j$ with costs as defined below. We show that the resulting matching $M_j$ is a rank-maximal matching among all matchings having cardinality $\abs{M}+j$ and total deviation $d_{tot}(M)$. Since a rank-maximal matching among all minimum total deviation matchings need not have maximum cardinality, we retain the most rank-maximal matching obtained during the augmentation procedure. That is, initially we set $M^*=M_0$. After computing a matching $M_j$, if $\sigma(M_j)\succ_{\mathcal{R}} \sigma(M^*)$, we update $M^*=M_j$. The augmentation procedure continues until the graph $H_j$ does not admit an $s$-$t$ path. At the end of Step~(B), we return $M^*$.

We now describe the construction of the directed graph $H_j$. The vertices of $H_j$ are defined as follows:
\begin{align*}
    V(H_j)=\{s,t\} \cup \A \cup \B.
\end{align*}

Let $\mathcal{A}'$ denote the set of applicants that are unmatched in $M_{j-1}$, and let $\mathcal{P}'$ denote the set of posts with $\ell(p)\leq \abs{M_{j-1}(p)}<u(p)$. Note that every post $p \in \mathcal{P}'$ satisfies $d_{M_{j-1}}(p)=0$, and $\mathcal{P}'$ is exactly the set of posts to which one additional applicant can be matched without increasing the total deviation. We claim that it is sufficient to restrict our attention to posts in $\mathcal{P}'$ because:
\begin{itemize}
    \item[(i)] For any iteration $j$, if we have a post $p$ with $\abs{M_{j-1}(p)} < \ell(p)$, then all the applicants in the neighbourhood of $p$ must be matched in $M_{j-1}$. Suppose not. Let $a$ be an unmatched applicant in the neighbourhood of $p$ with respect to the matching $M_{j-1}$. Then, by adding the edge $(a,p)$ to $M_{j-1}$, we could reduce the deviation of $p$, and thus reduce the total deviation. For $j=0$, this would contradict the fact that the matching $M_0=M$ is a matching with minimum total deviation in both $G'$ and $G$. 

    \item[(ii)] In an optimal matching, no post incurs a deviation from the upper target.
\end{itemize}
 Now, we define our edge set $E(H_j)$ as follows:
\begin{align*}
E(H_j)=&\mathcal{E}_{sA'} \cup \mathcal{E}_{AP} \cup \mathcal{E}_{PA} \cup \mathcal{E}_{P't}  \ \text{where,}\\
    \mathcal{E}_{sA'}=&\{(s,a) \ : \ a \in \A'\} \\%\hfill 
        \mathcal{E}_{AP}=&\{(a,p) \ : \ (a,p) \in E \setminus M_{j-1} \}\\
        \mathcal{E}_{PA}=&\{(p,a) \ : \ (a,p) \in E \cap M_{j-1}\}\\
        \mathcal{E}_{P't}=&\{(p,t) \ : \ p \in \B'\}
\end{align*}

\noindent {\bf Cost of the edges in $H_j$.} 
For an edge $(a,p) \in E$ of rank $i$, 
\begin{itemize}
    \item  if $(a,p) \in \mathcal{E}_{AP}$,   cost of the edge is $-(n+1)^{r-i}$
    \item  if $(p,a) \in \mathcal{E}_{PA}$,  cost of the edge is $(n+1)^{r-i}$ 
\end{itemize}
All other edges in $H_j$ have a cost of $0$.

The following observation establishes the correspondence between the alternating components in the symmetric difference of two matchings and the directed graph $H_j$.
\begin{obs} \label{lem:alternating}
    Let $N$ be a matching in $G_x$ having cardinality $\abs{M}+j$ and total deviation $d_{tot}(M)$. Then every alternating cycle in $M_{j-1} \oplus N$ corresponds to a directed cycle in $H_j$, and every augmenting path in $M_{j-1} \oplus N$ corresponds to a directed $s$-$t$ path in $H_j$.
\end{obs}
% \begin{proof}
%     Consider an alternating cycle in $M_{j-1} \oplus N$. Since matched and unmatched edges alternate on the cycle, the orientation of the edges alternate between applicant to post and post to applicant in $H_j$. Hence the alternating cycle becomes a directed cycle in $H_j$. Now consider an augmenting path in $M_{j-1} \oplus N$. The path starts at an unmatched applicant in $M_{j-1}$ and ends at a post $p \in \mathcal{P}'$. Suppose not, if an augmenting path ended at a post $p$ with $\abs{M_{j-1}(p)}<\ell(p)$, augmenting along the path would reduce the total deviation, contradicting the optimality of $M_{j-1}$. Since there is an edge from source $s$ to every unmatched applicant, and every post in $\mathcal{P}'$ has an edge to the sink $t$, the corresponding directed path in $H_j$ is an $s$-$t$ path.
% \end{proof}

We now prove an important property of the directed graph $H_j$.
\begin{claim} \label{cl:no-negative-cycle}
    There are no negative cost directed cycles in $H_j$.
\end{claim}
\begin{proof}
   We prove the claim by induction on the number of iterations. Base case is trivial as the initial matching $M_0=M$ is a rank-maximal matching among all matchings having cardinality $\abs{M}$ and total deviation $d_{tot}(M)$. Assume inductively that $M_{j-1}$ is a rank-maximal matching among all matchings having cardinality $\abs{M}+j-1$ and total deviation $d_{tot}(M)$. Suppose, for contradiction, that $H_j$ contains a negative cost directed cycle $C$. By the construction of $H_j$, the directed cycle $C$ corresponds to an alternating cycle with respect to $M_{j-1}$.
   
   Let $M'=M_{j-1} \oplus C$. Since $C$ is an alternating cycle, $\abs{M'}=\abs{M_{j-1}}=\abs{M}+j-1$. Further, switching along an alternating cycle does not change the number of applicants assigned to any post. Hence, $d_{tot}(M')=d_{tot}(M_{j-1})=d_{tot}(M)$. Since $C$ has negative cost, $cost(M') < cost(M_{j-1})$. By  \cref{lem:cost-rmm}, $\sigma(M')\succ_{\mathcal{R}} \sigma(M_{j-1})$, contradicting the induction hypothesis. Hence, $H_j$ does not contain any negative cost directed cycle.
\end{proof}
%

%

% We observe the following about our constructed directed graph $H_j$.
% \begin{claim}
%     There are no negative cost cycles in $H_j$.
% \end{claim}
% \begin{proof}
%     Suppose not; let $C$ be a negative-cost directed cycle in $H_j$. Note that this directed cycle $C$ is an alternating cycle with respect to the matching $M_{j-1}$ in the underlying undirected graph. Let $M'=M_{j-1} \oplus C$ be the matching obtained by switching the matched and unmatched edges along the cycle $C$. Note that $\abs{M'}=\abs{M_{j-1}}=\abs{M}+j-1$. It is easy to see that, by our cost assignment, $\sigma(M') \succ_{\mathcal{R}} \sigma(M_{j-1}$). This contradicts the fact that $M_{j-1}$ is a rank-maximal matching among all matchings in $G_x$ with cardinality $\abs{M}+j-1$. 
% \end{proof}

Now we describe our augmentation process. Let $T'_j$ denote a min-cost $s$-$t$ path in $H_j$, and let $T_j=T'_j \cap E$ be the underlying undirected path in $G_x$. From the construction of the directed graph $H_j$, it is easy to see that $T_j$ is an augmenting path with respect to $M_{j-1}$. We augment the matching $M_{j-1}$ using $T_j$, and the resulting matching is $M_j=M_{j-1}\oplus T_j$.

For a set $S\subseteq E$, let $\#_i(S)$ denote the number of rank $i$ edges in $S$. We require Definition~\ref{def:sig-path} to prove correctness.
\begin{definition}[Signature of a path $T_j$]
\label{def:sig-path}
 We define the signature of a path $T_j$ with respect to the matching $M_{j-1}$   as an $r$-tuple $(x_1,x_2, \ldots,x_r)$ denoted by $\sigma(T_j)$, where  
 \begin{align*}
     \textcolor{black}{ x_i=\text{$\#_i ( T_j \cap M_{j-1}) \ - \ $}\text{$\#_i (T_j \setminus M_{j-1})$}} 
 \end{align*}
\end{definition}
Note that, unlike the signature of a matching, which contains non-negative values, $\sigma(T_j)$ may contain negative values. It is useful to note the connection between the signature of a path $\sigma(T_j)$ and the cost of the path $cost(T_j)$.
 \[cost(T_j)=\sum\limits_{i=1}^r x_i (n+1)^{r-i}.\]

We observe the following:
\begin{obs}
\label{obs:1}
$\sum\limits_{i=1}^r x_i=-1.$
\end{obs}
Since $T_j$ is an  augmenting path, the number of unmatched edges in $T_j$ is one more than the number of matched edges in $T_j$. Thus, we have  Observation \ref{obs:1}. 
Let $\sigma(M_{j-1})=(y_1,y_2, \ldots,y_r)$.
For any $i$, the value of $x_i$ cannot be greater than the number of matched rank $i$ edges in $M_{j-1}$. This is stated as Observation \ref{obs:2}.
\begin{obs}
\label{obs:2}
For any $i\in [r]$, 
$x_i \leq y_i$.
\end{obs}
Since $M_j=M_{j-1}\oplus T_j$, it is easy to observe that $\sigma(M_j)=\sigma(M_{j-1})-\sigma(T_j)$. 
\begin{obs}
\label{obs:3}
Every component in $\sigma(M_j)$ is non-negative.
\end{obs}
  Observation \ref{obs:3} follows from Observation \ref{obs:2}. We state a series of claims to prove the correctness of Step~(B). We ensure that our augmentations do not change the value of total deviation of the resulting matching, and
  this is stated in Claim~\ref{cl:mtd1}.

\begin{restatable}{claim}{cltwentyfive}
\label{cl:mtd1}
If $d_{tot}(M)=k$, then $d_{tot}(M_j)=k$.
\end{restatable}
\begin{proof}
We need to show that at each iteration $j$, the augmentation preserves the value of total deviation. 
We use induction to prove the claim. The base case is trivial, that is, when $j=0$, the initial matching $M_0=M$ has a total deviation of $k$. We assume that the total deviation of matching $M_{j-1}$ computed in iteration $j-1$ is $k$.
Let $T_j'$ be a minimum cost $s$-$t$ path in $H_j$, and let $T_j$ be the corresponding  augmenting path in iteration $j$. The path $T_j$ starts at an unmatched applicant $a$ in $M_{j-1}$ and ends at a post $p$ with $\ell(p) \leq \abs{M_{j-1}(p)} < u(p)$. After augmenting $M_{j-1}$ with $T_j$,
a post $p'\ne p $ in $\B$ has $\abs{M_j(p')}=\abs{M_{j-1}(p')}$, and thus the deviation of this post remains unchanged. For the post $p$, $\abs{M_j(p)}=\abs{M_{j-1}(p)}+1$, and since $\ell(p) \leq \abs{M_{j-1}(p)} < u(p)$,  the deviation of post $p$ remains $0$.  Hence, the total deviation $d_{tot}(M_j)=k$. Thus, a  matching computed in any iteration of Step~(B) has a total deviation of $k$. 
\end{proof}

Next, we show that as the cost of path decreases, signature of the path with respect to rank-maximality also decreases. 
\begin{restatable}{claim}{cltwentyseven}
\label{cl:mtd3}
Let $T'_j$ and $Q'_j$ be two $s$-$t$ paths in $H_j$, and let $M_T=M_{j-1} \oplus T_j$, and $M_Q=M_{j-1} \oplus T_Q$. If $cost(T'_j)<cost(Q'_j)$, then, $\sigma(M_T)\succ_{\mathcal{R}}\sigma(M_Q)$. Equivalently, $\sigma(T_j)\prec_{\mathcal{R}}\sigma(Q_j)$.
\end{restatable}
\begin{proof}
By our cost assignment, $cost(T_j')=cost(M_T)-cost(M_{j-1})$, and $cost(Q_j')=cost(M_Q)-cost(M_{j-1})$. Therefore, $cost(T_j')<cost(Q_j')$ implies $cost(M_T)<cost(M_Q)$. Since $M_T$ and $M_Q$ are matchings, by \cref{lem:cost-rmm}, $\sigma(M_T) \succ_{\mathcal{R}} \sigma(M_Q)$. Since, $\sigma(M_T)=\sigma(M_{j-1})-\sigma(T_j)$, and $\sigma(M_Q)=\sigma(M_{j-1})-\sigma(Q_j)$, $\sigma(M_T) \succ_{\mathcal{R}} \sigma(M_Q)$ is equivalent to $\sigma(T_j)\prec_{\mathcal{R}}\sigma(Q_j)$.
\end{proof}
% \begin{proof}
% Suppose not. Then $\sigma(T_j)\succeq_{\mathcal{R}} \sigma(Q_j)$. First, suppose that $\sigma(T_j)=\sigma(Q_j)$. Then, by the definition of cost, 
% $cost(T_j)=cost(Q_j)$, which is a contradiction. Let $\sigma(T_j)=(x_1,x_2, \ldots, x_r)$ and $\sigma(Q_j)=(z_1,z_2, \ldots, z_r)$ be the signatures of $T_j$ and $Q_j$, respectively. Suppose that $\sigma(T_j) \succ_{\mathcal{R}} \sigma(Q_j)$. Then, by the definition of rank-maximality, there exists an index $i$ such that $x_k=z_k$ for $k < i$ and $x_i > z_i$. 
% The cost difference is
% \[cost(T_j)-cost(Q_j)=\sum\limits_{k=1}^r(x_k-z_k)(n+1)^{r-k}\]

% Since $x_k=z_k$  for $k < i$, and $x_i - z_i \ge 1$, we have
% \begin{align*}
%     cost(T_j)-cost(Q_j)&\ge (n+1)^{r-i} - \sum\limits_{k=i+1}^r z_k(n+1)^{r-k} \\
%     &\ge (n+1)^{r-i}-\left( \sum\limits_{k=i+1}^r z_k\right) (n+1)^{r-i-1}.
% \end{align*}
%  Since $T_j$, and $Q_j$ are augmenting paths, $\sum\limits_{k=1}^r z_k=-1$, and hence $\sum\limits_{k=i+1}^r z_k \le n$. Therefore, \[cost(T_j)-cost(Q_j) \ge (n+1)^{r-i} - n(n+1)^{r-i-1} =(n+1)^{r-i-1}>0.\]
% Thus, $cost(T_j) > cost(Q_j)$, which is a contradiction.
% % Since the $i$-th component of signature has a cost of $n^{r-i}$, this implies that $cost(T_j) > cost(Q_j)$, which is a contradiction.
% \end{proof}

%%%%%
To prove the correctness of the augmentation step, we assume inductively that $M_{j-1}$ is a rank-maximal matching among all matchings having cardinality $\abs{M_{j-1}}$ and total deviation $d_{tot}(M)$, and  establish the following lemma.

\begin{lemma}\label{lem:structure}
    Let $N$ be a matching in $G_x$ such that $\abs{N}=\abs{M_{j-1}}+1$, and $d_{tot}(N)=d_{tot}(M_{j-1})$. Then there exists an $M_{j-1}$ augmenting path $Q$ in $M_{j-1} \oplus N$, corresponding to a directed $s$-$t$ path $Q'$ in $H_j$, such that $cost(N)-cost(M_{j-1}) = cost(Q')$.
\end{lemma}
\begin{proof}
    The symmetric difference $M_{j-1} \oplus N$ can be decomposed into alternating path, alternating cycle and augmenting path components. Since, $\abs{N}=\abs{M_{j-1}}+1$, the decomposition contains at least one $M_{j-1}$ augmenting path. Let $Q$ be such an augmenting path, terminating at a post $p$. 
     Augmenting $M_{j-1}$ along $Q$ increases the number of applicants matched to $p$  by exactly one and does not change the number of applicants matched to any other post. The post $p$ must be unsaturated in $M_{j - 1}$ and cannot be undersubscribed in $M_{j - 1}$; otherwise augmenting along $Q$ would decrease the total deviation, contradicting the minimum total deviation of $M_{j-1}$. Thus, $\ell(p) \leq \abs{M_{j-1}(p) <u(p)}$, and $p \in \mathcal{P'}$. By Observation~\ref{lem:alternating}, $Q$ corresponds to a directed $s$-$t$ path $Q'$ in $H_j$.
     % Moreover, this post must be unsaturated in $M_{j - 1}$, and cannot be a undersubscribed in $M_{j - 1}$. Thus, the total deviation cannot change.

    Let $H = (M_{j - 1} \oplus N) \setminus Q$. Consider the graph $H'$ obtained by augmenting $M_{j - 1}$ along $H$. Note that augmenting $H'$ further by $Q$ would give $N$. Thus, it must be the case that $\abs{H'} = \abs{M_{j - 1}}$ and $d_{tot}(H') = d_{tot}(M_{j - 1})$. Consequently, by the optimality of $M_{j - 1}$, it must be the case that $cost(M_{j - 1}) = cost(H') = cost(N) -cost(Q)$. By rearranging, we obtain $cost(N)-cost(M_{j-1}) = cost(Q')$.
    % the required result.
\end{proof}

    Now we prove the following property about the matching $M_j$ computed in iteration $j$.
    \begin{lemma} \label{lem:mj-rmm}
        The matching $M_j$ computed in iteration $j$ is rank-maximal among all matchings with cardinality $\abs{M}+j$ and total deviation $d_{tot}(M)$.
    \end{lemma}
    \begin{proof}
        We prove this by induction on the number of iterations. The base case $j=0$ follows from Step~(A). Assume that $M_{j-1}$ is a rank-maximal matching among all matchings having cardinality $\abs{M}+j-1$ and total deviation $d_{tot}(M)$.  Now consider the graph $H_j$. By Claim~\ref{cl:no-negative-cycle}, there are  no negative cost directed cycles in $H_j$.
        Let $T_j'$ be the minimum cost directed $s$-$t$ augmenting path in $H_j$ with respect to $M_{j-1}$ and let $M_j=M_{j-1} \oplus T_j$. The cardinality of $M_j$ is $\abs{M}+j$. By Claim~\ref{cl:mtd1},  $d_{tot}(M_j)=d_{tot}(M)$.  Consider any matching $N$ such that $\abs{N}=\abs{M}+j$, and $d_{tot}(N)=d_{tot}(M)$. 
        
        By \cref{lem:structure}, there exists an $M_{j-1}$ augmenting path $Q$ in $M_{j-1} \oplus N$, corresponding to a directed $s$-$t$ path $Q'$, and  by \cref{{lem:structure}}, 
        \[cost(N)-cost(M_{j-1})= cost(Q').\] Since, $T_j'$ is a minimum cost directed $s$-$t$ path, $cost(T_j') \le cost(Q')$. Therefore,  \[cost(M_j)-cost(M_{j-1})=cost(T_j') \le cost(Q') = cost(N)-cost(M_{j-1}).\]
        Thus, $cost(M_j) \le cost(N)$. By \cref{lem:cost-rmm}, $\sigma(M_j) \succeq_{\mathcal{R}} \sigma(N)$. Since the choice of $N$ was  arbitrary, $M_j$ is rank-maximal among all matchings with cardinality $\abs{M}+j$ and total deviation $d_{tot}(M)$.
    \end{proof}

%%%%%%%%%%

Our iterative  procedure halts in iteration $j$ if the graph $H_j$ does not admit any  $s$-$t$ path. Recall that we begin by setting $M^*=M_0$, and in the iteration $j$, after computing $M_j$, if $\sigma(M_j) \succ_{\mathcal{R}} \sigma(M^*)$, we set $M^*=M_j$. We return the matching $M^*$  at the end of Step~(B). 
\begin{restatable}{lemma}{cltwentyeight}
\label{cl:mtd4}
The matching $M^*$ returned by Step~(B) is a solution to \RMMTD in $G$.
\end{restatable}
\begin{proof}
From Claim~\ref{cl:mtd1},  every matching $M_j$ computed in Step~(B) has minimum total deviation. Let $N$ be a minimum total deviation matching in $G$. By \cref{lem:cardinality}, there exists an integer $q \ge 0$ such that $\abs{N}=\abs{M}+q$. By \cref{lem:mj-rmm}, the matching $M_q$ is rank-maximal among all minimum total deviation matchings of cardinality $\abs{M}+q$. Thus, $\sigma(M_q) \succeq_{\mathcal{R}} \sigma(N)$. The algorithm retains a matching $M^*$ whose signature is at least as good as that of every matching $M_j$ computed in Step~(B). Thus, $\sigma(M^*) \succeq_{\mathcal{R}} \sigma(N)$. Since, $N$ is an arbitrary minimum total deviation matching in $G$, the matching $M^*$ is rank-maximal among all minimum total deviation matchings in $G$.
 \end{proof}

\paragraph{Illustration of algorithm for \RMMTD}
For an illustration, consider the soft quota setting instance $G$ given in \cref{fig:target}, and  consider the instance $G'$. The matching obtained by invoking Step~(A) is $M=\{(a_1, p_1), (a_3,p_2),(a_5,p_4)\}$ with $\sigma(M)=(1,2,0)$ (here we omit the number of unmatched applicants) and $d_{tot}(M)=2$, which is a solution to \RMMTD in $G'$.
To compute a solution to \RMMTD  in $G$, we apply our iterative algorithm as described in Step~(B). Observe that there are two $s$-$t$ paths in $H_1$, and we select the min-cost $s$-$t$  path $T'_1=\left<s,a_6,p_6,t\right>$ such that $cost(T_1)=-(n+1)^2$ and $\sigma(T_1)=(-1, 0, 0)$. After the first augmentation, we get a matching, $N_1=\{(a_1, p_1), (a_3,p_2),(a_5,p_4), (a_6,p_6)\}$ that has a signature of $(2,2,0)$. Observe that $N_1$ is a rank-maximal matching among all matchings with cardinality $4$ in $G$. In the next iteration, we select the $s$-$t$ path   $T'_2=\left<s,a_4,p_7,t\right>$ such that $cost(T_2)=-1$ and $\sigma(T_2)=(0, 0, -1)$. After the second augmentation, we get a matching 
$N_2=\{(a_1, p_1), (a_3,p_2), (a_4,p_7), (a_5,p_4), (a_6,p_6)\}$ that has a signature of $(2,2,1)$. Observe that, in the next iteration, the directed graph $H_3$ does not admit any  $s$-$t$ path, and thus $N_2$ is a solution to \RMMTD problem in $G$. Note that the matching $N_2$ is the same as the matching $M_1$ listed in \cref{tab:my_label1}.

\subsection{Modification for Fairness}

For the total deviation variant of the problem \FMMTD, we begin by constructing the network $H$ in Step~(A) as described in \cref{sec:alg-min-tot-d}. We modify the cost assignment on the edges in $H$ as follows. 

\noindent {\bf Cost of the edges in $H$.} 
\begin{itemize}
    \item For every edge $(a,p) \in \mathcal{E}_{AP}$  with rank $i$, we set a cost of $(n+1)^i$. 
    \item For all the other edges in $H$, we set the cost as $0$.
\end{itemize}
Due to the cost assignment, any min-cost flow in $H$  corresponds to a matching  $M$, which is a solution  to \FMMTD problem in the restricted instance $G'$.

In Step~(B), we augment the matching $M$ in $G_x$  to compute a solution to \FMMTD problem in $G$. We construct a directed graph $H_j$ with costs on edges as described in Step~(B). However, we modify the cost assignment as follows.

\noindent {\bf Cost of the edges in $H_j$.} For an edge $(a,p) \in E$ of rank $i$, 
\begin{itemize}
    \item  if $(a,p) \in \mathcal{E}_{AP}$   cost of the edge is $(n+1)^{i}$
    \item  if $(p,a) \in \mathcal{E}_{PA}$  cost of the edge is $-(n+1)^{i}$ 
\end{itemize}
The procedure of augmentation is the same as that described in Step~(B).

%% file: 5-MIN-MAX.tex
\section{MIN-MAX Variants}\label{minmax}
To solve the {\sf MIN-MAX} type problems in the soft quota setting, namely, \OSMMD and \OMMD, we exploit their connection  to related problems in the fixed quota setting.
 In particular, to obtain our solutions, we compute  optimal matchings (with respect to preferences) among all the matchings that are \textit{feasible} in the fixed quota setting instance.

 We consider an instance of fixed quota setting $G$, where each  post $p$ has an associated  lower quota $q^-(p)$ and an upper quota $q^+(p)$. For clarity, we denote the lower quota and upper quota of a post $p$ in $G$ by $q_G^-(p)$, and $q_G^+(p)$, respectively. When the instance is clear from the context, we omit the subscript $G$.

Assuming that a feasible matching exists in the instance, we first investigate the problem  {\sf OPT-LQ}, which aims to compute  a matching that is optimal with respect to the notion of OPT among all feasible matchings of the instance.
Note that here, OPT can be set to either rank-maximality or fairness. Recall that in the fixed quota setting, where the lower quota of every post is zero, both these notions are guaranteed to exist, and optimal matchings are efficiently computable by the algorithms given in \cite{irving2006rank,paluch2013capacitated,huang2016fair}. Now, we extend these algorithms  to compute an optimal matching among all feasible matchings by adapting the algorithm  in \cite{huang2016fair}. Assuming that feasible matchings exist in $G$, we present the algorithm for \RMMLQ.

Given an instance  $G$ of the \RMMLQ problem, we construct a fixed quota setting instance $G'$, where the lower quota of every post is set to $0$. 
For an instance of the fixed quota setting, where the lower quota of every post is $0$, we denote by {\sf MAX-CARD-RMM}  the problem of computing a rank-maximal matching among all maximum matchings in the instance.
 We reduce the problem of computing  \RMMLQ in $G$ to the problem of computing  {\sf MAX-CARD-RMM} in $G'$.

\subsection{ Reduction from \RMMLQ to {\sf MAX-CARD-RMM}}
Given an instance $G$ of \RMMLQ, we construct an instance $G'$ of the fixed quota setting, where the lower quota of every post is $0$, as follows.
In the instance $G'$, we retain the applicants $\A$, posts $\B$, and the preferences of applicants over posts from the instance $G$. For every post $p \in \B$, we set the lower quota $q_{G'}^-(p)=0$, and upper quota $q_{G'}^+(p)=q_{G}^+(p)$. For a post $p \in \B$, we introduce $q_{G}^+(p)-q_{G}^-(p)$ many dummy applicants and add a rank $r+1$ edge incident to $p$ for each such applicant. Let $U=\sum\limits_{p \in \B} q_{G'}^+(p)$.
We show a one-to-one correspondence between a matching with cardinality $U$ in $G'$ and a feasible matching in $G$, as stated in Claim~\ref{cl:rmmlq-mcrmm1}.
\begin{claim}
\label{cl:rmmlq-mcrmm1}
$G'$ admits a matching of cardinality $U$ if and only if $G$ admits a feasible matching.
\end{claim}
\begin{proof}
Suppose that $G$ admits a feasible matching, say $M$. We construct a matching $M'$ in $G'$ from  $M$ such that $\abs{M'}=U$. In the matching $M'$, we retain all the edges from $M$. For every post $p$, we match as many dummy applicants as possible so that the post becomes fully-subscribed in $M'$ with respect to its upper quota $q^+_{G'}(p) = q^+_G(p)$. Observe  that $\abs{M'}=U$.

On the other hand, suppose that $G'$ admits a matching $M'$ such that $\abs{M'}=U$. We construct a matching $M$ in $G$ by removing all the dummy applicants matched to a post $p \in \B$ in $M'$. Note that, for a post $p \in \B$, the number of dummy applicants matched to $p$ in $M'$ is at most $q_{G}^+(p)-q_{G}^-(p)$. Hence, the number of true applicants matched to $p$ in $M'$ is at least  $q_{G}^-(p)$. Thus, for every post $p$, we have $\abs{M(p)}=q_G^-(p)$, and therefore  $M$ is a feasible matching in $G$.
\end{proof}
We state the following claim about a feasible matching  in $G$ and  maximum matchings  in $G'$ with cardinality $U$.
\begin{claim}\label{cl:combined}
Let $G$ and $G'$ be the instances as defined above. Then, 
\begin{enumerate}[(i)]
    \item For any feasible matching $M$ in $G$ with $\sigma(M)=(x_1, x_2, \ldots, x_r)$ and cardinality $S=\sum\limits_{i=1}^r x_i$, there exists a corresponding maximum matching $M'$ in $G'$ with  $\sigma(M')=(x_1, x_2, \ldots, x_r, U-S)$. \label{cl:rmmlq-mcrmm2}
    \item For a maximum matching $M'$   in $G'$ with $\sigma(M')=(x_1, x_2, \ldots, x_r, x_{r+1})$, and $\abs{M'}=U$, there exists  a corresponding feasible matching $M$ in $G$ with $\sigma(M)=(x_1, x_2, \ldots, x_r)$. \label{cl:rmmlq-mcrmm3}
\end{enumerate}
\end{claim}
The proof of the above  claim follows directly from  the proof of Claim \ref{cl:rmmlq-mcrmm1}.

\begin{restatable}{lemma}{rmmlq}
\label{lem:rmmlq}
Let $M'$ be a solution to {\sf MAX-CARD-RMM} problem in $G'$ such that $\abs{M'}=U$, and let $M$ be the corresponding matching in $G$. Then $M$ is a solution to \RMMLQ problem in $G$. 
\end{restatable}
\begin{proof}
Since $\abs{M'}=U$, it follows from Claim~\ref{cl:rmmlq-mcrmm1} that $G$ admits a feasible matching. Let the signature of $M'$ be $\sigma(M')=(x_1, x_2, \ldots, x_r, x_{r+1})$. Then, by Claim~\ref{cl:combined}(\ref{cl:rmmlq-mcrmm3}), the signature of the corresponding matching $M$ in $G$ is  $\sigma(M)=(x_1, x_2, \ldots, x_r)$.
 Suppose for the sake of contradiction, that $M$ is not a solution to \RMMLQ problem in $G$. Then there exists another matching,  say ${M_1}$, which is a solution to \RMMLQ problem in $G$. Let $\sigma({M_1})=(y_1, y_2, \ldots, y_r)$. Now, consider the matching $M_1'$ in $G'$ that corresponds to $M_1$ in $G$. By Claim~\ref{cl:combined}(\ref{cl:rmmlq-mcrmm2}), $\sigma({M_1'})=(y_1, y_2, \ldots, y_r, U-S)$, where $S=\sum\limits_{i=1}^r y_i$. Observe that, since $\sigma(M_1)\succ_{\mathcal{R}} \sigma(M)$ in $G$, it follows that $\sigma(M_1')\succ_{\mathcal{R}} \sigma(M')$ in $G'$, which contradicts the fact that  $M'$ is a solution to {\sf MAX-CARD-RMM} problem in $G'$.
\end{proof}

\paragraph{ Modifications for fairness} Given an instance  $G$ of \FAIRLQ, we construct the same instance $G'$ as described above. But instead of {\sf MAX-CARD-RMM}, we reduce the problem of computing \FAIRLQ to a problem of computing fair matching on $H$.
By applying the techniques used in \cite{huang2016fair}, both problems \RMMLQ and \FAIRLQ admit an $\tilde{\mathcal{O}}(rmn)$ algorithm.

\subsection{Minimizing maximum deviation in the soft quota setting}
Having described the algorithms for the problems \RMMLQ and \FAIRLQ in the fixed quota setting, we now go back to our original problem of minimizing maximum deviation in the soft quota setting.
To solve the {\sf MIN-MAX} type problems in the  soft quota setting, namely {\sf OPT-MIN-MAX} and {\sf OPT-SIGN-MIN-MAX},  we use a reduction to problems in the fixed quota setting and apply the algorithms for the problems \RMMLQ and \FAIRLQ. We describe our algorithms for the preference optimality as rank-maximality. Analogously, we can solve the problems for the notion of fairness. 

We begin by presenting a useful observation about an optimal solution to the problem {\sf RMM-MIN-MAX}. 

\begin{obs} \label{obs:sq1} 
    Let $M^*$ be an optimal solution to {\sf RMM-MIN-MAX}, and let $\Delta^*$ be the optimal maximum deviation. Then, there is a post $p$ in $M^*$, which has a lower deviation exactly equal to $\Delta^*$.
    \end{obs}
    \begin{proof}
    Suppose not. Then the lower deviation  of all posts with respect to $M^*$ is strictly less than $\Delta^*$. Let the maximum lower deviation  of any post with respect to $M^*$ be $\Delta'$. Since $d_{max}(M^*)=\Delta^*$, there exists at least one post $p \in \B$ whose upper deviation is  $\Delta^*$.
    Let $\B_u$ be the set of posts that deviate from their upper target with respect to $M^*$, such that $\abs{M^*(p)} >u(p)+\Delta'$. 
    
    We construct a matching $M'$ from $M^*$
as follows. For every post $p \in \B_u$, we arbitrarily  select $\abs{M^*(p)} - u(p) - \Delta'$ matched edges incident on $p$ and delete them. We retain all other matched edges in $M^*$. Let the resulting matching be $M'$. Observe that for any post $p$, the upper deviation with respect to $M'$ is upper bounded by $\Delta'$. Since we retained the matched edges corresponding to any post $p \not\in \B_u$ in $M'$,  the lower deviation of $p$ with respect to $M'$ is also upper bounded by $\Delta'$. Thus, $d_{max}(M')=\Delta' < \Delta^*$, which contradicts the fact that $M^*$ is a matching with minimum maximum deviation.
    \end{proof}

Next, we describe the reduction from  soft quota setting to the fixed quota setting.
\paragraph{ Reduction from soft quota setting to the  fixed quota setting} Given an instance $G$ of the soft quota setting, and a fixed value of maximum deviation $\Delta$, we  construct a fixed quota setting instance $H_{\Delta}$ as follows. In $H_{\Delta}$,  the applicants, posts, and applicant preferences remain the same as those in  $G$. For a post $p$, we set the  lower quota as $q^-(p)=\max\{0,\ell(p)-\Delta\}$, and the upper quota  as $q^+(p)=u(p)+\Delta$. Note that if $H_{\Delta}$ admits a feasible matching, then 
the same matching is feasible for a fixed quota setting instance $H_{\Delta'}$ for any deviation $\Delta'\ge \Delta$. This is because an increase in the deviation widens the range of acceptable numbers of applicants for a post.

     Next, we present the algorithms for the {\sf RMM-MIN-MAX} and {\sf RMM-SIGN-MIN-MAX} problems.
\subsection{Algorithm for \RMMMD}\label{sec:alg-min-max-d}
Given an instance $G$ of the soft quota setting, our goal is to compute a rank-maximal matching among all matchings that have a minimum value of maximum deviation.     Under our assumption on $\ell(p)$, the maximum deviation of any post is upper bounded by $\abs{\A}$. From Observation \ref{obs:sq1}, the maximum deviation value of any matching is  also upper bounded by $\abs{\A}$. Therefore, the maximum deviation  of any matching lies in the range $[0, \abs{\A}]$. 
To compute the minimum value of maximum deviation, we start by guessing the maximum deviation value $\Delta$ and checking whether the corresponding  fixed quota setting instance $H_{\Delta}$ admits a feasible matching.  To check feasibility, we can ignore the preferences and use the standard flow network for the bipartite matching with demands on the edges.  We perform a binary search on $\Delta$ in the range $[0, \abs{\A}]$ to compute the smallest $\Delta$ such that there exists  a feasible solution. Let $\Delta^*$ be the minimum value of $\Delta$ obtained. Then, corresponding to $\Delta^*$, we compute a rank-maximal matching among all feasible matchings (\RMMLQ) in the fixed quota setting instance $H_{\Delta^*}$.
    By the definition of \RMMLQ, the resulting matching has the best possible signature with respect to the notion of rank-maximality while achieving  the minimum value for the maximum deviation.

\subsection{Algorithm for \RSMMD} \label{sub:rsmmd}
Given an instance $G$ of the soft quota setting, and  an input signature $\rho$, our goal is to compute a matching $M$ in $\mathcal{M}_{\mathcal{R}}$ (recall that $\mathcal{M}_{\mathcal{R}}$ is the set of matchings whose signatures are at least as good as $\rho$ with respect to rank-maximality)  that has a minimum value of maximum deviation.
 Under the assumption about $\ell(p)$,  lower deviation of any post is upper bounded by $\abs{\A}$. Similarly, the  upper deviation of any post is upper bounded by $\abs{\A}$.  Therefore, the maximum deviation of any matching lies in the range $[0, \abs{\A}]$. To compute the minimum value of maximum deviation for a matching in $\mathcal{M}_{\mathcal{R}}$, we start by guessing the maximum deviation value $\Delta$ and checking whether  the corresponding  fixed quota setting instance $H_{\Delta}$ admits a feasible matching. If the instance $H_{\Delta}$ admits a feasible matching, we compute a rank-maximal matching (\RMMLQ) $M$ among all feasible matchings in $H_{\Delta}$. If $\sigma(M)\succeq_{\mathcal{R}} \rho$, then $M$  satisfies the signature requirement with respect to rank-maximality. 
      We perform a binary search on $\Delta$ in the range $[0, \abs{\A}]$ to compute the smallest $\Delta$  such that there exists  a matching that meets the signature requirement.

Below, we illustrate the execution of our algorithms for \RMMMD and \RSMMD  for the example instance in \cref{fig:target}.
\paragraph{Illustration of algorithm for \RMMMD}
 Given the instance $G$, for a given $\Delta$, we construct the corresponding fixed quota setting instance $G_{\Delta}$ as described in our algorithm. Using binary search, we obtain the minimum value of the maximum deviation as $\Delta=1$. We then compute an \RMMLQ in the fixed quota setting instance $G_{1}$. It is easy to verify that $M_2$, listed in \cref{tab:my_label1} is a solution to the \RMMLQ problem in $G_{1}$, and hence a solution to \RMMMD problem in $G$.

\paragraph{Illustration of algorithm for \RSMMD} Given the instance $G$, for a fixed value of $\Delta$, we construct the corresponding fixed quota setting instance $G_{\Delta}$ as described in our algorithm.  We then apply binary search on the maximum deviation value $\Delta$ and solve the \RMMLQ problem in the fixed quota setting instance $G_{\Delta}$. For $\Delta=2$, we construct fixed quota setting instance $G_{2}$, and solve the \RMMLQ problem. It is easy to verify that $M_4$, listed in \cref{tab:my_label1}, is a solution to the \RMMLQ problem in $G_{2}$. Since $\sigma(M_4)\succeq_{\mathcal{R}} (4,0,0,3)$, the matching is a feasible solution for the problem \RSMMD.  It is easy to verify that  any matching that meets the signature requirement has a maximum deviation of at least $2$, and thus the matching $M_4$ is an optimal solution for the \RSMMD problem.

\subsection{Modification for Fairness in the soft quota setting} 
In this section, we state the necessary modifications required to solve the problems \FMMMD and \FSMMD.

\noindent{\bf Algorithm for \FMMMD.}
Given an instance $G$ of the soft quota setting, our goal is to compute a fair matching among all matchings that have a minimum value of maximum deviation.   As in the algorithm for the \RMMMD problem given in \cref{sec:alg-min-max-d}, we first compute the minimum value of maximum deviation, say $\Delta^*$. Then, corresponding to $\Delta^*$, we compute a fair matching among all feasible matchings (\FAIRLQ) in the fixed quota setting instance $H_{\Delta^*}$.
    By the definition of \FAIRLQ, the resulting matching has the best possible signature with respect to the notion of fairness while achieving  the minimum value for the maximum deviation.

\noindent{\bf Algorithm for \FSMMD.}
Given an instance $G$ of the soft quota setting, and  an input signature $\rho$, our goal is to compute a matching $M$ in $\mathcal{M}_{\mathcal{F}}$ (recall that $\mathcal{M}_{\mathcal{F}}$ is the set of matchings whose signatures are at least as good as $\rho$ with respect to fairness)  that has a minimum value of maximum deviation.
 As in the algorithm for the problem \RSMMD given in \cref{sub:rsmmd}, we guess a maximum deviation value $\Delta$ and check whether  the corresponding  fixed quota setting instance $H_{\Delta}$ admits a feasible matching. If the instance $H_{\Delta}$ admits a feasible matching, we compute a fair matching (\FAIRLQ) $M$ among all feasible matchings in $H_{\Delta}$. If $\sigma(M)\succeq_{\mathcal{F}} \rho$, then $M$  satisfies the signature requirement with respect to fairness. 
      We perform a binary search on $\Delta$ in the range $[0, \abs{\A}]$ to compute the smallest $\Delta$ such that there exists  a matching that meets the signature requirement.

%%%%%%%%%%%%%%%%%%%%%%%%%%%%%%%%%%%%%%%%%%%%%%%%%%%%%%%%%%%%%%%%%%%%%%%%
\section{Optimizing both total deviation and maximum deviation} \label{sec:tm-both}
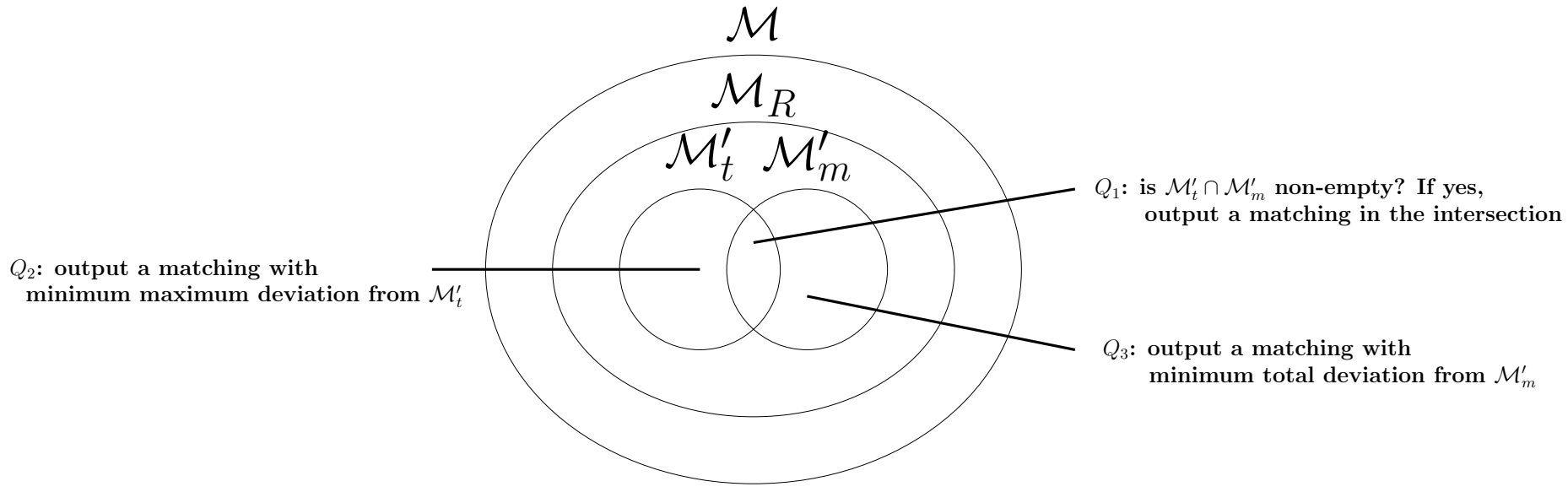
\begin{figure}
    \centering
    \scalebox{0.42}{
    \begin{tikzpicture}[
roundnode/.style={circle, draw=black!100,  inner sep=0pt, minimum size=3cm},
farrow/.style={-Latex}
]
\node[roundnode] (s) at (0,0) [label=above:{\huge $\mathcal{M}_t'$}] {};
\node[roundnode] (t) at (2,0) [label=above:{\huge $\mathcal{M}_m'$}] {};
\draw \boundellipse{1,0}{3.75}{2.75};
\draw \boundellipse{1,0}{5}{4};
 \node (a3) at (1,4) [label=above:{\huge $\mathcal{M}$}] {};
 \node (a4) at (1,3.25) {\huge{$\mathcal{M}_{R}$}};

\draw[thick, line width=0.5mm, -] (1,0.5) -- node [] {} (7,1.5);
\node (a5)  at (11,1.5)  {\bf $Q_1$: is $\mathcal{M}_t' \cap \mathcal{M}_m'$ non-empty? If yes,};
\node (a6)  at (12.2,1)  {\bf  output a matching in the intersection};
\draw[thick, line width=0.5mm, -] (2,-0.5) -- node [] {} (7,-1.5);
\node (a7)  at (10.4,-1.5)  {\bf $Q_3$: output a matching with  };
\node (a8)  at (12,-2)  {\bf  minimum total deviation from $\mathcal{M}_m'$};
\draw[thick, line width=0.5mm, -] (0,0) -- node [] {} (-5,-0);
\node (a9)  at (-10,0)  {\bf $Q_2$: output a matching with  };
\node (a10)  at (-8.5,-0.5)  {\bf  minimum maximum deviation from $\mathcal{M}_t'$};

    \end{tikzpicture}}
    \caption{Optimizing both total deviation and maximum deviation for the problem {\sf RMM-SIGN-MIN-Y}, where $Y \in \{TOT,MAX\}$.}
        \label{fig:osmydqn}
\end{figure}

  Our problems so far have been defined  to optimize either  the total deviation or the maximum deviation, but  not both simultaneously. A natural question is  whether there exists a matching that optimizes both the total deviation and the maximum deviation. To define our problems concretely, we fix the optimality notion as rank-maximality.
In \cref{sec:mmtpof}, we consider preference optimality first type of problems followed by deviation first type of problems in \cref{sec:mmtdf}.

\subsection{Preference optimality first} \label{sec:mmtpof}
Recall that, in the preference optimality first type of the problem, we are given an instance of the soft quota setting along with a signature. The  goal is to compute a matching that satisfies the signature with respect to rank-maximality while optimizing either the total deviation or the maximum deviation. The set of matchings that meet the signature requirement with respect to rank-maximality is denoted  by $\mathcal{M}_R$. Within $\mathcal{M}_R$, the set of matchings that minimize the total deviation is denoted by $\mathcal{M}_t'$, while the set of matchings that minimize the maximum deviation is denoted by $\mathcal{M}_m'$. In the \RSMTD problem, the goal is to output any matching from  $\mathcal{M}_t'$, whereas in the \RSMMD problem, the goal is to output any matching from    $\mathcal{M}_m'$ (see \cref{osmydomyd}).

 Having seen efficient algorithms for the \RSMTD and \RSMMD problems,
 we now  define a problem that aims to optimize both total deviation and maximum deviation simultaneously. That is, given an instance $G$ of the soft quota setting and an input signature $\rho$, our goal is to decide whether the intersection of $\mathcal{M}_t'$ and $\mathcal{M}_m'$ is non-empty, and, if so, computing a matching from this intersection. This problem is denoted by  \RSMTM (marked as $Q_1$ in \cref{fig:osmydqn}), where the {\sf TM} in the name indicates that our goal is to optimize both total and maximum deviations simultaneously. 
 If the intersection of $\mathcal{M}_t'$ and $\mathcal{M}_m'$ is empty,  we consider the following two optimization problems:
 \begin{itemize}
     \item \RSMTMD: compute a matching that has minimum value of maximum deviation from the set $\mathcal{M}_t'$.  This problem is denoted by  \RSMTMD (marked as $Q_2$ in \cref{fig:osmydqn}), where the {$\textsf{T}_\textsf{M}$}  indicates that the primary objective is  to minimize the total deviation. Within the set of matchings having minimum total deviation, denoted by $\mathcal{M}_t'$, the secondary objective is to minimize the maximum deviation.
     
     \item \RSMMTD:  compute a matching that has minimum value of total deviation from the set $\mathcal{M}_m'$.  This problem is denoted by \RSMMTD (marked as $Q_3$ in \cref{fig:osmydqn}), where the {$\textsf{M}_\textsf{T}$}  indicates that
     the primary objective is  to minimize the maximum deviation. Within  the set of matchings having minimum value of maximum deviation, denoted by $\mathcal{M}_m'$, the secondary objective is to minimize the total deviation.
 \end{itemize}

We solve the above problems in polynomial time by adapting our algorithm for the problem \RSMTD. Recall that, for a given  total deviation $\Delta$ and  a split $(ld,ud)$ such that $\Delta=ld+ud$, if the network $H_{(ld,ud)}$ (shown in \cref{fig:network1}) admits a feasible flow, then the total deviation of the corresponding matching is at most $\Delta$. However, there are no guarantees on the maximum deviation of the computed matching. In the network depicted in \cref{fig:network1},  the lower deviation for a post $p$  is controlled by the capacity of the edge $(d',p')$, while   the upper deviation is constrained by the capacity of the edge $(p'',d'')$. By modifying the capacities of these edges, we can bound the maximum deviation. To solve our problems, we modify the network so that we can restrict our attention to matchings with specific values of total deviation and maximum deviation.

Suppose that, for a given instance $G$ and an input signature $\rho$,  the optimal values of the total deviation and the maximum deviation are $\Delta'_t$ and $\Delta'_m$, respectively. That is, every matching in $\mathcal{M}'_t$ has total deviation $\Delta'_t$, whereas every matching in $\mathcal{M}'_m$ has maximum deviation $\Delta'_m$. We now  describe our algorithms.

 \paragraph{\bf Algorithm for \RSMTM}
 To determine whether the set    $\mathcal{M}_t' \cap \mathcal{M}_m'$ is non-empty  we make use of our network in \cref{fig:network1}. 
 We note that, since the optimal value of total deviation is $\Delta'_t$, the network $H_{\Delta'_t} $ for a particular split as 
 $\Delta'_t=ld+ud$ admits a feasible flow, and the corresponding matching satisfies the signature requirement. 
 To restrict our attention to matchings with  maximum deviation $\Delta_m'$, we apply the following modifications to  the network.
  \begin{itemize}
      \item For every post $p \in \B$, we set the capacities of the edges $(d',p')$ and $(p'',d'')$ to $\Delta_m'$ instead of $ld$ and $ud$, respectively. 
      \item We retain the demands and capacities of all other edges as they are in the original network given in \cref{fig:network1}. 
  \end{itemize}
  Let the modified network be $H'$. For each split $(ld,ud)$ of $\Delta_t'$, we compute a min-cost flow in the network $H'_{(ld,ud)}$, and check whether the corresponding matching satisfies the signature requirement.  As in the algorithm for \RSMTD, we consider all possible splits of $\Delta_t'$.  If for any  split, say $(ld^*,ud^*)$, the matching corresponding to the min-cost flow, say $M$ meets the signature requirement, then the intersection of $\mathcal{M}_t'$ and $\mathcal{M}_m'$ is non-empty and we have found the required matching. 
  If none of the splits  lead to a matching that satisfies the signature requirement, then the intersection of $\mathcal{M}_t'$ and $\mathcal{M}_m'$ is empty. 

 \paragraph{\bf Algorithm for \RSMTMD}
 In the problem \RSMTMD, the goal is to compute a matching that minimizes the maximum deviation from the set $\mathcal{M}_t'$. 
Observe that any matching in the set $\mathcal{M}_t'$ has a maximum deviation of at least $\Delta_m'$ and at most $\Delta_t'$.
Therefore, we  perform a linear search within this range to find the optimal value of the maximum deviation.
For a specific value of maximum deviation, say $i \in [\Delta_m', \Delta_t']$, we apply the following modifications to  the network given in \cref{fig:network1}.
\begin{itemize}
  \item Fix the value of  total deviation $\Delta$ as $\Delta_t'$. Thus, we restrict our attention to matchings with total deviation $\Delta_t'$ (matchings in the set $\mathcal{M}_t'$).
      \item For every post $p \in \B$, we set the capacities of the edges $(d',p')$ and $(p'',d'')$ to $i$ instead of $ld$ and $ud$, respectively, where $i \in [\Delta_m', \Delta_t']$. 
      \item We retain the demands and capacities of all other edges as they are in the original network given in \cref{fig:network1}.
  \end{itemize} 
  Let the modified network be $H'$. For each split $(ld,ud)$ of $\Delta_t'$, we compute a min-cost flow in the network $H'_{(ld,ud)}$, and check whether the corresponding matching satisfies the signature requirement. If for any split, say $(ld^*,ud^*)$, the matching corresponding to the min-cost flow, say $M$ meets the signature requirement, then $i$ is a candidate value for the  maximum deviation. We perform a linear search on $i$ to compute a matching in $\mathcal{M}_t'$ having the smallest value of maximum deviation. With $O({\Delta_t'}^2)$ flow computations in the modified network $H'$, we can compute a solution for the problem. Since the value of $\Delta_t'$ is polynomial in $\abs{\A}$, the algorithm runs in polynomial time.
  \paragraph{\bf Algorithm for \RSMMTD}
In the problem \RSMMTD, the goal is to compute a matching  that minimizes total deviation from the set $\mathcal{M}_m'$.
Observe that any matching in the set $\mathcal{M}_m'$ has a total deviation of at least $\Delta_t'$ and at most $\min\{ \abs{\B} \cdot \Delta_m', \sum\limits_{p \in \B} \ell(p)+\abs{\A} \}$.  Therefore,  we perform a linear search within this range to find the optimal value of total deviation. We apply the following modifications to  the network given in \cref{fig:network1}.
\begin{itemize}
      \item For every post $p \in \B$,  we set the capacities of the edges $(d',p')$ and $(p'',d'')$ to $\Delta_m'$ instead of $ld$ and $ud$, respectively.
      \item We retain the demands and capacities of all other edges as they are in the original network given in \cref{fig:network1}.
  \end{itemize}
  Let the modified network be $H'$. Now, we apply our algorithm for \RSMTD on the modified network. For a given value of $\Delta$,  we consider all possible splits corresponding to $\Delta$. If for any split, say $(ld^*,ud^*)$, the matching corresponding to the min-cost flow, say $M$ meets the signature requirement, then $\Delta$ is a candidate value for the  total deviation. We perform a linear search on $\Delta$  to compute a matching in $\mathcal{M}_m'$ having the smallest value of total deviation.
Note that the time complexity of the above algorithm is same as that of the algorithm for \RSMTD. 

\subsection{Deviation first} \label{sec:mmtdf}
Now, we consider analogous problems in the deviation first variant.
Recall that, in the deviation first type of the problem, we are given an instance of the soft quota setting, and our goal is to compute a rank-maximal matching among all matchings that  optimize either the total deviation or the maximum deviation. We denote the set of matchings that minimize total deviation (or max  deviation respectively)  by $\mathcal{M}_t$ (respectively, $\mathcal{M}_m$). In the \RMMTD problem, the goal is to output a rank-maximal matching from the set   $\mathcal{M}_t$, while in the \RMMMD problem the goal is to output a rank-maximal matching from the set   $\mathcal{M}_m$.

We define analogous problems to optimize the total deviation and maximum deviation simultaneously as described in \cref{sec:mmtpof}.
Given an instance $G$ of the soft quota setting, our goal is to decide whether the intersection of $\mathcal{M}_t$ and $\mathcal{M}_m$ is non-empty, and if so, compute a matching from the intersection. We denote this problem by \RMMTM (marked as $Q_4$ in \cref{fig:osmydqn2}), where the {\sf TM} in the problem name indicates that our goal is to optimize both the total and the maximum deviation simultaneously. If the intersection of $\mathcal{M}_t$ and $\mathcal{M}_m$ is empty, we focus on the following two optimization problems:

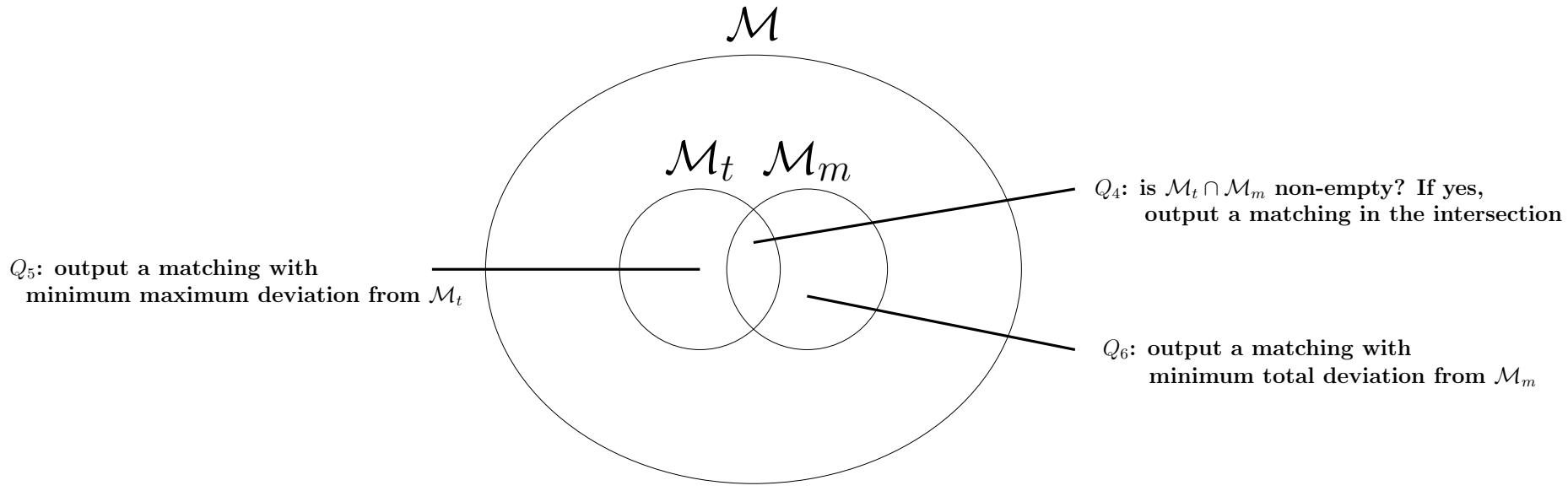
\begin{figure}
    \centering
    \scalebox{0.42}{
    \begin{tikzpicture}[
roundnode/.style={circle, draw=black!100,  inner sep=0pt, minimum size=3cm},
farrow/.style={-Latex}
]
\node[roundnode] (s) at (0,0) [label=above:{\huge $\mathcal{M}_t$}] {};
\node[roundnode] (t) at (2,0) [label=above:{\huge $\mathcal{M}_m$}] {};
\draw \boundellipse{1,0}{5}{4};
 \node (a3) at (1,4) [label=above:{\huge $\mathcal{M}$}] {};

\draw[thick, line width=0.5mm, -] (1,0.5) -- node [] {} (7,1.5);
\node (a5)  at (11,1.5)  {\bf $Q_4$: is $\mathcal{M}_t \cap \mathcal{M}_m$ non-empty? If yes,};
\node (a6)  at (12.2,1)  {\bf  output a matching in the intersection};
\draw[thick, line width=0.5mm, -] (2,-0.5) -- node [] {} (7,-1.5);
\node (a7)  at (10.4,-1.5)  {\bf $Q_6$: output a matching with  };
\node (a8)  at (12,-2)  {\bf  minimum total deviation from $\mathcal{M}_m$};
\draw[thick, line width=0.5mm, -] (0,0) -- node [] {} (-5,-0);
\node (a9)  at (-10,0)  {\bf $Q_5$: output a matching with  };
\node (a10)  at (-8.5,-0.5)  {\bf  minimum maximum deviation from $\mathcal{M}_t$};

    \end{tikzpicture}}
    \caption{Optimizing both total deviation and maximum deviation for the problem {\sf RMM-MIN-Y}, where $Y \in \{TOT,MAX\}$.}
    \label{fig:osmydqn2}
\end{figure}

 \begin{itemize}
     \item \RMMTMD: compute a matching that has minimum value of maximum deviation from the set $\mathcal{M}_t$.  This problem is denoted by \RMMTMD (marked as $Q_5$ in \cref{fig:osmydqn2}), where the {$\textsf{T}_\textsf{M}$} indicates that the primary objective is  to minimize the total deviation. Within  the set of matchings that have minimum value of total deviation, denoted by $\mathcal{M}_t$, the secondary objective is to minimize the  maximum deviation.
     
     \item \RMMMTD: compute a matching that has minimum value of total deviation from the set $\mathcal{M}_m$.  This problem is denoted by \RMMMTD (marked as $Q_6$ in \cref{fig:osmydqn2}), where {$\textsf{M}_\textsf{T}$} indicates that the primary objective is to minimize the maximum deviation. Within the set of matchings that have minimum value of maximum deviation, denoted by $\mathcal{M}_m$, the secondary objective is to minimize the total deviation.
 \end{itemize}
We solve the above problems in polynomial time by adapting our algorithm for the problem \RSMTD. 
Suppose that, for a given instance $G$, the optimal values of the total deviation and the maximum deviation are $\Delta_t$ and $\Delta_m$, respectively. That is, every matching in $\mathcal{M}_t$ has total deviation $\Delta_t$, whereas every matching in $\mathcal{M}_m$ has maximum deviation $\Delta_m$. Now we describe our algorithms.
\paragraph{Algorithm for \RMMTM } Recall that, in an optimal matching for the  {\sf RMM-MIN-TOT} problem, no post  incurs a deviation with respect to the upper target. This implies that any matching in the intersection of  $\mathcal{M}_t$ and $\mathcal{M}_m$  if it exists, has an upper deviation of $0$. 
As described in \cref{sec:mmtpof},  we  restrict the maximum deviation of a post $p$  to $\Delta_m$. To solve the \RMMTM problem, we modify the network given in \cref{fig:network1} as follows. 
  \begin{itemize}
  \item  We set $ud=0$ and $ld=\Delta_t$. Thus, we restrict our attention to matchings with total deviation   $\Delta_t$ (matchings in the set $\mathcal{M}_t$).
      \item For every post $p \in \B$, we set the capacity of the edge $(d',p')$ as  $\Delta_m$ instead of $ld$. 
      \item We retain the demands and capacities of all other edges as they are in the original network given in \cref{fig:network1}.
  \end{itemize}
  Let the modified network be $H'$.
If the network $H'_{(\Delta_t,0)}$ admits a feasible flow, we  compute a matching $M$ corresponding to a min-cost flow. By the correctness of our algorithm  for \RSMTD, $M$ is a rank-maximal matching among all matchings with total deviation $\Delta_t$. By the construction of the network  $H'_{(\Delta_t,0)}$, it follows that $d_{tot}(M)=\Delta_t$, and $d_{max}(M)=\Delta_m$. Thus, $M$ is a solution to the problem \RMMTM.   If the network $H'_{(\Delta_t,0)}$ does not admit any feasible flow, then $\mathcal{M}_t$ and $\mathcal{M}_m$ are disjoint.

\paragraph{Algorithm for \RMMTMD }
In the problem \RMMTMD, the goal is to compute a matching that minimizes the  maximum deviation from the set $\mathcal{M}_t$. Observe that any matching in the set $\mathcal{M}_t$ has a maximum deviation of at least $\Delta_m$ and at most $\Delta_t$. Therefore, we perform  a linear search in this range to find the optimal value of maximum deviation.
For a given value of maximum deviation, say $i \in [\Delta_m, \Delta_t]$, we apply the following modification to  the network given in \cref{fig:network1}.
\begin{itemize}
  \item We set $ud=0$ and $ld=\Delta_t$. Thus, we restrict our attention to matchings with total deviation   $\Delta_t$ (matchings in the set $\mathcal{M}_t$).
      \item For every post $p \in \B$, we set the capacity of the edge $(d',p')$ as $i$ instead of $ld$,  where $i \in [\Delta_m, \Delta_t]$. 
      \item We retain the demands and capacities of all other edges as they are in the original network given in \cref{fig:network1}.
  \end{itemize}
  Let the modified network be $H'$. For a particular value of $i$, if the network $H'_{(\Delta_t,0)}$ admits a feasible flow, then $i$ is a candidate value for the  maximum deviation. We perform a linear search on $i$ to compute a matching in $\mathcal{M}_t$ having the smallest value of maximum deviation. With $O({\Delta_t}^2)$ flow computations in the modified network $H'$, we can compute a solution to the problem. Since the value of $\Delta_t$ is polynomial in $\abs{\A}$, the algorithm runs in polynomial time.

\paragraph{Algorithm for \RMMMTD }
In the problem \RMMMTD, the goal is to compute a matching  that has the minimum value of total deviation from the set $\mathcal{M}_m$. Observe that any matching in the set $\mathcal{M}_m$ has a total deviation of  at least $\Delta_m$ and at most $\min\{ \abs{\B} \cdot \Delta_m, \sum\limits_{p \in \B} \ell(p) \}$.  Therefore, we perform a linear search within this range to find the optimal value of total deviation. We apply the following modification to  the network given in \cref{fig:network1}.
\begin{itemize}
  \item We set $ud=0$ and $ld=\Delta$.
      
      \item For every post $p \in \B$, we set the capacity of the edge $(d',p')$  to $\Delta_m$ instead of $ld$.
      \item We retain the demands and capacities of all other edges as they are in the original network given in \cref{fig:network1}.
  \end{itemize}
  Let the modified network be $H'$. We now apply our algorithm for \RSMTD on the modified network.  For a given value of $\Delta$,
  if the network $H'_{(\Delta,0)}$ admits a feasible flow, then $\Delta$ is a candidate value for the total deviation. We perform a linear search on $\Delta$  to compute a matching in $\mathcal{M}_m$ having the smallest value of total deviation.
Note that the time complexity of the above algorithm is same as that of the algorithm for    {\sf RMM-SIGN-MIN-TOT}.

%% file: 7-generalizedSQS.tex
\section{Generalized Soft Quota Setting} \label{gsqs}

Having considered the problems in the soft quota setting, we now generalize our framework.
 In the soft quota setting, deviating  by one unit from the given targets incurs a uniform  deviation {\em cost} across all the posts. A natural generalization of the soft quota setting involves associating costs  with these deviations.  In an instance  $G$ of the { \em generalized soft quota setting},  every post $p$ is associated with two non-zero integral costs -- a lower deviation cost $c_{\ell}(p)$, and an upper deviation cost $c_{u}(p)$. These values represent  the cost of  deviating by one unit from the lower target and the upper target, respectively.  

As in the soft quota setting, our goal in the  generalized soft quota setting is to compute an assignment that has two orthogonal optimization objectives:

\noindent\textbf{(i) Minimizing deviation cost.} For a matching $M$, we define the total deviation cost $d_{tc}(M)$, and the maximum deviation cost $d_{mc}(M)$ as below, and our goal is to minimize one of them. Let $\B_{\ell}$ and $\B_u$ denote the posts that deviate from their lower targets and upper targets, respectively, in $M$. Recall that a post $p$ can deviate either from the lower target or from the upper target, but  not both. Thus, the sets $\B_{\ell}$ and $\B_u$ are disjoint.
\begin{align*}
    d_{tc}(M)=\sum\limits_{p \in \B_{\ell}} d_M(p) \cdot c_{\ell}(p) +\sum\limits_{p \in \B_{u}} d_M(p) \cdot c_u(p)
\end{align*}
 and 
 \begin{align*}
     d_{mc}(M)=\max\{\max_{p \in \B_{\ell}} \ d_M(p) \cdot c_{\ell}(p), \ \max_{p \in \B_u} \ d_M(p) \cdot c_u(p)\}
 \end{align*}
 
\noindent\textbf{(ii) Optimality with respect to preferences.} As done earlier, we use rank-maximality and fairness as our notions of optimality.

\subsection{Our Problems and Results}
In the generalized soft quota setting, we care about deviation cost as well as preference optimality, and this defines our problems.

\noindent\textbf{Deviation cost first}. In this type of problem, we give precedence to minimizing the deviation cost first, and subject to that, we optimize the preferences. Let $\mathcal{M}$ be the set of all matchings in an instance $G$ of a generalized soft quota setting. Let $\mathcal{M}_{tc}$ (respectively $\mathcal{M}_{mc}$) denote the set of matchings that have a minimum value of total deviation cost (respectively, the minimum value of maximum deviation cost). We define our first set of problems:
\begin{itemize}
    \item {\sf OPT-MIN-TOT-DC}: The goal is to compute a matching that satisfies the notion $OPT$ with respect to the preferences in the set $\mathcal{M}_{tc}$.
    \item {\sf OPT-MIN-MAX-DC}: The goal is to compute a matching that satisfies the notion $OPT$ with respect to the preferences in the set $\mathcal{M}_{mc}$.
\end{itemize}

\noindent\textbf{Preference optimality first}. This type of problem has a signature requirement $\rho$ as part of the input, and we consider all matchings in the generalized soft quota setting that satisfy the signature requirement with respect to the preference optimality under consideration. Let $\mathcal{M}_{OPT}$ be the set of all matchings that meet the signature requirement. Let $\mathcal{M}_{tc}'$ (respectively $\mathcal{M}_{mc}'$) denote the set of matchings that have a minimum value of total deviation cost (respectively, the minimum value of maximum deviation cost) in $\mathcal{M}_{OPT}$. We define the following problems:
\begin{itemize}
    \item {\sf OPT-SIGN-MIN-TOT-DC}: The goal is to output any matching in $\mathcal{M}_{tc}'$.
    \item {\sf OPT-SIGN-MIN-MAX-DC}: The goal is to output any matching in $\mathcal{M}_{mc}'$.
\end{itemize}

For the maximum deviation cost variants, as well as for the {\sf OPT-MIN-TOT-DC} problem, we show that our algorithms in the soft quota setting can be extended to solve the analogous problems in the generalized soft quota setting. For the problem {\sf OPT-SIGN-MIN-TOT-DC}, we are not aware of any hardness or algorithmic results. We show polynomial time algorithms for solving special cases of the problem {\sf OPT-SIGN-MIN-TOT-DC}.

\subsection{Minimizing maximum deviation cost}

To solve the {\sf MIN-MAX-DC} type problems in the generalized soft quota setting, namely {\sf OPT-MIN-MAX-DC} and {\sf OPT-SIGN-MIN-MAX-DC}, we use a reduction to problems in the fixed quota setting. We describe the algorithms for the notion of rank-maximality. 

We begin by presenting a useful observation about an optimal solution to the problem {\sf RMM-MIN-MAX-DC}. 
\begin{obs} \label{obs:gsq1}
    Let $M^*$ be an optimal solution to {\sf RMM-MIN-MAX-DC}, and let $C^*$ be the optimal maximum deviation cost. Then, there exists a post $p$ such that, in $M^*$, the lower deviation cost of $p$ is exactly equal to $C^*$.
    \end{obs}
    \begin{proof}
    Suppose not. Then the lower deviation cost of every post with respect to $M^*$ is strictly less than $C^*$. Let the maximum lower deviation cost of any post with respect to $M^*$ be $C'$. Since $d_{mc}(M^*)=C^*$, there exists at least one post $p \in \B$ whose upper deviation cost is $C^*$.  Let $\B_u$ denote the set of posts that deviate from their upper target with respect to  $M^*$ such that $\abs{M^*(p)}>u(p) + \Big\lfloor \dfrac{C'}{c_{u}(p)} \Big\rfloor$. 
    
    We construct a matching $M'$ from $M^*$
as follows. For every post $p \in \B_u$, we arbitrarily select $\abs{M^*(p)} - u(p) - \Big\lfloor \dfrac{C'}{c_{u}(p)} \Big\rfloor$ matched edges incident on $p$ and delete them. We retain all other matched edges in $M^*$. Let the resulting matching be $M'$. Observe that, for any post $p$, the upper deviation cost with respect to $M'$ is at most $C'$. Since we retained all the matched edges incident on  a post $p \not\in \B_u$ in $M'$,  the lower deviation cost of $p$ with respect to $M'$ is also at most $C'$. Thus, $d_{mc}(M')=C' < C^*$, which contradicts the fact that $M^*$ is a matching with minimum maximum deviation cost.
    \end{proof}
    
Next, we describe the reduction from generalized soft quota setting to the fixed quota setting.

\paragraph{ Reduction from generalized soft quota setting to the fixed quota setting} Given an instance $G$ of the generalized soft quota setting, and a fixed value of cost $C$, we  construct a fixed quota setting instance $H_C$ as follows. In $H_C$, the applicants, posts, and applicant preferences remain the same as those of $G$. For a post $p$, we set the  lower quota as $q^-(p)=\max\{0,\ell(p)-\Big\lfloor \dfrac{C}{c_{\ell}(p)} \Big\rfloor\}$, and the upper quota  as $q^+(p)=u(p)+\Big\lfloor \dfrac{C}{c_{u}(p)} \Big\rfloor$. Note that if $H_C$ admits a feasible matching, then 
the same matching is feasible for a fixed quota setting instance $H_{C'}$ for any cost $C'\ge C$. This is because an increase in the cost widens the range of acceptable numbers of applicants for a post.
     
     Let $C_{\ell}^*= \max\limits_{p \in \B}c_{\ell}(p)$ and $C_{u}^*= \max\limits_{p \in \B}c_{u}(p)$. Next, we present the algorithms for {\sf RMM-MIN-MAX-DC} and {\sf RMM-SIGN-MIN-MAX-DC}.
   
    \paragraph{Algorithm for {\sf RMM-MIN-MAX-DC}} Given an instance $G$ of the generalized soft quota setting, our goal is to compute a rank-maximal matching among all matchings that have a minimum value of maximum deviation cost. 
    Under the assumption about $\ell(p)$, the maximum deviation of any post is upper bounded by $\abs{\A}$. From Observation \ref{obs:gsq1}, the maximum deviation cost of any matching is  upper bounded by $\abs{\A} \cdot C_{\ell}^*$. Therefore, the maximum deviation cost of any matching lies in the range $[0, \abs{\A} \cdot C_{\ell}^*]$.
    To compute the minimum value of maximum deviation cost, we start by guessing the maximum deviation cost value $C$ and checking whether the corresponding  fixed quota setting instance $H_C$ admits a feasible matching.  To check feasibility, we can ignore the preferences and use the standard flow network for the bipartite matching with demands on the edges.  We perform a binary search on $C$ in the range $[0, \abs{\A} \cdot C_{\ell}^*]$ to compute the smallest $C$ such that there exists  a feasible solution. Let $C^*$ be the minimum value of $C$ obtained. Then, corresponding to $C^*$, we compute rank-maximal matching among all feasible matchings (\RMMLQ) in the fixed quota setting instance $H_{C^*}$.
    By the definition of \RMMLQ, the resulting matching has the best possible signature with respect to the notion of rank-maximality while achieving  the minimum value for the maximum deviation cost.

   \paragraph{Algorithm for {\sf RMM-SIGN-MIN-MAX-DC}} Given an instance $G$ of the generalized soft quota setting and an input signature $\rho$, our goal is to compute a matching $M$ in $\mathcal{M}_{\mathcal{R}}$ (recall that $\mathcal{M}_{\mathcal{R}}$ is the set of matchings whose signatures are at least as good as $\rho$ with respect to rank-maximality) that has a minimum value of maximum deviation cost. Under the assumption about $\ell(p)$,  lower deviation of any post is upper bounded by $\abs{\A}$. Similarly, the  upper deviation of any post is upper bounded by $\abs{\A}$.  Therefore, the maximum deviation cost of any matching lies in the range $[0, \abs{\A} \cdot (\max \{C_{\ell}^*, C_u^*\})]$. To compute the minimum value of maximum deviation cost for a matching in $\mathcal{M}_{\mathcal{R}}$, we
 start by guessing the maximum deviation cost value $C$ and check whether  the corresponding  fixed quota setting instance $H_C$ admits a feasible matching. If the instance $H_C$ admits a feasible matching, we compute a rank-maximal matching $M$ (\RMMLQ) among all feasible matchings in $H_C$. If $\sigma(M)\succeq_{\mathcal{R}} \rho$, then $M$  satisfies the signature requirement with respect to rank-maximality. 
      We perform a binary search on $C$ in the range $[0, \abs{\A} \cdot (\max \{C_{\ell}^*, C_u^*\})]$ to compute the smallest $C$ such that there exists   a matching that meets the signature requirement. 

\subsection{Algorithm for {\sf RMM-MIN-TOT-DC}}
Given an instance $G$ of the generalized soft quota setting, our goal is to compute a rank-maximal matching among all matchings having the minimum value of total deviation cost. We observe that in an optimal matching $M^*$, no post incurs a deviation cost on the upper target. 

For a given instance $G$ of the generalized soft quota setting, let  $G_x$ denote the corresponding fixed quota setting instance, where for a post $p$, we set the lower quota $q^-(p)=0$ and upper quota $q^+(p)=u(p)$. Let  $G'$ be the generalized soft quota setting instance obtained from $G$, where for every post $p$, we restrict $\ell'(p)=u'(p)=\ell(p)$.
Recall that in the case of uniform costs for all posts, any maximum matching in $G'_x$ is a minimum total deviation matching in $G'$ by Lemma~\ref{lem:mmsq}. 
However, the analogous claim need not be true in the generalized soft quota setting, as illustrated in the following example.

\begin{figure*}[ht]
     \centering
     \begin{subfigure}[b]{0.15\textwidth}
         \centering
          \begin{align*}
    a_1: \ & p_1, \ p_2 \\
    a_{2}: \ &p_1, \ p_2
\end{align*} 
         % }
         % \vspace{-0.35in}
         \caption{}
         \label{fig:pref-gsq}
     \end{subfigure}     
     % \hspace{0.25in}
     \begin{subfigure}[b]{0.3\textwidth}
         \centering
         \begin{align*}
         p_1 \ &[2,2,1,y]\\
         p_2 \ &[1,1,10,y']
    % \ell(p_1)=u(p_1)=2, && c_{\ell}(p_1)&=1 \\
    % \ell(p_2)=u(p_2)=1,  && c_{\ell}(p_2)&=10
\end{align*}
         \caption{ }
         \label{fig:cost-gsq}
     \end{subfigure}
     \begin{subfigure}[b]{0.45\textwidth}
         \centering
         \begin{align*}
    M_1=\{(a_1,p_2), (a_2,p_1)\},&& d_{tc}(M_1)&=1\\
    M_2=\{(a_1,p_1), (a_2,p_1)\},&& d_{tc}(M_2)&=10
\end{align*}
         \caption{ }
         \label{fig:match-gsq}
     \end{subfigure}
        \caption[Example to illustrate a maximum matching in $G'_x$ need not be a minimum total deviation cost matching in $G'$]{An instance $G$ of generalized soft quota setting  with two applicants and two posts. \cref{fig:pref-gsq} shows the edge set of $G$. Each post has four values -- lower target,  upper target, lower deviation cost and the upper deviation cost as shown in \cref{fig:cost-gsq}. Two matchings $M_1$ and $M_2$ in $G$ are shown in \cref{fig:match-gsq} along with their total deviation cost.  }
        \label{fig:eg-tdc-gsq}
\end{figure*}

An instance $G$ of generalized soft quota setting  with two applicants $\{a_1, a_2\}$ and two posts $\{p_1,p_2\}$ is given in \cref{fig:eg-tdc-gsq}. The edge set of $G$ forms a complete bipartite graph, as shown in \cref{fig:pref-gsq}. The applicant preferences are shown in Figure~\ref{fig:pref-gsq}; however, they are not needed for the illustration. Each post in $G$ has four values -- lower target,  upper target, lower deviation cost and the upper deviation cost as shown in \cref{fig:cost-gsq}. While the upper deviation cost values $y$ and $y'$ can take any non-zero integer values, they are not required for the illustration.  Since,  for every post $p$, we have $\ell(p)=u(p)$ in $G$, it follows that $G=G'$.  Consider the  matchings $M_1$ and $M_2$ in $G'$, along with their respective total deviation costs, as given in \cref{fig:match-gsq}. Both $M_1$ and $M_2$ are  maximum matchings in the corresponding fixed quota setting instance $G'_x$. Although the deviations of both the matchings are the same, their deviation costs are different.  

Recall that in the soft quota setting, the minimum values of total deviation in $G'$ and $G$ are the same, as established in Lemma~\ref{cl:clgg'}. Using the same arguments, we obtain an  analogous result in the generalized soft quota setting stated as Lemma~\ref{lem:gsq4}. 
\begin{lemma}\label{lem:gsq4}
Minimum value of total deviation cost in $G'$ and $G$ is the same.
\end{lemma}

%Proof of Lemma~\ref{lem:gsq4} is a straightforward generalization of the proof of Lemma~\ref{cl:clgg'}.
% \begin{proof}
% Let $k'$ be the total deviation cost of a matching that achieves the minimum value of total deviation cost in $G'$. Note that $k'$ is the sum of lower deviation costs of all the posts in $G'$. Suppose that, the minimum value of the total deviation cost in $G$ be $k$ where $k<k'$. Let $M$ be a matching that achieves the minimum value of total deviation cost in $G$. Note that $k$ is the sum of lower deviation costs of all the posts in $G$.

% In the generalized soft quota setting instance $G'$, some posts may incur violation on the upper target with respect to the matching $M$. This is because, in $G'$, the upper target of a post $p$ is $u'(p)=\ell(p)$ which is at most the corresponding upper target $u(p)$ in $G$. Now, we construct a matching $M'$ in $G'$, where for every post, we remove $\max\{0, \abs{M(p)}-\ell(p)\}$ number of applicants which are matched to $p$. Note that in $G'$, no post incurs a violation on the upper target with respect to the matching $M'$. Thus, the total deviation cost of $M'$ in $G'$ is $k$, which is the total deviation cost of $M$ in $G$. Thus, it contradicts the fact that $k'$ is the minimum value of total deviation cost in $G'$
% \end{proof}

As illustrated by the example in \cref{fig:eg-tdc-gsq}, the approach in  \cref{sec:alg-min-tot-d} cannot be applied directly, and we propose the following modification. To compute a {\sf RMM-MIN-TOT-DC} matching in $G$, our idea is the following.
\begin{itemize}
    \item We construct an instance $G'$ from $G$, where for every post $p$, $\ell'(p)=u'(p)=\ell(p)$. While the construction  of $G'$ remains unchanged,  we focus on the set of matchings that minimize total deviation cost, and select a rank-maximal matching  $M$ from this set. Note that $M$ is a solution to {\sf RMM-MIN-TOT-DC}  in $G'$.
    
    \item We augment the matching $M$ in the corresponding fixed quota setting instance $G_x$ (where $q^-(p)=0$, and $q^+(p)=u(p)$) to obtain an {\sf RMM-MIN-TOT-DC} matching $M^*$ in $G$. This procedure is the same as step~(B) of the algorithm for the \RMMTD problem given in \cref{sec:alg-min-tot-d}.
\end{itemize}

Next, we describe the procedure for computing a rank-maximal matching among all matchings with the minimum total deviation cost.
\subsubsection{Computing {\sf RMM-MIN-TOT-DC} in \texorpdfstring{$G'$}{G'}}
Given the instance $G'$, we reduce the problem to computing a min-cost max-flow in a suitably designed flow network $H$. A matching $M$ corresponding to a min-cost max-flow in $H$ has the following two properties:
\begin{itemize}
    \item $M$ is a matching having the minimum value of total deviation cost in $G'$ and also in $G$.
    \item $M$ is a rank-maximal matching among all matchings with minimum total deviation cost in $G'$. However, $M$ need not be {\sf RMM-MIN-TOT-DC} in $G$.
\end{itemize}
The construction of the network $H$ is similar to that described in step~(A) of the algorithm for the \RMMTD problem in the soft quota setting, except for the cost assignment on the edges (see \cref{sec:alg-min-tot-d} for the network description). The costs are assigned as follows:
\begin{itemize}
    \item For an edge $(a,p)$ in $G'$ of rank $i$, we set cost of the corresponding edge in $\mathcal{E}_{AP}$ to $-((n+1)^r+(n+1)^{r-i})$. This cost assignment is the same as that used in the \RMMTD problem.
    \item  For an edge $(d,p)$ in $\mathcal{E}_{DP}$, we set the cost as $(n+1)^{r+1}\cdot c_{\ell}(p)$, where $c_{\ell}(p)$ is the corresponding lower deviation cost. Recall that, in the \RMMTD problem, we set the cost of these edges as $0$.
    \item For the remaining edges in $H$, we set the cost to $0$. This is the same as in the \RMMTD problem.
\end{itemize}
There is a natural correspondence between a flow $f$ in $H$ and a matching $M$ in $G'$ as follows:
\begin{align*}
    M=\{(a,p)\ | \ f(a,p)=1, \ (a,p) \in  \mathcal{E}_{AP}\}.
\end{align*}  
 Note that, by our construction, a max-flow in $H$ has value $\sum\limits_{p \in \B} \ell(p)$.
We define the total deviation cost of a max-flow $f$, which is different from the total cost $cost(f)$ of the max-flow. 
\begin{definition}[Total deviation cost of a max-flow ] Total deviation cost of a max-flow $f$ in $H$ is defined as
\begin{align*}
    d_{tc}(f)=\dfrac{\sum\limits_{p \in \B}d_f(p)}{(n+1)^{r+1}}
\end{align*}
where $d_f(p)=\sum\limits_{d \in D_p} f(d,p)\cdot (n+1)^{r+1} \cdot c_{\ell_p}$ is the total deviation cost via $p$ through the dummy applicants.    
\end{definition}
We now focus on matchings in $G'$ that do not deviate from the upper target. Such matchings correspond to  feasible matchings in  $G'_x$. For any matching $M$ in $G'_x$, we can always construct a max-flow $f$ in $H$ such that $d_{tc}(M)=d_{tc}(f)$. Similarly, for any max-flow $f$ in $H$, we can construct a matching $M$ in $G'$ that does not deviate from the upper target. Therefore, we can restrict our attention to max-flows in $H$.

\paragraph{Cost of a max-flow in $H$ and total deviation cost of a matching in $G'$} By the design of our network, we ensure that the cost of a max-flow in $H$ corresponds to the total deviation cost of the corresponding matching in $G'$. This is proved in Lemma~\ref{lem:gsq1}.

\begin{lemma} \label{lem:gsq1}
    Let $f,f'$ be two max-flows in $H$, and let $M_f,M_{f'}$ be the corresponding matchings in $G'$. If $d_{tc}(M_f)<d_{tc}(M_{f'})$, then $cost(f)<cost(f')$.
\end{lemma}
\begin{proof}
For a post $p$, the incoming flow is via edges in $\mathcal{E}_{DP}$, and $\mathcal{E}_{AP}$. In $H$, only the edges in  $\mathcal{E}_{AP}$ and $\mathcal{E}_{DP}$ have a non-zero cost.
Due to the multiplicative factor $(n+1)^{r+1}$ in the cost of edges in $\mathcal{E}_{DP}$, and the fact that $d_{tc}(M_f)=d_{tc}(f)$,
the contribution of edges in $\mathcal{E}_{DP}$ to the cost of the flow $f$ is $d_{tc}(M_f) \cdot n^{r+1}$.
Recall the cost of an edge in $\mathcal{E}_{AP}$. 
Observe that cost of the flow $f$ can be upper bounded by
$$cost(f) \leq \abs{M_f}(-(n+1)^r) - \abs{M_f}+d_{tc}(M_f)\cdot (n+1)^{r+1},$$ where the equality is achieved  when all edges are of rank $r$. Similarly, the  cost of the flow $f'$ can be lower bounded by  $$cost(f') \geq \abs{M_{f'}}(-(n+1)^r) - \abs{M_{f'}}((n+1)^{r-1})+d_{tc}(M_{f'})\cdot (n+1)^{r+1},$$ where the equality is achieved when all edges are of rank $1$. 

We show that $\abs{M_f} \ge \abs{M_{f'}}$. Suppose, for the sake of contradiction, assume that $\abs{M_f} < \abs{M_{f'}}$. Consider the contribution of edges in $\mathcal{E}_{DP}$ to the costs of the flows $f$ and $f'$. Observe that there are $\sum\limits_{p \in \B} \ell(p)-\abs{M_f}$ terms with a coefficient of $(n+1)^{r+1}$ in   $d_{tc}(M_f) \cdot (n+1)^{r+1}$. Similarly, there are $\sum\limits_{p \in \B} \ell(p)-\abs{M_{f'}}$ terms with a coefficient of $(n+1)^{r+1}$ in   $d_{tc}(M_{f'}) \cdot (n+1)^{r+1}$. Since $\abs{M_f} < \abs{M_{f'}}$, we have $d_{tc}(M_f)\cdot (n+1)^{r+1}> d_{tc}(M_{f'})\cdot (n+1)^{r+1}$, which contradicts the fact that $d_{tc}(M_f)<d_{tc}(M_{f'})$. 

Using the facts   $d_{tc}(M_f)<d_{tc}(M_{f'})$, and $\abs{M_f} \ge \abs{M_{f'}}$,  together with the same arguments as in the proof of \cref{cl:td-rmmlq4-1}, we conclude  that $cost(f)<cost(f')$.
\end{proof}
  % Let $f,f'$ be two max-flows in $H$ with $cost(f)<cost(f')$, then $d_{tc}(f)\leq d_{tc}(f')$
  
\begin{lemma}\label{lem:gsq2}
    Let f be a min-cost flow in $H$, and let $M_f$ be a corresponding matching in $G'$. Then $M_f$ is a matching with minimum total deviation cost in $G'$. 
\end{lemma}
\begin{proof}
    Suppose not. Let $M^*$ be a matching with minimum total deviation cost in $G'$, and let $f^*$ be the corresponding flow. Since $d_{tc}(M^*)<d_{tc}(M_f)$, it follows from  \cref{lem:gsq1} that $cost(f^*)<cost(f)$, which contradicts the fact that $f$ is a min-cost flow.
\end{proof}

Given the connection between the total deviation cost of $G$ and $G'$ established in Lemma~\ref{lem:gsq4}, which is independent of the preferences, we know that the  minimum total deviation cost in $G$ is equal to that in $G'$. This leads to the following corollary. 
\begin{corollary}\label{cor:gsq1}
Any min-cost max-flow in $H$ corresponds to a matching with minimum total deviation cost in $G$.
\end{corollary}
We now prove the preference optimality of the matching obtained by step~(A).
\begin{lemma}
\label{lem:gsq5}
Any min-cost max-flow in $H$ corresponds to a solution to the {\sf RMM-MIN-TOT-DC} problem in $G'$.
\end{lemma}
Proof of Lemma~\ref{lem:gsq5} is a direct generalization of the proof of Lemma~\ref{cl:cl22}.

% Proof of \cref{lem:gsq5} is similar to the proof of \cref{cl:cl22} by replacing total deviation by total deviation cost.
% \begin{proof}
% Let $f$ be a min-cost max-flow in $H$ and let $M$ be the corresponding matching in $G'$. By \cref{cor:gsq1},  $M$ is a matching with minimum total deviation cost in $G'$.
% Suppose $M$ is not an {\sf RMM-MIN-TOT-C} matching in $G'$, then there exists another matching $M'$ which is a rank-maximal matching among all matchings with minimum total deviation cost in $G'$. Note that  $d_{tc}(M)=d_{tc}(M')$.
% Since $\sigma(M')\succ_{\mathcal{R}} \sigma(M)$,  by the definition of rank-maximality, there exist an index $i$ such that $\sigma(M')_i > \sigma(M)_i$ and $\sigma(M')_j = \sigma(M)_j$ for $j< i$. Recall that, corresponding to a rank $i$ edge in $G'$, we have an edge in $\mathcal{E}_{AP}$ with a cost of $-n^{r-i}$ in $H$. Thus the max-flow $f'$ corresponding to the matching $M'$ has $cost(f')<cost(f)$ which is a contradiction.  
%  \end{proof}
 
 In step~(A), we have computed a matching $M$ that is a solution to {\sf RMM-MIN-TOT-DC} problem in $G'$. Using careful augmentations, as described in the Step (B) of the \RMMTD problem in the soft quota setting, we can compute a rank-maximal matching among all matchings with minimum total deviation cost in $G$.

\subsection{Special case of {\sf RMM-SIGN-MIN-TOT-DC} }

Given an instance $G$ of the generalized soft quota setting and an input signature $\rho$, our goal is to compute a matching that satisfies the signature requirement and has a minimum value of the total deviation cost. For the general case of this problem, we are not aware of any efficient algorithm. As a starting point, we first consider a simpler scenario in which all posts have the same lower deviation cost $c_{\ell}$ and the same upper deviation cost $c_u$. By modifying the network given in \cref{fig:network1}, we can solve this problem efficiently. We then extend this approach to a special case, where posts are divided into a constant number of partitions, with each partition having a uniform lower and upper deviation cost across all its posts. We provide an efficient algorithm for this case, which we describe below.

We consider the case of the {\sf RMM-SIGN-MIN-TOT-DC} problem, where posts in $\B$ are partitioned into $\B_1 \cup \B_2 \cup \ldots \cup \B_k$. For each post $p\in \B_i$, the  cost per lower deviation is $x_i$, and the cost per upper deviation is $y_i$. To solve the {\sf RMM-SIGN-MIN-TOT-DC} problem in this case, we modify the flow network in \cref{fig:network1} as follows. 
\begin{itemize}
    \item We remove all the outgoing edges from $d'$, and  introduce nodes $d_i'$, $i \in [k]$. We add an edge from  $d'$ to $d_i'$ with vector $[ld_i,ld_i]$ (demand and capacity). 
    \item For each $i \in [k]$, we add an edge from $d_i'$  to every post $p'$ such that $ p \in \B_i$, with vector $[0,ld_i]$.
    \item We remove all incoming edges to $d''$, and  introduce nodes $d_i''$, for $i \in [k]$. We add an edge from  $d_i''$ to $d''$ with vector $[ud_i,ud_i]$. 
    \item For each $i \in [k]$, we add an edge from $p''$ to $d_i''$, for  $ p \in \B_i$,    with vector $[0,ud_i]$.
\end{itemize}

In this network, for each $i \in [k]$, the values  $ld_i$ and $ud_i$ represent the total lower deviation and the total  upper deviation, respectively, for all posts in the partition $\B_i$. A min-cost max-flow in this modified network corresponds to  a rank-maximal matching in which the total lower  and upper deviations for each partition are fixed by the values $\{(ld_i, ud_i)\}_{1 \leq i \leq k}$.

Observe that, for a given $i \in [k]$, the total deviation of all posts in $\B_i$ is at most $|\A| \cdot |\B_i|$.
Thus, the total number of possible values for $(ld_i, ud_i)$ is at most $(|\A| \cdot |\B_i| + 1)^2$. Consequently, the total number of possible combinations of deviations $\{(ld_i, ud_i)\}_{1 \leq i \leq k}$ is $\mathcal{O}(n^{4k})$ (assuming that $|\A| = \mathcal{O}(|\B|)$). Each such combination corresponds to a specific network, and for each network, if it admits a feasible flow, then we compute a rank-maximal matching, say $M$, among all feasible matchings. If $\sigma(M)\succeq_{\mathcal{R}} \rho$, then $M$ is a candidate solution. By iterating through all combinations, we compute a solution that satisfies the signature requirement and has the minimum  total deviation cost.
The time complexity of the algorithm is $\mathcal{O}(n^{4k} \cdot T(n))$, where $T(n)$ is the time taken to solve the min-cost max-flow problem.